\documentclass[12pt]{elsarticle}

\usepackage{latexsym,amsmath,amscd,amssymb,graphics,ulem}
\usepackage{enumerate}
\usepackage{cancel}
\usepackage{graphicx}
\usepackage{framed}
\biboptions{sort&compress}
\usepackage{color}
\usepackage[all]{xy}

\usepackage{etoolbox}

\makeatletter
\def\@mkboth#1#2{}
\newlength\appendixwidth
\preto\appendix{\addtocontents{toc}{\protect\patchl@section}}
\newcommand{\patchl@section}{%
  \settowidth{\appendixwidth}{\textbf{Appendix }}%
  \addtolength{\appendixwidth}{1.5em}%
  \patchcmd{\l@section}{1.5em}{\appendixwidth}{}{\ddt}%
}
\makeatother

\newcommand{\K}{\mathcal{K}}

\newtheorem{theorem}{Theorem}

\newtheorem{lemma}[theorem]{Lemma}

\newtheorem{remark}[theorem]{Remark}

\numberwithin{theorem}{section}

\def\be{\begin{equation}}
\def\ee{\end{equation}}
\def\bea{\begin{eqnarray}}
\def\eea{\end{eqnarray}}
\def\ba{\begin{array}}
\def\ea{\end{array}}

\def\bOm{\boldsymbol{\Omega}}

\def\d{\delta}

\let\<\langle
\let \>\rangle

\newcommand{\rem}[1]{}

\newcommand{\de}{\delta}

\newcommand{\bPsi}{\boldsymbol{\Psi}}
\newcommand{\bv}{\boldsymbol{v}}

\newcommand{\br}{\boldsymbol{r}}
\newcommand{\bGam}{\boldsymbol{\Gamma}}
\newcommand{\bom}{\boldsymbol{\omega}}
\newcommand{\bgam}{\boldsymbol{\gamma}}
\newcommand{\bsigma}{\boldsymbol{\Sigma}}
\newcommand{\bpsi}{\boldsymbol{\Psi}}

\newcommand{\bY}{\boldsymbol{Y}}

\newcommand{\pp}[2]{\frac{\partial #1}{\partial #2}}

\newcommand{\dede}[2]{\frac{\delta #1}{\delta #2}}
\newcommand{\prt}{\partial}

\newcommand{\lp}{\left(}
\newcommand{\rp}{\right)}

\newcommand{\mse}{\mathfrak{se}}

\newcommand{\ds}{\displaystyle}

\newcommand{\todo}[1]{\vspace{5 mm}\par \noindent
\framebox{\begin{minipage}[c]{0.95 \textwidth}
\tt #1 \end{minipage}}\vspace{5 mm}\par}
\newcommand{\revision}[2]{#2} 

\begin{document}

\begin{frontmatter} 
\title{Stability of helical tubes conveying fluid}
\author[FGB]{Fran\c{c}ois Gay-Balmaz} 
\ead{gaybalma@lmd.ens.fr }
\author[DG]{Dimitri Georgievskii} 
\ead{georgiev@mech.math.msu.edu }
\author[VP]{Vakhtang Putkaradze\corref{cor1} }
\ead{putkarad@ualberta.ca}
\cortext[cor1]{Corresponding author}

\address[FGB]{CNRS - LMD - Ecole Normale Sup\'erieure de Paris, France} 
\address[DG]{Chair of Elasticity,  Moscow State University, Leninskiye Gory, 1, Moscow, Russia, 119991}
\address[VP]{Department of Mathematical and Statistical Sciences, University of Alberta, Edmonton, AB   T6G 2G1 Canada}

\begin{abstract}
We study the linear stability of elastic collapsible tubes conveying fluid, when the equilibrium configuration of the tube is helical.   A particular case of such tubes, commonly encountered in applications,  is represented by quarter- or semi-circular tubular joints used at pipe's turning points. 
The  stability theory for pipes with non-straight equilibrium configurations, especially for collapsible tubes, allowing dynamical change of the cross-section, has been elusive as it is difficult to accurately develop the dynamic description via traditional methods. We develop a methodology for studying the three-dimensional dynamics of collapsible tubes based on the geometric variational approach. We show that the linear stability theory based on this approach allows for a complete treatment for arbitrary three-dimensional helical configurations of collapsible tubes by reduction to an equation with constant coefficients. We discuss new results on stability loss of straight tubes caused by the cross-sectional area change.  Finally, we develop a numerical algorithm for computation of the linear stability using our theory and present the results of numerical studies for both straight and helical tubes.  
\end{abstract}

\begin{keyword} 
Elastic tubes conveying fluid \sep collapsible tubes \sep helical equilibria \sep variational methods
 \sep linear stability
\end{keyword} 

\end{frontmatter} 
\normalem

\tableofcontents
\section{Background of the studies in dynamics of flexible tubes conveying fluid} 

The dynamics of tubes conveying  fluid poses many interesting problems in both applied and fundamental mechanics, in addition to its practical importance for engineering applications. For such systems, an instability appears when the flow rate through the tube exceeds a certain critical value. While this phenomenon has been known for a very long time, the quantitative research in the field started around 1950 \cite{AsHa1950}. Benjamin \cite{Be1961a,Be1961b} was perhaps the first to formulate a quantitative theory for the 2D dynamics  of the {\it initially straight tubes} by considering a linked chain of tubes conveying fluids and using an augmented Hamilton principle of critical action that takes into account the momentum of the jet leaving the tube. A continuum equation for the linear disturbances was then derived as the limit of the discrete system. This linearized equation for the initially straight tubes was further considered by Gregory and Pa\"idoussis \cite{GrPa1966a}. 

These initial developments formed the basis for further stability analysis of this problem for finite, initially straight tubes \cite{Pa1970,PaIs1974,Pa1998,ShMi2001,DoLa2002,PaLi1993,Pa2004,AkIvKoNe2013,AkGeNe2015,AkGeNe2016}. The linear stability theory has shown a reasonable agreement with experimentally observed onset of the instability \cite{GrPa1966b,Pa1998,KuSa2005,CaCr2009,Cr-etal-2012}. Nonlinear deflection models  were also considered in \cite{SeLiPa1994,Pa2004,MoPa2009,GaPaAm2013}, and the compressible (acoustic) effects in the flowing fluid in \cite{Zh2008}. Alternatively,  a more detailed 3D theory of motion was developed in \cite{BeGoTa2010} and extended in \cite{RiPe2015}, based on a modification of the Cosserat rod treatment for the description of elastic dynamics of the tube, while keeping the cross-section of the tube constant and orthogonal to the centerline. In particular, \cite{RiPe2015} analyzes several non-straight configurations, such as tube hanging under the influence of gravity,  both from the point of view of linear stability and nonlinear behavior. Unfortunately, this Cosserat-based theory could not easily incorporate the effects of the cross-sectional changes in the dynamics. Some authors have treated the instability from the point of view of the {\it follower force approach}, which treats the system as an elastic beam, ignoring the fluid motion,  with a force that is always tangent to the end of the tube. Such a force models the effect of the jet leaving the nozzle \cite{BoRoSa2002}. However, once the length of the tube becomes large, the validity of the follower force approach has been questioned, see  \cite{El2005} for a lively and thorough discussion.
For the history of the development of this problem in the Soviet/Russian literature, we refer the reader to the monograph \cite{Sv1987} (still only available in Russian). To briefly touch upon the developments in Russian literature that have been published in parallel with their western counterparts, we refer the reader to the selection of papers \cite{Mo1965,Mu1965,Il1969,Anni1970,VoGr1973,Sv1978,Do1979,Ch1984,So2005,AmAl2015}.

Because of its importance for practical applications, the theory of curved pipes conveying fluid has been considered in earlier works in some detail. The equations of motion for the theory were derived using the balance of elastic forces from tube's deformation and  fluid forces acting on the tube when the fluid is moving along a curved line in space.
In the western literature, we shall mention the earlier work \cite{Ch1972}, followed with more detailed studies \cite{MiPaVa1988a,MiPaVa1988b,DuRo1992} which developed the theory suited for both extensible and inextensible tubes and discussed the finite-element method realization of the problem. We shall also mention \cite{DoMo1976,AiGi1990} deriving a variational approach for the planar motions of initially circular tubes, although the effect of curved fluid motion was still introduced as extra forces through the Lagrange-d'Alembert principle.  In the Soviet/Russian literature,   \cite{Sv1978} developed the rod-based theory of oscillations and \cite{So2005} considered an improved treatment of forces acting on the tubes. Most of the work has been geared towards the understanding of the planar cases with in-plane vibrations as the simplest and most practically relevant situations (still, however, leading to quite complex formulas). 

In spite of considerable progress and understanding achieved so far, we believe that there is still much room for improvement in the theoretical treatment of the problem. In particular, the derivation of the theory based on the balance of forces is not variational and the approximations of certain terms tend to break down the intrinsic variational structure of the problem. In contrast, the theory of flexible tubes conveying fluid as developed in \cite{FGBPu2014,FGBPu2015} is truly variational and all the forces acting on the tube and the fluid, as well as the boundary forces are derived automatically from the variational principle. More importantly, it is very difficult (and perhaps impossible)  to extend the previous theory to accurately take into account the changes in the cross-sectional area of the tube, also called the collapsible tube case. In fact, we are not aware of any studies on the subject of stability for initially curved collapsible tubes, especially undertaken from a variational point of view. 

In previous works, the effects of cross-sectional changes have  been considered through the quasi-static approximation: if $A(s,t)$ is the local cross-section area, and $u(s,t)$ is the local velocity of the fluid, with $s$ being the coordinate along the tube and $t$ the time,  then the quasi-static assumption states that  $uA=$ const, \cite{SeLiPa1994,Pa2004,MoPa2009,GhPaAm2013}. Unfortunately, this simple law is not correct in general and should  only be used for steady flows. This problem has been addressed by two of the authors of this paper in \cite{FGBPu2014,FGBPu2015}, where a geometrically exact setting for dealing with a variable cross-section was developed and studied, showing the important effects of the cross-sectional changes on both linear and nonlinear dynamics. 
The nonlinear theory was derived from a variational principle in a rigorous geometric setting and for general Lagrangians. It can incorporate general boundary conditions and arbitrary deviations from equilibrium in the three-dimensional space.  From a mathematical point of view, the Lagrangian description of these systems involves both left-invariant (elastic) and right-invariant (fluid) quantities.  
The theory derived in \cite{FGBPu2014,FGBPu2015} further allowed consistent variational approximations of the solutions, both from the point of view of deriving simplified reduced models and developing structure preserving numerical schemes \cite{FGBPu2016}. 

In this work, we undertake a detailed study of the fully three dimensional vibrations for the problem when the 
equilibrium  spatial configuration of the centerline for the tube is helical, and the cross-sectional area of the tube is allowed to change. Since a circular arc is a particular case of a helix, the linear stability of a tube with centerline having a circular arc can be considered as a particular case of our studies. The geometric approach underpinning the theory developed in \cite{FGBPu2014,FGBPu2015} considers the dynamics in the framework of the group of rotations and translations. This, in turn, allows for the complete analysis of the stability of an initially helical tube by reducing it to a system of equations with constant coefficients.  To put it in simpler terms, the geometric framework unifies the concept of the stability analysis of the initially helical and straight tubes. Of course, the stability analysis of the helical tubes is much more complicated as compared to the straight ones; nevertheless, a substantial analytic progress can still be achieved in the more complex case of initially helical tube as well, which is precisely the focus of this paper. 

{\section{Mathematical preliminaries and background of the variational method }}
\label{sec:prelim} 

\subsection{Introduction to geometric variational methods} 

In this Section, we shall outline the background of the method and introduce some useful notations. We will try to make this Section self-consistent so the reader unfamiliar with the variational methods can follow the derivation of Section~\ref{sec:varintro} below without difficulty. We believe that such an introduction is important, as the notations employed in this paper differ from those employed in previous literature on the subject, even though in spirit we are following the variational approach already employed by Benjamin \cite{Be1961a}. However, the three-dimensionality of the motion of the tube and the conservation law of fluid volume necessitates some new notations and ideas that, as far as we are aware of, have not been previously discussed in the literature, apart from our papers \cite{FGBPu2014,FGBPu2015}. While one can get quite far using the common approach of balancing forces and torques acting on the tube for the consideration of simpler situations and geometries, the case of cross-sectional changes, in our opinion, cannot be reliably treated in this way. On the contrary, variational methods provide automatically the force and torque balances through a well-established formal procedure. As we outline in this paper, minimum assumptions are needed  for derivation of the equations of motion, such as the existence of a Lagrangian describing the flow without the necessity to specify the forms of elastic energy and types of deformations. The most crucial advantage of variational methods lies in the ability to consistently treat the three-dimensional dynamics and incorporate the changing cross-section for time-dependent flow. We do not believe that such a result is possible using the force and torque balances, as the terms arising from the changing cross-section involve a pressure-like contribution with a form that is impossible to guess a priori. 
 This Section provides a pedagogical introduction to our method, introduces some notations, and explains the differences between our approach and the one used before by other authors.

 \subsection{Rigid body equation} 
Consider a mechanical system with a configuration space $Q$, position and velocity coordinates $(q,\dot q)$, and with a Lagrangian function $L(q,\dot q)$. It is well-known that the equations of motion, \textit{i.e.}, the Euler-Lagrange equations, for such a mechanical system can be derived through the Hamilton critical action principle 
\begin{equation} 
\de \int_{t_0}^{t_1} L(q,\dot q) \mbox{d} t =0 \quad 
\Longleftrightarrow \quad 
\frac{\mbox{d} }{\mbox{d} t } \pp{L}{\dot q} - \pp{L}{q} =0 \, , 
\label{Crit_action} 
\end{equation} 
for variations $\de q$ satisfying $\de q (t_0) = \de q (t_1) =0$. Non-conservative forces $F$ (for example, friction forces), can also be introduced by addition of the term $\int_{t_0}^{t_1} F \cdot \de q\,\mbox{d} t$ into the variation \eqref{Crit_action}, called the Lagrange-d'Alembert principle for external forces, which should not be confused with the Lagrange-d'Alembert used for nonholonomic constraints \cite{Ho2008,Bloch2003}.

While the method described by equations \eqref{Crit_action} is elegant and widely used, it often needs appropriate extensions and developments to become practical. In order to illustrate this point, let us start with the derivation of perhaps the simplest possible mechanical model, namely, the rigid body moving about its fixed center of mass in space. While such a model may seem quite detached from the scope of the paper, the reader will note that our approach uses essentially the same method in spirit, so the understanding of this problem is useful for further study. A rigid body position is described by a $3 \times 3$ orientation matrix $\Lambda$ satisfying $\Lambda^T \Lambda=\Lambda \Lambda^T={\rm Id}_{3 \times 3}$, or, in other words, the configuration space $Q$ of a rigid body is the group $SO(3)$ of rotation matrices. A Lagrangian depending on the configurations and velocities can be constructed and has the form $L(\Lambda, \dot \Lambda)$. A naive application of the method \eqref{Crit_action} will lead to the Euler-Lagrange equations for $9$ matrix coordinates of $\Lambda$, coupled with $6$ constraints coming from $\Lambda \Lambda^T={\rm Id}_{3 \times 3}$. While the total number of equations is $3$, as expected, the equations of motions obtained by this method are excessively complex. One can parameterize the group $SO(3)$ using, for example, three Euler angles, in which case \eqref{Crit_action} will give highly non-intuitive equations for these angles. It is however known, since the time of Euler, that such an approach is not fruitful. Instead, Euler has derived elegant equations of motion by going to the variables of angular velocity which we today call the symmetry-reduced variables. In 1901, Poincar\'e \cite{Po1901} has carried out a modern derivation of these equations which we will briefly outline here. 
 
The key to Poincar\'e's method is to notice that since the whole system is invariant with respect to arbitrary rotations of space, the Lagrangian should also be invariant with respect to such rotations. More precisely, for any fixed rotation matrix $A \in SO(3)$, we have $L(A \Lambda, A \dot \Lambda)= L(\Lambda, \dot \Lambda)$. The fact that $\Lambda$ is multiplied from the left by $A$ comes from physics; as a rule, the dynamics of elastic and rigid bodies is left invariant. Then, the Lagrangian can be brought to a form that depends on the single variable $ \omega =\Lambda^{-1} \dot \Lambda$, called the angular velocity \emph{in the body frame}. 
 
 \subsection{Notation: vectors as antisymmetric matrices and vice versa} 
\label{sec:hatmap}

A careful reader has noticed that the object $\omega=\Lambda^{-1} \dot \Lambda$, that we have called the angular velocity,  is an antisymmetric $3 \times 3$ matrix. This can be seen by differentiating the identity for orientation matrices: 
 \begin{equation} 
 \frac{\mbox{d}}{\mbox{d} t} \Lambda^T \Lambda = {\rm Id}_{3 \times 3} \quad \Rightarrow \quad 
\dot  \Lambda^T \Lambda + \Lambda^T \dot \Lambda =0 \quad \Rightarrow \quad \omega^T + \omega=0 \, . 
 \label{SO3_identities} 
 \end{equation} 
As it turns out, these matrices are equivalent to vectors in three-dimensional space through the so-called hat map, which is defined as follows. To a given antisymmetric $3 \times 3$ matrix $\omega$, we associated a vector $\bom$ according to the following rule: 
 \begin{equation} 
 \label{mapping_matr_vec} 
 \omega = \left( 
 \begin{array}{ccc} 
 0 & - \omega_3 & \omega_2 
 \\ 
 \omega_3 & 0 & - \omega_1 
 \\ 
 -\omega_2 & \omega_1& 0 
 \end{array}
 \right) 
 \quad 
 \Rightarrow 
 \quad 
 \bom=\left( 
 \begin{array}{c} 
 \omega_1 
 \\
 \omega_2 
 \\ 
 \omega_3 
 \end{array}
 \right) .
 \end{equation} 
 Then, for any column vector $\mathbf{v}=(v_1,v_2,v_3)^T \in \mathbb{R}^3$, we have 
 \begin{equation} 
 \omega \mathbf{v} = 
 \left(
  \begin{array}{c} 
 \omega_2 v_3 - \omega_3 v_2 
 \\ 
 \omega_3 v_1 - \omega_1 v_3 
 \\ 
\omega_1 v_2 - \omega_2 v_1 
 \end{array}
 \right) 
 = \bom \times \mathbf{v} \,.
 \label{hatmap_dep} 
 \end{equation} 
Thus, to every antisymmetric $3 \times 3$ matrix $\omega$ we can associate a vector $\bom$ through the rule \eqref{mapping_matr_vec}. The mapping from vectors to antisymmetric matrices is called the \emph{hat map}, and we use the notation $\widehat{\bom} = \omega$. The inverse procedure, taking an antisymmetric matrix and producing a vector, is called the \emph{inverse hat map} and is denoted as 
$\omega^\vee = \bom$. In coordinates we have $\omega_{ij}=- \epsilon_{ijk} \omega_k$ where $\epsilon_{ijk}$ is the completely antisymmetric tensor with $\epsilon_{123}=1$. Because of this property, the notation $\widehat{\bom} = \bom \times$ is also used, although we will not employ it here. Another useful property of the hat map relates the commutator of matrices $a$ and $b$ to the cross product of vectors $\mathbf{a}=a^\vee$ and $\mathbf{b}=b^\vee$ as 
\begin{equation} 
\left( a b - b a \right)^\vee =[a,b]^\vee = \mathbf{a} \times \mathbf{b} 
\quad 
\Leftrightarrow 
\quad 
a b - b a = [a,b]=\widehat{\mathbf{a} \times \mathbf{b}}\,.
\label{commutator_matr} 
\end{equation} 
Thus, we can treat the angular velocity $\omega$ to be both an antisymmetric matrix when it is defined as $\omega=\Lambda ^{-1}  \dot \Lambda$, and, in the same time,  a 3-vector using 
 $\bom = \omega^\vee= ( \Lambda ^{-1}  \dot \Lambda )^\vee$ through the hat map. These representations are completely equivalent and are fundamental for our further discussions. 
 
In addition, it is also useful to review the concept of differentiation with respect to vectors and matrices, in order to make the meaning of equations more precise. Clearly, the derivative of a scalar function, such as the Lagrangian, with respect to a column vector is a row vector, and their product can be computed using either the dyadic algebra or scalar product. In other words, for column vectors $\mathbf{a}$ and $\mathbf{b}$, and a function $F(\mathbf{a})$, we have 
 \begin{equation} 
\pp{F}{\mathbf{a}} \mathbf{b}= \left(\pp{F}{\mathbf{a}}\right)^T \cdot \mathbf{b} = \sum \pp{F}{a_i} b_i = (\mbox{row}) (\mbox{vector}) = (\mbox{scalar}). 
\label{deriv_vec}
\end{equation}
The equivalent representation of derivatives in terms of matrices is less straightforward. First, we need to introduce the pairing (scalar product) between two $3 \times 3$ matrices $A$ and $B$
\begin{equation} 
 \big< A\, , \, B \big> = \frac{1}{2} {\rm tr} \big( A^T B \big) \, . 
 \label{pairing} 
 \end{equation} 
 We will typically take derivatives of functions of the type $F(a) =\frac{1}{2} \big< \mathbb{D} a, a \big>$ for antisymmetric matrices $a$ and a
diagonal matrix $\mathbb{D} = {\rm diag} (d_1,d_2,d_3)$, having the physical meaning of the inertia matrix.  One can readily check that the matrix $\pp{F}{a}=\mathbb{D} a$ is, in general, not antisymmetric so it cannot be directly interpreted as a vector. However, for any antisymmetric matrix $b$, the product 
$\left<\pp{F}{a} \, , \, b  \right> $ only depends on the antisymmetric part of $\pp{F}{a}$. Thus, the following quantity is readily interpreted as a vector 
\begin{equation} 
\label{vec_deriv_matr}
\pp{F}{\mathbf{a}} = \frac{1}{2} \left[ \pp{F}{a} - \left(\pp{F}{a}  \right)^T \right]^\vee \, . 
\end{equation} 
Because of the apparent complexity of \eqref{vec_deriv_matr}, we shall always use vector derivatives \eqref{deriv_vec} in the formulas in this paper.

\medskip

\subsection{Euler-Poincar\'e variational theory}

Let us now return to the question of a rigid body dynamics and consider a left-invariant Lagrangian $L(\Lambda, \dot \Lambda)$ with respect to arbitrary rotations of space. As we mentioned, we can rewrite this Lagrangian as a function of the angular velocity only, \textit{i.e.}, we have $L( \Lambda , \dot \Lambda )= \ell \big( ( \Lambda^{-1} \dot \Lambda)^\vee \big)  =\ell( \boldsymbol{\omega} )$ for a function $\ell$ defined on $3$-vectors and given by the kinetic energy: $\ell(\bom)=\frac{1}{2} \mathbb{I} \bom \cdot \bom$.  How do we write the analogue of the Euler-Lagrange equations for the Lagrangian $\ell(\bom)$? If we write the variations of the action as
\[ 
\de \int_{t_0}^{t_1}  L( \Lambda , \dot\Lambda )  \mbox{d} t = \de \int_{t_0}^{t_1}  \ell(\bom) \mbox{d} t = \int _{t_0}^{t_1}  \pp{\ell}{\bom} \cdot \de \bom  \mbox{d} t\, , 
 \] 
we need to compute the variations $\de \bom$ that are induced by the variations $ \delta \Lambda $. Defining $\Sigma=\Lambda^T \de \Lambda$ which is also an antisymmetric matrix or, equivalently, its associated vector $\bsigma=\Sigma^\vee$, we compute
\begin{equation}
\begin{aligned} 
\de \omega & = \de \Lambda^{-1} \dot \Lambda =\de \left(  \Lambda^{-1} \right)  \dot \Lambda + \Lambda^{-1} \de \dot \Lambda = - \Lambda^{-1} \de \Lambda \Lambda^{-1} \dot \Lambda + \Lambda^{-1} \de \dot \Lambda = - \Sigma \Omega + \Lambda^{-1} \de \dot \Lambda
\\
\dot \Sigma & = \frac{\mbox{d}}{\mbox{d} t} \left( \Lambda^{-1} \de \Lambda \right)=  \frac{\mbox{d}}{\mbox{d} t} \left( \Lambda^{-1} \right) \de \Lambda + \Lambda^{-1} \de \dot \Lambda
\\ & \qquad \qquad = - \Lambda^{-1} \dot \Lambda \Lambda^{-1} \de \Lambda + 
\Lambda^{-1} \de \dot \Lambda  =- \Omega \Sigma + \Lambda^{-1} \de \dot \Lambda.
\end{aligned} 
\label{vardef} 
\end{equation}
In \eqref{vardef}, we have used the fact that the $ \delta $ derivative and the time derivative commute and 
\[ 
 \frac{\mbox{d}}{\mbox{d} t} A^{-1} = - A^{-1} \dot A A^{-1} \, , \quad \mbox{consequently}, \quad \de A^{-1} = - A^{-1} \,  ( \de A)  \, A^{-1} \, , 
\] 
since the variation $\delta$ is, formally, the derivative with respect to some parameter before setting the value of that parameter to $0$. Subtracting the equations \eqref{vardef} to eliminate the cross-derivatives $\de \dot \Lambda$, we obtain the expression for the variation of $\omega$ in terms of $\Sigma$ as 
\begin{equation} 
\de \omega = \dot \Sigma + \big[ \omega, \Sigma \big] \quad \Leftrightarrow \quad 
\de \bom = \dot \bsigma + \bom \times \bsigma \, . 
\label{var_calc} 
\end{equation} 
Substitution of \eqref{var_calc} into the variational principle, integrating by parts once and using that $\bsigma(t_0)=\bsigma(t_1)=0$ as a consequence of $ \delta \Lambda (t_0)= \delta \Lambda (t_1)=0$, gives 
\begin{equation} 
\begin{aligned} 
\de \int_{t_0}^{t_1}  \ell(\bom) \mbox{d} t & = \int_{t_0}^{t_1}  \pp{\ell}{\bom} \cdot \de \bom \mbox{d} t = \int_{t_0}^{t_1}  \pp{\ell}{\bom} \cdot \left(  \dot \bsigma + \bom \times \bsigma \right) \mbox{d} t 
\\ 
& =  - \int_{t_0}^{t_1} \left( \frac{\mbox{d}}{\mbox{d} t}  \pp{\ell}{\bom} + \bom \times \pp{\ell}{\bom} \right) \cdot  \bsigma \mbox{d} t \,.
\end{aligned} 
\label{var_derivation}
\end{equation} 
Since $\bsigma(t)$ is an arbitrary function of time, the equations of motion are 
\begin{equation} 
\frac{\mbox{d}}{\mbox{d} t}  \pp{\ell}{\bom} + \bom \times \pp{\ell}{\bom} = \mathbf{0} 
\quad 
\Rightarrow 
\quad 
\frac{\mbox{d}}{\mbox{d} t}  \mathbb{I} \bom = \mathbb{I} \bom \times  \bom \, , 
\label{Euler_eq_RB} 
\end{equation} 
which are the well-known Euler equations for the motion of a rigid body. Of course, one could have derived \eqref{Euler_eq_RB} using the balance of angular momentum, as Euler himself has done. The example of a rigid body dynamics is too simple to demonstrate the full prowess of the method yet, which will be done in the derivation of our equations in Section~\ref{sec:varintro} below. For now, we would like to draw the attention of the reader to the fact that the function multiplying $ \boldsymbol{\Sigma} $ in \eqref{var_derivation} is \emph{exactly the angular momentum balance}. Thus, the advantage of the variational derivation is that the angular and, as we shall see, the linear momentum balance are computed automatically through a well-defined procedure, no matter how complex the Lagrangian may be. In contrast, trying to compute the angular and linear momentum balance equations by equating terms from Newton's laws is, in our opinion, extremely difficult if not impossible when the system is highly complex, like in the case studied in this paper.

\subsection{ Exact geometric rod: extension to two independent and two dependent variables}

Having reviewed the general variational principle on the simple example of the rigid body, let us turn our attention to the variational description of Cosserat, or geometrically exact, rod theory \cite{SiMaKr1988}. While the variational principle is the same in spirit as it is for the rigid body, there are two fundamental differences. 
\begin{enumerate} 
\item There are two independent variables, one being the time $t$ and another being the parameter along the rod $s$, not necessarily the arc length. 
\item 
{The configuration of the tube deforming in space is defined by: (i) the position of its line of centroids given by the map $(s,t) \mapsto  \br(s,t)\in \mathbb{R}  ^3 $, and (ii) the orientation of the cross sections of the tube at the points $\br(s,t)$, defined by using  a moving orthonormal basis $ \mathbf{d}_i  (s,t)$, $i=1,2,3$.
The moving basis is described by an orthogonal transformation $\Lambda(s,t)\in {S O}(3)$ such that $ \mathbf{d} _i (s,t)= \Lambda(s,t) \mathbf{E} _i $, where $ \mathbf{E} _i$, $i=1,2,3$ is a fixed material frame.} 
\end{enumerate} 
\revision{R3Q4}{Note that the local frame $ \mathbf{d} _i$, $i=1,2,3$, is not related to the Frenet-Serret frame of the moving curve. Indeed, the latter associates a frame to a curve based exclusively on the information about the curve itself, which in our case is a centerline.  In contrast, in Cosserat theory, one considers the rod as a geometric object including a curve in space, each point of the curve having a frame attached to it.}
The combined element $(\Lambda, \br)$ belongs to the group of rotations and translations in space, denoted $SE(3)$ and called the special Euclidean group. While a consistent theory can be derived using the new group in complete analogy to $SO(3)$ described above \cite{ElGBHoPuRa2010}, it is easier and more transparent to limit ourselves to the rotation-invariant variables. Since there are two independent variables $s$ and $t$ and two dependent variables $(\Lambda, \br)$, four rotation-invariant variables can be defined:
\begin{equation} 
\begin{aligned} 
\bom & =\big(  \Lambda^{-1} \partial_t \Lambda \big)^\vee \,  , \qquad 
\bgam  = \Lambda^{-1} \partial_t \br \, , 
\\
\bOm & =\big(  \Lambda^{-1} \partial_s \Lambda \big)^\vee \,  ,  \qquad 
\bGam  = \Lambda^{-1} \partial_s \br \, . 
\end{aligned} 
\label{variables_def} 
\end{equation} 
{The meaning of the variables is the following: 
\begin{enumerate} 
\item $\bom$ is the angular velocity of the frame in the body frame for a given $s$;
\item $\bgam$ is the linear velocity of the frame in the body frame for a given $s$;
\item $\bOm$ is the Darboux vector, \textit{i.e.} the angular strain of the frame rotation computed as the frame is being slid along the rod at a fixed time;
\item $\bGam$ is the local stretch of the rod elements computed in the body frame.
\end{enumerate} 
{It is also worth to note that $\bGam$ can take arbitrary vector values, since its physical meaning is the derivative $\partial_s \br(s,t)$ expressed in the body frame. This is in contrast with the inextensible and unshearable rod where $\bGam$ is constrained as $\bGam=\bGam_0 = \mathbf{E}_1$. Derivation of such and equation is done in \cite{FGBPu2015} and  involves another Lagrange multiplier for the constraint, denoted $\mathbf{z}$ in that paper. } 
}
\paragraph{Kinematic compatibility conditions} 
The compatibility constraints are coming from the equality of cross-derivatives in $s$ and $t$, \textit{i.e.} $\Lambda_{st}=\Lambda_{ts}$ and 
$\br_{st}=\br_{ts}$. Written in terms of the variables in \eqref{variables_def}  these conditions read: 
\begin{equation} 
 \partial _t \boldsymbol{\Omega} =  \boldsymbol{\Omega} \times \boldsymbol{\omega} +\partial _s  \boldsymbol{\omega} \, , 
 \qquad 
\partial_t \bGam+ \bom \times \bGam=\partial_s \boldsymbol{\gamma} + \bOm \times \bgam \,.
\label{compatibility_1} 
\end{equation} 
Note that equations \eqref{compatibility_1} have no physics in it, and are equally valid for a rod made out of steel, wood, rubber or any other material, as long as the motion of the rod is differentiable in space and time. 

\paragraph{Dynamic equations} 
The kinetic energy of the rod depends on the velocities $\bom$ and $\bgam$ (and possibly $\bOm$ and $\bGam$ for some cases), whereas the potential energy depends on the deformations $\bOm$ and $\bGam$. The symmetry-reduced Lagrangian thus depends on all the variables \eqref{variables_def} and is of the form $f (\bom,\bgam,\bOm,\bGam)$. Since the length of the segment between $s$ and $s+\mbox{d} s$ is given by $|\bGam|\mbox{d} s=|\partial_s \br| \mbox{d} s$, the critical action principle is written as 
\begin{equation}
\de \int_{t_0}^{t_1}\ell(\bom,\bgam,\bOm,\bGam)   \mbox{d} t = 
 \de \int_{t_0}^{t_1}\!\!  \int_0^L  f (\bom,\bgam,\bOm,\bGam) |\bGam|\,  \mbox{d}s   \mbox{d} t=0 \, , 
\quad \ell := \int_0^L f|\bGam|\,  \mbox{d}s \, . 
\label{crit_action_rod} 
\end{equation} 
We now need to reproduce the computation of the variations \eqref{var_calc} for the case of two variables, $\Lambda$ and $\br$. We therefore introduce two variations which are both $3$-vectors:
\begin{equation} 
\bsigma = \big( \Lambda^{-1} \de \Lambda \big)^\vee 
\, , \quad 
\bpsi = \Lambda^{-1} \de \br \,.
\label{sigma_psi_def}
\end{equation} 
A short calculation completely analogous to \eqref{vardef} gives the following expression for variations of the quantities \eqref{variables_def} 
 \begin{equation}\label{constrained_variations_rod}
\begin{aligned} 
\vspace{-2mm} 
& \delta \bom = \partial _t \boldsymbol{\Sigma}  + \bom \times \boldsymbol{\Sigma} , \qquad\, \,\delta \boldsymbol{\gamma} =\partial _t  \bpsi + \boldsymbol{\gamma} \times \boldsymbol{\Sigma} + \bom \times \bpsi, 
\\
\vspace{-2mm} 
& \delta \boldsymbol{\Omega} = \partial _s  \boldsymbol{\Sigma}  +\boldsymbol{\Omega} \times \boldsymbol{\Sigma} , \qquad \delta \boldsymbol{\Gamma} =\partial _s  \bpsi + \boldsymbol{\Gamma} \times \boldsymbol{\Sigma} + \boldsymbol{\Omega} \times \bpsi\,.
\vspace{-2mm} 
\end{aligned}  
\end{equation}
The equations of motion are obtained by using \eqref{crit_action_rod} and the variations \eqref{constrained_variations_rod}. We thus get 
\begin{equation} 
\label{exact_rod_deriv}
\begin{aligned} 
0&=\de \int_{t_0}^{t_1} \ell(\bom,\bgam,\bOm,\bGam)\, \mbox{d} t 
\\ 
& = \int _{t_0}^{t_1}\!\! \int_0^L   \left[   \dede{\ell}{\bom} \cdot \de \bom + \dede{\ell}{\bgam} \cdot \de \bgam + 
\dede{\ell}{\bOm} \cdot \de  \bOm + \dede{\ell}{\bGam} \cdot \de  \bGam \right] \,\mbox{d} s   \mbox{d} t 
\\ 
&= \mbox{Use \eqref{constrained_variations_rod} and integrate by parts in $s$ and $t$ } 
\\ 
& = \int _{t_0}^{t_1}\!\!  \int_0^L \big( \mbox{Angular momentum balance} \big) \cdot \bsigma \, \mbox{d} s \mbox{d} t 
\\ & \qquad  \qquad \qquad + 
\big( \mbox{Linear momentum balance} \big) \cdot \bpsi \,  \mbox{d} s \mbox{d} t   .
\end{aligned} 
\end{equation}
 In \eqref{exact_rod_deriv} we have made use of the variational derivatives of the Lagrangian $\ell$, which are defined in terms of the $L ^2 $ pairing on the interval $[0,L]$ as follows:
\begin{equation}\label{var_der} 
\left.\frac{d}{d\varepsilon}\right|_{\varepsilon=0} \ell( \bom + \varepsilon \delta \bom,\bgam,\bOm,\bGam )= \int_0^L\frac{\delta \ell}{\delta\bom } \cdot\delta \bom \, \mbox{d}s,
\end{equation} 
similarly for the other variables. 
Note that we do not need to explicitly find the terms in the balance angular and linear momentum equations, these terms emerge automatically through the variational principle. It was proven in \cite{SiMaKr1988,ElGBHoPuRa2010} that the resulting equations, called the \emph{exact geometric rod equations}, are equivalent to Cosserat rod equations, as we illustrate in \ref{app:rod}. Note also that this method is valid for arbitrary Lagrangians defining the rod. This is, in our opinion, a drastic advantage over theories relying on a particular (\emph{e.g.}, linear) form of certain elasticity terms. {In addition, the derivation of \eqref{exact_rod_deriv} is algorithmic and straightforward, whereas one has to be extremely careful when balancing terms in Cosserat-like rod theory, especially when applied to a tube conveying a moving fluid \cite{RiPe2015,BeGoTa2010}. Thus, in our opinion, the variational approach is advantageous over the Newton-Euler approach of direct force balance for complex problems such as the one considered here.}

\section{Derivation of main equations for the motion of collapsible tube in 3D} 
\label{sec:varintro} 

Having reviewed the variational theory of elastic rods, we are now ready to derive the equations of motion for a collapsible elastic tube conveying fluid. This derivation follows the general theory \cite{FGBPu2014,FGBPu2015} and plays a fundamental role in the present paper. The interested reader may consult these articles for the complete treatment of the variational approach, as well as for detailed discussions on boundary conditions, linearized stability of straight tubes, and fully nonlinear traveling solutions.

\subsection{Physical assumptions} 
\paragraph{Elastic rod dynamics} We assume that the part of the Lagrangian describing the elastic tube is completely described by the variables $(\bom,\bgam,\bOm,\bGam)$ introduced in \eqref{variables_def}. The treatment of the tube as an elastic rod is well-established in the literature. For constant fluid velocity and constant cross-section, our derivation would correspond to that of \cite{BeGoTa2010}.

{\paragraph{Change of the cross-section} 
We assume that the cross-sectional area $A$ depends on the instantaneous tube configuration, \textit{i.e}, is determined by $\Lambda$, $\partial _s\Lambda$ and $\partial _s\br$, but not on the tube's dynamic variables or fluid motion. Since the scalar function defining the cross-sectional area $A$ has to be invariant with respect $S O(3)$ rotations,  we can  posit a real-valued function  $A=A(\bOm,\bGam)$ which we consider arbitrary, but given. 
The variations in $A$ thus come  from the bending, twisting and stretching of the local element of the tube. 
Such assumption is valid unless the walls of the tube are excessively stretchable and lead to varicose- and aneurism-like instabilities. For example, for a typical pressures of $\sim$ 2 atm in the tube, corresponding to a practical household situations like a garden hose, the cross-sectional deformations are negligible unless the tube is made out of  flexible material, such as toy balloon latex.  In other words, the approximation we use here corresponds to the normal component of stress tensor on tube's wall being balanced by the wall's reaction force without any noticeable additional deformation, and the tangential stress component vanishing due to fluid's lack of viscosity. 
}

{\paragraph{Fluid flow approximation and its limitations} As we see below, to describe the fluid flow, we utilize a single velocity function $u(s,t)$ corresponding to the mean velocity of the fluid in a given cross-section. Mathematically, this approximation assumes the simplest possible flow of fluid at a given time $t$ and position $s$, since a single function $u(s,t)$ is assumed to provide a sufficient description of the fluid motion. This model is consistent with most literature on the subject, but certainly represents a simplification of the flow. Indeed, one can imagine a flow where part of the kinetic energy of the fluid is going into the inner swirling motion. 
In particular, the concept of entrance length $L_e$ is useful here:  after traveling such length from the entrance of the tube,  the flow takes on fully developed profile, laminar or turbulent. For a given Reynolds number $R_d$ based on the diameter of the flow $d$, the laminar entrance length is usually estimated as $L_e \simeq 0.05 {\rm Re_d} d $ and the estimates for the turbulent entrance length vary rather strongly in the literature, one of the estimates being  $L_e  \simeq {\rm Re_d}^{1/4} d$. Another way to estimate the generation of vortices in developed flow is through Dean's number which can be written as 
${\rm De} = {\rm Re_{d}} \sqrt{d/(2 r)}$, $r$ being the typical radius of curvature. For a helical basic state, $r^{-1} \simeq |\bOm_0 \times \bGam_0|$. However, Dean's theory is applicable to developed flow only, and is not known to be accurate for very large values of Reynolds numbers. In any case, the validity condition of the plug fluid flow approximation is $L \ll L_e$, where $L$ is the length of the tube, before the flow becomes fully developed inside the tube.  }
\\
{\paragraph{On vorticity generation and its role in the dynamics} A special note should be given here about possible effects of the swirl in the flow. Such presence of the swirl, even when the one-dimensional  approximation for the fluid is used, would change the Lagrangian and correspondingly change the dynamical behavior, as direct numerical simulations indicate for moderate Reynolds numbers \cite{Xie-etal-2016}.  In terms of theoretical modeling, vorticity appears from the interaction of the boundary with the fluid through the viscous terms, and is brought about by intricate interaction of the boundary layer with the bulk of the flow. The limit of viscosity tending to zero is intricate and does not necessarily lead to the lack of vortex generation, especially for the curved pipes and non-steady flow. While our current approach on neglecting the swirl is consistent with most of the literature of the subject, a consistent model of swirl would be highly useful. We do not know of a consistent theoretical method to incorporate vorticity generation in the pipes at high Reynolds numbers and will explore this interesting question in future work.  }

{\paragraph{Advantages of the theory} Finally, it is useful to note what approximations or assumptions on the flow are \emph{not} needed for our theory. Namely, we do not need to assume a particular type of elasticity laws, or restriction of the motion to only certain types (say, only stretching or only bending), or particular law of change of cross-sectional area $A(\bOm,\bGam)$ with deformations. Equation \eqref{full_3D} is valid for all Lagrangians, all cross-sectional area laws change and any motion of the tube in three dimensions. In addition, our equations utilize the correct conservation law, see \eqref{fluid_vol_cons}, rather than the law $A u= \text{const}$ which is not accurate for time-dependent motions.  Finally, our theory, being variational in nature, allows to develop fully variational and structure-preserving numerical schemes for this problem of fluid-structure interaction \cite{FGBPu2016}, which is something that is not possible in theories based on balance of forces and torques.} 

\paragraph{On the extension of the theory to include varicose instabilities} There has been a great interest in studying the dynamics of tubes with easily deformable walls, especially for physiological applications like blood and air flow. Our theories will be most readily applicable to the fully filled tubes 
\cite{LuPe1998,KoMa1999,JuHe2007,Tang-etal-2009,StWaJe2009}, see also recent review article \cite{HeHa2011} for more references and discussion.  As we discussed above, easily flexible walls bounding the flow will violate the assumption of the cross-sectional area $A$ depending on the deformations $(\bOm,\bGam)$. As it turns out, one cannot simply incorporate the pressure $ \mu $ as $A=A(\mu,\bOm,\bGam)$ for both mathematical and physical reasons. The solution lies in the development of variational methods, where a shape parameter, such as the radius of the tube $R$, is taken to be a new dependent variable $R=R(s,t)$. The corresponding variational treatment gives a new Euler-Lagrange equation for  the radius $R(s,t)$ in addition to the angular and linear momenta and fluid momenta \eqref{full_3D} below. \revision{R2Q6d}{We shall note, at this point, that it would be quite premature to get into a more detailed exposition of this theory. Here, we just note that the typical values of $\mu$ should be of the order of $\rho u_0^2$. For fluid being water and a typical velocity of $1$m/s, we get $\mu \sim 1$kPa$\sim 0.01$ atm. Such pressures will not result in noticable deformations of the tubes with walls made out  of the latex in party air balloon.
If we consider speeds $u_0 \sim 10$m/s, then $\mu \sim 1$ atm. A typical party air balloon would expand considerably at these pressures, however, something like a  medical tube with thicker walls ($\sim 1$mm thickness), which we have used for our experiments, or garden hose with walls lined with steel wires, will not experience any expansion whatsoever. 
If $R$ is the typical radius of the
tube, $E$ is Young's modulus of the tube material, and $h$ the thickness of the wall, then the typical additional deformation is $\de R \sim \rho u_0^2  R^2/E h$. For the assumption of $A$ to depend only on the deformations to be valid, and not to be dependent on other variables, we  need $\de R \ll R$,  \emph{i.e.}  $\frac{h}{R} \gg \frac{\mu}{E}$. For example, for a very soft rubber tube with $E=10^7$Pa=$100$atm, the approximation is valid if  $h  \gg 0.01 R$. For higher values of $E$ coming from less compliant materials, or steel-wire reinforced walls, the relative deformation  of the walls will be even more negligible. 
}

\subsection{Derivation of equations of motion }

\paragraph{Fluid flow description}
We approximate the fluid motion by a one-dimensional mapping from the initial position of the fluid particle $S$ to its current position at time $t$ denoted as $s=\varphi(S,t)$. We will  refer to this description as the Lagrangian description of the fluid motion, as it expresses the movement of the fluid particles from their initial to their final positions. The velocity of the Lagrangian particle labeled $S$ relative to the tube is $\partial_t \varphi(S,t)$. In order to compute the velocity $u(s,t)$ of the same particle relative to the tube at the point $s$, which can be thought of as the Eulerian velocity, we need to map the point $S$ back to $s$ using the relationship $S=\varphi^{-1}(s,t)$, so  
\revision{R3Q9}{
\begin{equation} 
u(s,t) =\partial_t \varphi (\varphi^{-1} (s,t),t)  = \partial_t \varphi \circ \varphi^{-1} (s,t) = \left. \pp{}{t} \varphi(S,t)  \right|_{S=\varphi ^{-1} (s,t)} \, . 
\label{Eulerian_vel} 
\end{equation} 
We note that an alternative derivation is possible using the variations of the back-to-labels map $\varphi^{-1}(S,t)$, as was done in \cite{FGBPu2016} for the purpose of derivation a variational discretization of the problem. However, this derivation is only tractable if the tube has initially uniform cross-section, otherwise one would have to additionally treat the evolution equation for $\varphi^{-1}(S,t)$. We shall not delve on this technical point here and refer the interested reader to \cite{FGBPu2016}. 
}
Notice that the velocity $u$ only has one component along the tube. In reality, as we see below, $u$ encompasses the integrated flux of fluid through the cross section. 
We also need to compute the variation of velocity $u$ given by \eqref{Eulerian_vel}. In order to accomplish that, we introduce the variation 
$\eta (s,t) = \de  \varphi  \circ \varphi^{-1}(s,t)$ and proceed similarly to \eqref{vardef} to obtain 
\begin{equation} 
\de u = \partial _t  \eta + u  \partial _s \eta - \eta  \partial _s u \, . 
\label{uvar}
\end{equation} 
Variations with respect to $u$ will provide additional terms proportional to $\eta$, which will give the balance of fluid momentum equation integrated along the tube.

\paragraph{Mass conservation} 
The crucial part of the theory is the mathematical implementation of the conservation law, which we believe has not been adequately addressed in the literature. 
With the physical condition that the fluid is filling up the whole available area inside the tube, and assuming that the fluid inside the tube is incompressible (in 3D), the volume conservation along the tube reads
\begin{equation} 
\partial _t Q+\partial _s (Qu) =0 \, , \quad Q:=A(\bOm,\bGam)  | \bGam| \, , 
\label{fluid_cons} 
\end{equation} 
where the extra factor of $|\bGam|$ appears since $s$ is not assumed to be the arc length. Physically, $Q(\bOm,\bGam) \mbox{d} s$ is the volume of fluid in the interval $(s,s+\mbox{d} s)$. If $A=A(s)$ is independent of $t$, and $|\bGam|=$const, \eqref{fluid_cons} reduces to the conservation law $uA= \text{const}$.  This is the equation for velocity used in  \cite{SeLiPa1994,Pa2004,MoPa2009,GhPaAm2013}; however, this approach is inexact as it neglects the time variation of $A$ and stretch $\bGam$. We believe that it is impossible to accurately resolve this issue without adequately taking into consideration the incompressibility constraint which we can write as follows. 
\\
Consider the fluid volume in the interval $(s,s+\mbox{d} s)$. At $t=0$, this volume is $A_0(s) |\bGam_0(s)|  \mbox{d} s =Q_0(s)  \mbox{d} s $. Without loss of generality, we assume that the  labelling of the material particles of the tube at $t=0$ coincides with the arc length, so $| \bGam_0 (s)|=1$, so $Q_0(s)=A_0(s)$. Then, the fluid particle at time $t$, which has travelled from its initial point $\varphi^{-1}(s,t)$, carries the initial volume 
$Q_0(\varphi^{-1}(s,t) ) \partial_s \varphi^{-1}(s,t) \mbox{d} s$. This volume has to coincide with the volume at time $t$, which is equal to $A(\bOm,\bGam) |\bGam| \mbox{d} s=Q(\bOm,\bGam) \mbox{d} s$. Thus, the conservation law for fluid volume at time $t$ reads 
\begin{equation} 
Q_0(\varphi^{-1}(s,t) ) \partial_s \varphi^{-1}(s,t) = Q(\bOm,\bGam) \, . 
\label{fluid_vol_cons}
\end{equation} 
\begin{figure}
\centering
\includegraphics[width=0.75 \textwidth]{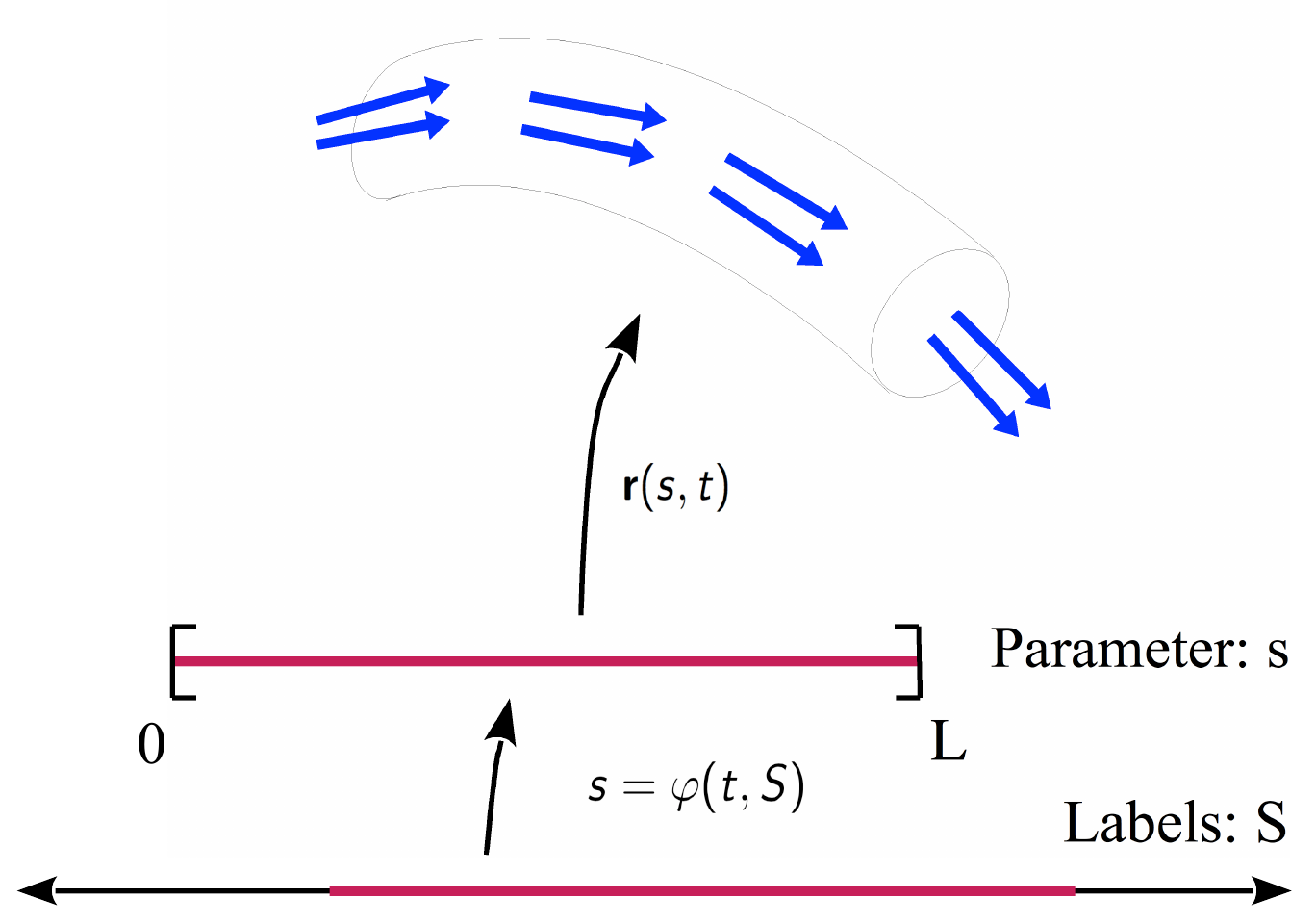}
\caption{A sketch illustrating the derivation of equation \eqref{fluid_vol_cons}. Lagrangian labels $S$ of the fluid are mapped into the tube's material points $s$ using $\varphi(s,t)$, and then into the point $\br(s,t)$ on the centerline. Conservation law follows from equating the initial volume available to the fluid in the $S$-space, to the corresponding volume in physical space.}
\end{figure} 
 \color{black} 

\paragraph{Equations of motion} 
The exact geometric variational approach taken in \cite{FGBPu2014,FGBPu2015} is based on the critical action principle
\vspace{-1.5mm} 
\begin{equation} 
\de  \int_{t_0}^{t_1} \left[ \ell \big( \bom, \bgam, \bOm, \bGam, u \big) + \int_0^L\mu \,\left( (Q_0 \circ \varphi ^{-1})\partial_s \varphi ^{-1}  - Q(\bOm,\bGam)   \, \right) \mbox{d} s  \right] \mbox{d} t =0 
\, , 
\label{min_action_constr} 
\vspace{-1.5mm} 
\end{equation} 
in which \eqref{fluid_vol_cons} is imposed with the help of a Lagrange multiplier $ \mu (t,s)$ and with respect to the variations \eqref{constrained_variations_rod} and \eqref{uvar} 
\cite{HoPu2009,ElGBHoPuRa2010}. 
The complete equations of motion for flexible tubes conducting fluid are:
\begin{equation}\label{full_3D} 
\hspace{-3mm} 
\left\lbrace\begin{array}{l}
\displaystyle\lp \prt_t + \bom\times\rp\dede{\ell}{\bom}+\bgam\times\dede{\ell}{\bgam} +\lp\prt_s + \bOm\times\rp\!\left( \dede{\ell}{\bOm} -\frac{\partial Q}{\partial \boldsymbol{\Omega} }\mu   \right)+\bGam\times\!\left( \dede{\ell}{\bGam}-\frac{\partial Q}{\partial \boldsymbol{\Gamma}  } \mu   \right) =0\\
\displaystyle\lp \prt_t + \bom\times\rp\dede{\ell}{\bgam} + \lp\prt_s + \bOm\times\rp\left( \dede{\ell}{\bGam}-\frac{\partial Q}{\partial \boldsymbol{\Gamma}  } \mu   \right)=0   \\
\displaystyle\vspace{0.2cm}\pp{m}{t}  + \partial _s\left(mu-\mu \right) =0 \, , \quad m:=\frac{1}{Q} \dede{\ell}{u}   \\
\displaystyle\vspace{0.2cm} \partial _t \boldsymbol{\Omega} = \boldsymbol{\Omega} \times \bom +\partial _s  \boldsymbol{\omega}, \qquad  \partial _t \boldsymbol{\Gamma} + \bom\times \boldsymbol{\Gamma} = \partial _s \boldsymbol{\gamma} + \boldsymbol{\Omega} \times \boldsymbol{\gamma}\\
\displaystyle  Q( \boldsymbol{\Omega} , \boldsymbol{\Gamma} )=(Q_0  \circ \varphi ^{-1} )(\partial _s \varphi  ^{-1} )\; \Rightarrow \; \partial _t Q+ \partial _s (Qu)=0\,.
\end{array}\right.
\end{equation}
 We recall that the variational derivatives $ \frac{\delta \ell}{\delta \boldsymbol{\omega} }$, $\frac{\delta \ell}{\delta \boldsymbol{\gamma} }$,... used here are defined relative to the $L ^2 $ pairing, see \eqref{var_der}.
These equations form a closed system of equations for the problem, with the terms proportional to $\mu$ describing the effect of the cross-sectional dynamics. They are valid for an arbitrary cross-sectional dependence $A(\bOm, \bGam)$ and an arbitrary Lagrangian  $\ell$.  
As explained in \cite{FGBPu2015}, the variational principle \eqref{min_action_constr} is rigorously justified by a reduction process applied to the Hamilton principle with holonomic constraint, written in terms of the Lagrangian variables $\Lambda , \dot \Lambda , \br , \dot{\br}, \varphi , \dot{ \varphi }$, with free variations $ \delta \Lambda $, $ \delta \br$, $ \delta \varphi $, vanishing at the temporal extremities.
\revision{R3Q1}{For the cross-sectional area being constant, \textit{i.e.} $\mu=0$, appropriate expressions for the kinetic energy and elasticity, and with the additional terms introducing gravity, the system \eqref{full_3D} reduces to the equations obtained by force and momentum balance for Cosserat rods \cite{BeGoTa2010,RiPe2015} under appropriate transformation of the forces described in \ref{app:rod}. \\
Equations \eqref{full_3D} represent  a general framework for the further analysis of elastic tubes conveying fluid, as long as the tube can be modelled by the general Cosserat rod theory, and the fluid's motion can be modelled as one-dimensional motion along the rod. These equations should not be understood as equations for one particular  geometry, or realization, of the tube. For particular choices of Lagrangians and cross-sectional profiles, these equations are capable of describing cases like: linear and nonlinear motion of the initially straight cantilever pipes, pipes with supported ends and pipes with complex nozzles, linear and nonlinear motion of initially curved (circular) pipes with constant cross-section, Timoshenko vs Euler beam dynamics of straight pipes, pipes  with elastic supports at the ends and/or intermediate points, extensible and inextensible theory of the motion of pipes with constant cross-section, nonlinear motion of pipes with  varying cross-section that is fixed along the tube, composite and/or biologically related tubes with high anisotropicity and nonlinear elasticity and others. In this manuscript, we concentrate on the motion of a helical tube with a dynamically varying cross-section, including, as a particular case, tube consisting of circular arcs. 
}

It is also worth discussing the boundary conditions in the system, especially for the free ends for the cantilever-type situations, when one of the extremities is fixed and the other one is free to move, which is the commonly observed instability of the garden hose. It is well known, see  \cite{Be1961a} and the follow-up works, that the tube conveying fluid does not form a closed Lagrangian system if there is a free boundary, as the fluid is leaving the tube at that free boundary and exerts a force onto that end. In \cite{FGBPu2015}, a detailed consideration of the boundary conditions in the general case was undertaken, and we refer the reader to that paper for details. To briefly summarize this theory, the generalized forces at the free end $\mathbf{F}_{\bOm}$ (torque), $\mathbf{F}_{\bGam}$ (force) and $F_u$ (fluid force) can be computed by tracking the terms proportional to $\bsigma$, $\bPsi$ and $\eta$ at that particular end, and by using the Lagrange-d'Alembert variational principle. These forces are given by the following expressions 
\begin{equation}
F_u:=\frac{\delta \ell}{\delta u}u - \mu Q\Big|_{s=L}, \qquad 
\mathbf{F} _{\bGam}:=\dede{\ell}{\bGam}- \mu \frac{\partial Q}{\partial \boldsymbol{\Gamma} }  \Big|_{s=L} , 
\qquad 
\mathbf{F} _{\bOm}:= \dede{\ell}{\bOm} - \mu \frac{\partial Q}{\partial  \bOm }   \Big|_{s=L},
\label{forcedef} 
\end{equation}
and have to be evaluated for a particular choice of boundary conditions on the dynamical variables $(\bom, \bgam,\bOm,\bGam,u,\mu)$ at the free end.

\section{Equations of motion for a particular choice of the Lagrangian and steady state helical solution}

In this section we describe the Lagrangian and cross-sectional dependence of a fluid conveying tube with helical equilibrium solution. Then we proceed to the linearisation around the helical state.

\subsection{A  particular choice of Lagrangian and cross-sectional dependence} 

To find particular helical steady states, let us consider the particular Lagrangian for linearly elastic tubes studied in \cite{FGBPu2014,FGBPu2015}: 
\begin{equation}\label{Lagrangian_fluid_tube}
\begin{split} 
\ell & ( \boldsymbol{\omega} ,\boldsymbol{\gamma} , \boldsymbol{\Omega} ,  \boldsymbol{\Gamma} ,u)  
= \frac{1}{2} \int_0^L\Big( \alpha | \bgam|^2 +  \mathbb{I} \bom\! \cdot\! \bom +\rho A( \boldsymbol{\Omega} , \boldsymbol{\Gamma} ) \left| \boldsymbol{\gamma} + \boldsymbol{\Gamma} u\right | ^2
\\
&   \qquad - \mathbb{J} (\bOm-\bOm_0) \!\cdot \! (\bOm-\bOm_0 ) 
-\lambda |\bGam-\bGam_0|^2 \Big)| \boldsymbol{\Gamma} | \mbox{d}s\,:=\int_0^L f( \boldsymbol{\omega} ,\boldsymbol{\gamma} , \boldsymbol{\Omega} ,  \boldsymbol{\Gamma} ,u) |\bGam| \mbox{d} s ,
\end{split} 
\end{equation}
with the shape function 
{
\begin{equation} 
\label{Aeq} 
A(\bOm,\bGam) = A_0 - \frac{K_{\bOm}}{2} \left| \bOm-\bOm_0 \right|^2 - D_{\bGam} \mathbf{E}_1 \cdot  \left( \bGam-\bGam_0 \right) -\frac{K_{\bGam}}{2} \left| \bGam-\bGam_0\right|^2 \, . 
\end{equation} 
}
\revision{R3Q14}{Here and below, we have used a shorthand notation $\mathbb{A} \mathbf{v} \cdot \mathbf{v} = (\mathbb{A} \mathbf{v}) \cdot \mathbf{v}$ for an arbitrary tensor $\mathbb{A}$ and vector $\mathbf{v}$ to avoid an excessive use of parentheses. } 
\paragraph{Justification of the formula for cross-sectional area change}
While the computation of an exact analogue of formula \eqref{Aeq} for a tube constructed from general material is rather complex, one can justify the terms in that formula on symmetry and incompressibility grounds.

Let us first consider the deformation of a tube where all cross-sections remain normal to the centerline during the dynamics, which is the case for the incompressible and unshearable tube. We denote by $A_{\rm deformed}$ the area function in this case. Because of the invariance with respect to rotations and translations in space, this function can depend only on the variables $\bOm$ and $\bGam$. Let us first consider the dependence on $\bOm$. For a uniform material, and straight initial configuration, \eqref{Aeq} cannot contain a term linear in $\bOm$. Indeed, the area must be invariant under a change of sign of the rotation  $\bOm \rightarrow - \bOm$ while keeping the deformation fixed, \textit{i.e.}, $A_{\rm deformed}(\bOm, \bGam) = A_{\rm deformed}(-\bOm, \bGam)$. In addition, the area function will in general also depend on the stretching of the material of the tube: for example, a uniform extension of a straight elastic tube along its axis will decrease its cross-sectional area, so its expression may contain terms that are both linear and nonlinear in $\bGam-\bGam_0$. To the lowest relevant (quadratic) order the assumption for cross-sectional area dependence on deformations is then
\begin{equation}
A_{\rm deformed}(\bOm, \bGam) =A_0 - \frac{1}{2}    \mathbb{K}_{\bOm} \bOm  \cdot \bOm - 
 \mathbf{M}_1\cdot (\bGam - \bGam_0 ) -  \frac{1}{2}   \mathbb{M}_2 (\bGam-\bGam_0) \cdot  (\bGam-\bGam_0) \, , 
\label{Aeq_simple}
\end{equation}
for some vector $\mathbf{M}_1$ and  positive symmetric tensor $\mathbb{M}_2$. 
The most obvious extension of this formula for a non-straight equilibrium is to consider a term quadratic in $\bOm - \bOm_0$ in \eqref{Aeq_simple}. 

Let us now allow the more general case when the cross-sections tilt with respect to the tangent to the centerline by an angle $\theta(s,t)$. 
\revision{R3Q6a\\R3Q12}{Since the effective area available for the fluid motion is reduced by $\cos \theta= \mathbf{d} _1 \cdot \br_s/|\br_s| =\mathbf{E}_1 \cdot \bGam/|\bGam|$, we need to modify \eqref{Aeq_simple} as
\begin{equation} 
A_{\rm deformed,tilted}(\bOm, \bGam) =A_{\rm deformed} (\bOm, \bGam) \mathbf{E}_1 \cdot \frac{\bGam}{|\bGam|}   = 
A_{\rm deformed} \frac{ \left( \mathbf{E}_1 \cdot \bGam_0 + \mathbf{E}_1 \cdot (\bGam-\bGam_0 ) \right) }{\left| \mathbf{E}_1 \cdot \bGam_0 + \mathbf{E}_1 \cdot (\bGam-\bGam_0 ) \right|} . 
\label{Aeq_tilted} 
\end{equation}
}
Let us assume for simplicity that the undisturbed configuration has $\bGam_0$ pointing along $\mathbf{E}_1$ direction and the normalization of $s$ is chosen such that $|\bGam_0|=1$, so that $ \boldsymbol{\Gamma} _0=\mathbf{E} _1$. While the formulas we derive will be valid for a general function $A(\bOm,\bGam)$, this assumption $ \boldsymbol{\Gamma} _0=\mathbf{E} _1$ will be used throughout the paper for the linear stability analysis of helical flows.
Using \eqref{Aeq_simple} combined with \eqref{Aeq_tilted}, we see that in general, up to and including the second order in $\bOm$ and $\bGam-\bGam_0$, the effective area change will have a quadratic term in $\bOm-\bOm_0$, as well as a linear and quadratic term in $\bGam-\bGam_0$. Moreover, assuming that the main change in cross-section due to stretching comes from the deformation along the tube's axis, as is the case for slender elastic tubes made out of isotropic materials, we have $\mathbf{M}_1 =D_0 \mathbf{E}_1  = D_0 \bGam_0$ in \eqref{Aeq_simple}. Then, the resulting equation, up to second order in $\bGam-\bGam_0$, and $\bOm-\bOm_0$, will be 
\begin{equation} 
\begin{aligned} 
A(\bOm, \bGam) =A_0 &-  \frac{1}{2}    \mathbb{K}_{\bOm} \left(\bOm-\bOm_0\right) \cdot \left(\bOm-\bOm_0\right)  
 -\left(D_0 - A_0\right)  \bGam_0 \cdot \left(\bGam-\bGam_0 \right) 
\\\qquad &-\frac{1}{2}   \mathbb{K}_{\bGam}  (\bGam-\bGam_0 ) \cdot (\bGam-\bGam_0) \,.
\end{aligned} 
\label{Aeq_tilted2} 
\end{equation}
\revision{R3Q6a}{Here, we introduced the tensor $ \mathbb{K}_{\bGam}$ which depends linearly on $\mathbb{M}_2$, as well as quadratically on the coefficients $D_0$. A more general form of equation \eqref{Aeq_tilted2} is studied below in \eqref{Aeq2}. 
}


{The equation \eqref{Lagrangian_fluid_tube} is valid in the assumption of a tube made out of a linearly elastic (but not necessarily isotropic) material. The inertia tensor $\mathbb{I}$ is always diagonal, and the properties of the tensor $\mathbb{J}$ depend on the elastic properties of the tube.  }

{The extensional and flexural rigidities are included in \eqref{Lagrangian_fluid_tube} through the coefficients $\lambda$ and the components of the tensor $\mathbb{J}$. More precisely, as it will be apparent from \eqref{Timoshenko_no_fluid} later, $\lambda=k A_t G$, where 
$k$ is the Timoshenko's coefficient, $A_t$ is the cross-sectional area of the tube and $G$ is the shear modulus. 
\\ 
For a tube made out of elastic isotropic material, the tensor $\mathbb{J}={\rm diag}(J_1,J_2,J_3)$ is diagonal and includes bending rigidities. For a tube that is initially rotationally symmetric about its axis, $J_2=J_3=J$ is the bending rigidity of the tube computed as $J=E I_A$, where $E$ is Young's modulus and $I_A$ is the second moment of area. The coefficient $J_1$ represents the twisting rigidity of the tube, which for elastic materials is proportional to the shear modulus $G$ and depends on the shape of the tube. 
}

{\paragraph{Non-pinching condition} We must also emphasize that \eqref{Aeq} is an approximation for the tube for rather small deviations from equilibrium and cannot be valid for all deformations. If $|\bOm - \bOm_0|$ or $|\bGam-\bGam_0|$ is large enough, 
then $A(\bOm,\bGam)$ becomes negative which is impossible. Thus, the condition for validity of \eqref{Aeq} is that every term on the right-hand side must be small compared to $A_0$. Remembering that $\bGam$ is dimensionless, $K_{\bOm}$ has dimensions of length$^4$, and $D_{\bGam}$ and $K_{\bGam}$ have dimension of area, the conditions of validity of \eqref{Aeq} are 
\begin{equation} 
\frac{K_{\bOm}}{A_0 } \left| \bOm-\bOm_0 \right|^2 \ll 1 \, , \quad 
\frac{D_{\bGam}}{A_0}   \left| \bGam-\bGam_0 \right| \ll 1 \, , \quad 
\frac{K_{\bGam}}{A_0} \left| \bGam-\bGam_0\right|^2 \ll 1 \, . 
\label{no_pinch} 
\end{equation}
The order of magnitude the coefficient $K_{\bOm}$ and $D_{\bGam}$, $K_{\bGam}$ are $R_0^4$ and $R_0^2$, respectively, with $R_0$ being the typical diameter of the tube.
}

This shape function generalizes the expression of $A(\bOm,\bGam)$ considered in \cite{FGBPu2014,FGBPu2015}, which can be obtained from \eqref{Aeq} by setting $\bOm_0=0$ and 
$D_{\bGam}=K_{\bGam}=0$. In what follows, we shall assume $K_{\bOm} \geq 0$, $D_{\bGam}\geq 0$, and $K_{\bGam} \geq0$ to be given parameters specified by the tube's physical properties. We can also choose the initial markers $s$ along the tube in such a way that $s$ becomes the arc length, thereby choosing $|\bGam_0|=1$.

\begin{remark}[On tensor properties of $K_{\bOm}$  and $K_{\bGam}$]
{\rm As we noted above in \eqref{Aeq_tilted2}, $K_{\bOm}$  and $K_{\bGam}$ in \eqref{Aeq} may be tensor quantities and in that case \eqref{Aeq} should be written as 
{
\begin{equation} 
\label{Aeq2} 
\begin{aligned} 
A(\bOm,\bGam)& = A_0 - \frac{1}{2}  \mathbb{K}_{\bOm} \left( \bOm-\bOm_0 \right) \cdot \left( \bOm-\bOm_0 \right) 
\\& \qquad\qquad \qquad -    \mathbf{D}_{\bGam} \cdot \left( \bGam-\bGam_0 \right) 
- \frac{1}{2}   \mathbb{K}_{\bGam} \left( \bGam-\bGam_0 \right) \cdot  \left( \bGam-\bGam_0 \right)   . 
\end{aligned} 
\end{equation}
We shall take for simplicity the tensors $ \mathbb{K}_{\bOm}$ $ \mathbb{K}_{\bGam}$, and vector $ \mathbf{D}_{\bGam}$ to be proportional to the identity matrix and $\mathbf{E}_1$, respectively. Physically, \eqref{Aeq2}  indicates that the cross-sectional area decreases under bending, and, provided that $ \mathbf{D}_{\bGam} \cdot \mathbf{E}_1>0$, the area decreases under stretching and increases upon compression.  We shall note that all calculations in this paper generalize in a straightforward fashion to the treatment of the more complex law  \eqref{Aeq2}. 
\\
\revision{R3Q6b}{
As an illustration, let us consider the example of a circular cylinder satisfying the property that $A |\bGam|=$const, which may be perceived as the incompressible volume available to the fluid. Under uniform axial extension $\bOm=0$ and $\bGam=\mathbf{E}_1 \Gamma$. Then, $A |\bGam| = A_0 |\bGam_0|=A_0$. Assuming $\Delta \bGam = \bGam - \bGam_0$ to be small and  $\bGam_0 = \mathbf{E}_1$, we obtain, up to the order $|\Delta \bGam|^3$
\[ 
\frac{1}{A_0} A(\bOm=0,\bGam)= |\bGam|^{-1} = 
 |\bGam_0 + \Delta \bGam|^{-1}\simeq   1- \mathbf{E}_1  \cdot  \Delta \bGam  - \frac{1}{2} |\Delta \bGam|^2  + \frac{3}{2}  \left(\mathbf{E}_1 \cdot  \Delta \bGam \right)^2. 
\] 
Comparing with \eqref{Aeq2} we see that this calculation yields 
\begin{equation} 
\label{DGam_KGam_example}
\mathbf{D}_{\bGam} = D_{\bGam} \mathbf{E}_1, \quad D_{\bGam} = A_0 \, , \quad 
\mathbb{K}_{\bGam}=A_0 
\left(
\begin{array}{ccc} 
-2 & 0 & 0  
\\ 
0 & 1 & 0 
\\ 
0 & 0 & 1
\end{array} 
\right) \, . 
\end{equation} 
}
In general, the coefficients $K_{\bGam}$, $D_{\bGam}$ and $K_{\bOm}$ will depend on the geometry of the tube and its material properties such as the Poisson ratio, linear vs nonlinear elasticity, \emph{etc}.}
}
\end{remark}

We shall use the variational derivatives of the Lagrangian $\ell$ in \eqref{Lagrangian_fluid_tube},  defined with respect to the $L ^2 $ pairing in \eqref{var_der}. These derivatives can, in our case, be computed as the partial derivatives of the integrand function $F$ as 
\begin{equation} 
\dede{\ell}{\bom}=\pp{F}{\bom}\, , \quad 
\dede{\ell}{\bgam}=\pp{F}{\bgam}\, , \quad
\dede{\ell}{\bOm}=\pp{F}{\bOm}\, , \quad
\dede{\ell}{\bGam}=\pp{F}{\bGam}\, , \quad 
\dede{\ell}{u}=\pp{F}{u}.
\label{dede_vsdd}
\end{equation} 
If the Lagrangian \eqref{Lagrangian_fluid_tube} depended on its arguments in a more complex way, such as their derivatives or integrals, then one would use the variational derivatives in the formulas below.
For later use, it is useful to write the equations of motion explicitly. 
Using \eqref{Lagrangian_fluid_tube} and \eqref{Aeq} in \eqref{full_3D}, we obtain for the derivatives 
\begin{equation} 
\begin{aligned} 
\pp{A}{\bGam} &=-D_{\bGam} \mathbf{E}_1 - K_{\bGam} (\bGam-\bGam_0),\, \quad 
\pp{A}{\bOm} =- K_{\bOm}  (\bOm - \bOm_0 )  
\\
\dede{\ell}{\bom} & = \mathbb{I}  \bom |\bGam|, \quad 
\dede{\ell}{\bgam} = \alpha  \bgam |\bGam| + \rho Q ( \bgam + u \bGam ) \\ 
\dede{\ell}{\bOm} & = -\mathbb{J} \left( \bOm - \bOm_0 \right)  |\bGam|  +\frac{1}{2} \rho \frac{\partial A}{\partial \bOm}| \boldsymbol{\gamma} + \boldsymbol{\Gamma} u| ^2 |\boldsymbol{\Gamma} |  \\ 
 \dede{\ell}{\bGam} &=  \frac{1}{2} \rho   \frac{\partial A}{\partial \bGam}   |\bgam+u \bGam|^2|\boldsymbol{\Gamma} | + \rho  Q  u \left( \bgam + \bGam u \right) - \lambda \left(\bGam - \bGam_0 \right) |\boldsymbol{\Gamma} |+ f(\bom,\bgam,\bOm,\bGam,u) \frac{\bGam}{| \boldsymbol{\Gamma} | } 
\\ 
 f&:= \frac{1}{2} \Big( \alpha | \bgam|^2 +  \mathbb{I} \bom\! \cdot\! \bom +\rho A( \boldsymbol{\Omega} , \boldsymbol{\Gamma} ) \left| \boldsymbol{\gamma} + \boldsymbol{\Gamma} u\right | ^2  
 \\
& \qquad \qquad \qquad -  \mathbb{J} (\bOm-\bOm_0) \!\cdot \! (\bOm-\bOm_0 ) -\lambda |\bGam-\bGam_0|^2 \Big)
\\ 
\dede{\ell}{u}& = \rho Q \bGam \cdot (\bgam + u \bGam )\, ,  \quad m=\frac{1}{Q} \dede{\ell}{u} = \rho \bGam \cdot (\bgam + u \bGam )\, .  
\end{aligned} 
\label{deriv_calc} 
\end{equation} 
With these expressions, it is possible to write the equations of motion \eqref{full_3D} for the particular choices of Lagrangian and shape function made above as 
\begin{equation}\label{full_3D_explicit} 
\hspace{-3mm} 
\left\lbrace
\begin{aligned}
&\displaystyle\lp \prt_t + \bom\times\rp 
\left( \mathbb{I} |\bGam| \bom \right) 
+ \rho Q  u\bgam \times  \bGam\\
 &\qquad +\lp\prt_s + \bOm\times\rp\!
\left( 
-\mathbb{J} \left( \bOm - \bOm_0 \right)  |\bGam|- K_{\bOm}  (\bOm - \bOm_0 ) \left( \frac{1}{2} \rho | \boldsymbol{\gamma} +\boldsymbol{\Gamma} u| ^2 - \mu \right)| \boldsymbol{\Gamma} |\right) 
\\
&\qquad +\bGam\times \lp  \displaystyle  \dede{\ell}{\bGam} - \mu  |\bGam| \pp{A}{\bGam}  \rp  =\mathbf{0}\\
&\displaystyle\lp \prt_t + \bom\times\rp \lp \alpha |\bGam|  \bgam + \rho Q ( \bgam + u \bGam ) \rp
\\ 
& \qquad  + \lp\prt_s + \bOm\times\rp
  \lp  \displaystyle  \dede{\ell}{\bGam} - \mu \lp |\bGam| \pp{A}{\bGam} + A \frac{\bGam}{|\bGam|}  \rp \rp=\mathbf{0}   \\
&\displaystyle\vspace{0.2cm}
\partial _t  \left(\rho \bGam \cdot (\bgam + u \bGam )  \right) 
  + \partial _s\left( \rho \bGam \cdot (\bgam + u \bGam ) u-\mu \right) =0    \\
&\displaystyle\vspace{0.2cm} \partial _t \boldsymbol{\Omega} = \boldsymbol{\Omega} \times \boldsymbol{\omega} +\partial _s  \boldsymbol{\omega}, \qquad  \partial _t \boldsymbol{\Gamma} + \boldsymbol{\omega} \times \boldsymbol{\Gamma} = \partial _s \boldsymbol{\gamma} + \boldsymbol{\Omega} \times \boldsymbol{\gamma}\\
&\displaystyle \vspace{0.2cm} \partial _t \big( A(\bOm,\bGam) |\bGam| \big) + \partial _s \big( A(\bOm,\bGam) |\bGam| u \big)=0\,
\\ 
& \dede{\ell}{\bGam}, \quad \pp{A}{\bGam}\quad  \mbox{given by \eqref{deriv_calc} }\,.
\end{aligned}\right.
\end{equation}

{These equations generalize those derived in \cite{RiPe2015} as they fully include all components of the inertia of the beam. As we show in Section~\ref{sec:stab_straight}, for a straight base configuration our equations \eqref{full_3D_explicit} reduce to the analogue of the Timoshenko beam equations with flowing fluid and changing cross-section, generalizing earlier works on the subject by other authors, as well as our previous results reported in \cite{FGBPu2014,FGBPu2015}.  }

\subsection{Helical equilibrium states} 
\begin{figure}[h]
\includegraphics[width=0.8\textwidth,angle=0]{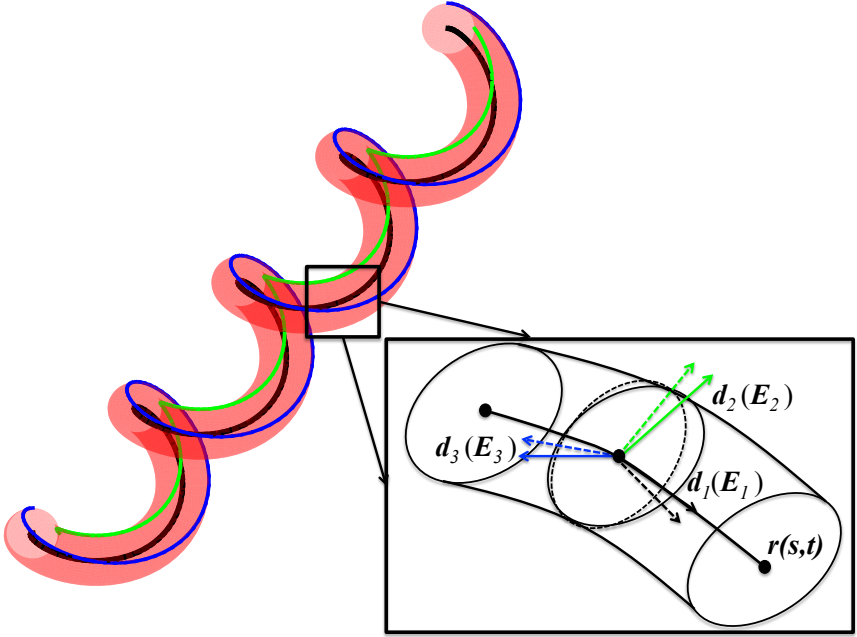}
\caption{\label{fig:setup}  The setup of a the helical tube with intrinsic coordinates indicated at several points of $s$, computed for $\bOm_0=4 \pi (1,1,1)^T/\sqrt{3}$ and $\bGam_0=(1,0,0)^T$.  At $s=0$, the basis $(\mathbf{E}_1,\mathbf{E}_2,\mathbf{E}_3)$ coincides with the spatial basis $(\mathbf{d}_1, \mathbf{d}_2, \mathbf{d}_3)$, by the choice of orientation: 
since the end $s=0$ is fixed, we can choose  $\Lambda={\rm Id}_{3 \times 3}$ at $s=0$. Also, at $s=0$,  $\mathbf{E}_1=\bGam_0$. The blue/green solid lines indicate, respectively, traces of the 'arrow tips' of the  vectors $\mathbf{d}_2$/$\mathbf{d}_3$
along the surface of the tube, while $\mathbf{d}_1=\br_s$, or $\mathbf{E}_1=\bGam_0$ is the unit tangent to the centerline of the helical tube, for all $s$.
The insert illustrates the frame vectors $\mathbf{d}_i$, $i=1,2,3$ (solid lines), measured in spatial frame, with the black color used for $\mathbf{d}_1$, green for $\mathbf{d}_2$ and blue for $\mathbf{d}_3$. These vectors are obtained from the rotation of the fixed material frame $(\mathbf{E}_1,\mathbf{E}_2,\mathbf{E}_3)$ deformed from its original position, shown by the dashed lines of the corresponding colors.  The centerline $\br(s,t)$ (solid black line)  deforms in space and time under the evolution of the elastic tube. Also shown the deformation of the cross-section from its original state (dashed line) to  its final state (solid line).
}
\end{figure}

Let us look for an helical equilibrium configuration of the tube, \textit{i.e.}, $\bom=\mathbf{0}$, $\bgam=\mathbf{0}$, $\bOm=\bOm_0$ and $\bGam=\bGam_0$. Indeed, if $\bOm_0$ and $\bGam_0$ are neither parallel nor orthogonal, then the configuration of the tube is a helix, {as is illustrated on Figure~\ref{fig:setup}.} In the degenerate cases, if these vectors are parallel, the centerline is a straight line with cross section spinning around the axis of the tube as a function of $s$. If these vectors are orthogonal, then the centerline for the tube traces out a circle in space. 
{A general condition on the Lagrangian  allowing for the existence of a helical equilibrium can be derived by computing the condition of existence of the equilibrium  solution 
\begin{equation} 
\label{helix_cond_ell} 
  (\bom,\bgam,\bOm,\bGam,u,\mu)=(\mathbf{0},\mathbf{0}, \bOm_0, \bGam_0,u_0,\mu_0 ) \, , 
\end{equation} 
where $\bOm_0$, $\bGam_0$, $u_0$ are given constants in space and time, and the constant $\mu_0$ is yet undetermined.} In this case, the fluid momentum equation, the compatibility conditions and the conservation law, \textit{i.e.}, the last three equations of \eqref{full_3D}, are satisfied identically. The angular and linear momentum equations, \textit{i.e.}, the first two equations of \eqref{full_3D}, are satisfied, provided some algebraic relations between the derivatives of $\ell$, $A(\bOm,\bGam)$ and $\mu$ at equilibrium hold. 

For the Lagrangian  \eqref{Lagrangian_fluid_tube} and shape function \eqref{Aeq}, the partial derivatives at the equilibrium are given by 
{
\[
\pp{A}{\bOm}=\mathbf{0} ,  \, 
\pp{A}{\bGam}=  - D_{ \boldsymbol{\bGam} } \boldsymbol{\Gamma}  _0 ,  \, 
\dede{\ell}{\bom}=\mathbf{0} ,  \, 
\dede{\ell}{\bgam}=\rho A_0 u_0 \bGam_0  ,  \, 
\dede{\ell}{\bOm}=\mathbf{0}   ,  \, 
\dede{\ell}{\bGam}=\rho u_0^2 \left( \frac{3}{2}  A_0  - \frac{1}{2} D_{\bGam} \right) \bGam_0 ,
\]
} {where we chose $ \boldsymbol{\Gamma} _0= \mathbf{E} _1$. 
From the definition of $Q(\bOm,\bGam)=A(\bOm,\bGam) |\bGam|$ we also conclude that at equilibrium 

\begin{equation} 
\label{Qderiv}
\pp{Q}{\bOm} =\mathbf{0} \, , \quad \mbox{however,} \quad \pp{Q}{\bGam} =\left( A_0 -D_{\bGam} \right) \bGam_0 \neq \mathbf{0}  \, . 
\end{equation} 
The angular momentum equation in \eqref{full_3D} vanishes identically. The linear momentum equation gives 
\begin{equation} 
\label{lin_mom_0} 
\bOm_0 \times \bGam_0 \left( \rho u_0^2\left( \frac{3}{2} A_0 - \frac{1}{2} D_{\bGam} \right) - \mu_0 (A_0  - D_ {\boldsymbol{\Gamma}}) \right) =\mathbf{0} \,.
\end{equation} 
If $\bOm_0$ and $\bGam_0$ are not parallel, then the equilibrium condition on $\mu_0$ is given by 
\begin{equation} 
\label{mu0}
\mu _0= \rho u_0^2\frac{ \frac{3}{2} A_0 - \frac{1}{2} D_{\bGam} }{A_0- D_{\boldsymbol{\Gamma} }}.
\end{equation}
Note that it is necessary to assume $A_0 \neq D_{\bGam}$ for the solution \eqref{mu0} to exist. For the straight equilibrium we have $ \boldsymbol{\Omega} _0=0$ so \eqref{lin_mom_0} is satisfied for any constant value of $ \mu _0$. \revision{R3Q5}{One should understand the choice of $\mu_0$ above as a selection of the Lagrange multiplier and not as the selection of the physical pressure, since the direct connection between $\mu$ and the physical pressure for the tube is still uncertain and will be undertaken in future studies. Note that the selection \eqref{mu0} guarantees that the helical shape of the tube is always preserved for any value of $u_0$. No additional approximations, such as neglecting deflections of a steady state from its equilibrium helical shape, are needed.}
\revision{R3Q15}{Physically, one may interpret  the case when $A_0=D_{\bGam}$ as corresponding to the case when the deformation of the tube's cross-section due to stretching results in precisely the right azimuthal contraction so that the volume within is conserved. Thus, when $A_0=D_{\bGam}$, there is no value of the Lagrange multiplier $\mu$ providing equilibrium helical configuration. } 

\revision{R3Q16}{\subsection{Linearization of equations of motion around the helical equilibrium} }

Let us now consider the linearization of the system of equations \eqref{full_3D} with Lagrangian \eqref{Lagrangian_fluid_tube} around the equilibrium. We take $\epsilon \ll 1$ and write
 \begin{equation} 
 \left\{
 \begin{array}{rlrl} 
 \bom &=\epsilon \bom_1 + \ldots, &\bgam&=\epsilon \bgam_1 + \ldots  \\
 \bOm &= \bOm_0 + \epsilon \bOm_1+ \ldots, & \bGam &=\bGam_0 + \epsilon \bGam_1+ \ldots\\ 
  u&=u_0 + \epsilon u_1 + \ldots ,  & \mu &= \mu_0 + \epsilon \mu_1 + \ldots 
 \end{array} 
 \right. 
 \label{linvar} 
 \end{equation} 
where by $\ldots$ we will denote terms that are order $\epsilon^2$ and higher.
Then, the partial derivatives of the Lagrangian are given by the following expansions:

\begin{equation} 
\label{linderiv}
\left\{
 \begin{array}{rll} 
\ds  \dede{\ell}{ \bom} & =\epsilon \mathbb{I} \bom_1 + \ldots & \vspace{2mm} \\ 
\ds \dede{\ell}{\bgam} &=\rho A_0 \bGam_0 u_0 + \epsilon \mathbf{N}_1 + \ldots, &
 \mathbf{N}_1:=   \left( \alpha+\rho A_0 \right) \bgam_1 -  \rho u_0 D_{\bGam} \bGam_0 \left( \bGam_0 \cdot \bGam_1 \right)  \\
& & \qquad + \rho A_0 \left(\bGam_1 u_0 + \bGam_0 u_1+ \boldsymbol{\Gamma} _0( \boldsymbol{\Gamma} _0 \cdot \boldsymbol{\Gamma} _1)u_0\right)
  \vspace{2mm}\\
\ds \dede{\ell}{\bOm} &= \epsilon \mathbf{P}_1 + \ldots ,  &
\mathbf{P}_1:=  -\left( \frac{1}{2} \rho K_{\bOm} u_0^2 \, {\rm Id}_{3 \times 3} + \mathbb{J} \right) \bOm_1 \vspace{2mm} \\ 
\ds \dede{\ell}{\bGam} &=\rho \left( \frac{3}{2} A_0  - \frac{1}{2} D_{ \boldsymbol{\Gamma} } \right)  u_0^2 \bGam_0 + \epsilon \mathbf{R}_1 +  \ldots,
&\ds \mathbf{R}_1 \mbox{ given by \eqref{Rdef} below} 
\vspace{2mm} \\ 
\ds \frac{1}{Q} \dede{\ell}{u}   &=\rho  \left( u_0 + \epsilon V_1 \right)  + \ldots, & 
V_1:= \bGam_0 \!\cdot\! \bgam_1 + u_1 + 2 u_0 \bGam_1 \!\cdot\! \bGam_0.
 \end{array} 
 \right. 
\end{equation}
For the sake of compactness of the exposition, and in order not to overburden the reader with many tedious but important technical details, we have moved most of the details of calculations into \ref{app:details}. As is derived in  \ref{app:details}, $\mathbf{R}_1$ is given by the expression  
\begin{equation} 
\begin{aligned} 
\mathbf{R} _1 =  & - \frac{1}{2} \rho   u_0 ^2 K_{ \boldsymbol{\Gamma} }\boldsymbol{\Gamma} _1 + \frac{3}{2} \rho u_0 ^2 A_0  \boldsymbol{\Gamma} _1 + 3\rho A_0 u_0 u_1 \boldsymbol{\Gamma} _0    \\
& \phantom{ \frac{1}{2} }
+\frac{3}{2}\rho A_0 u_0 ^2 (\boldsymbol{\Gamma} _0\cdot\boldsymbol{\Gamma} _1) \boldsymbol{\Gamma}  _0 + \rho A_0 u_0\boldsymbol{\gamma} _1 +\rho A_0 u_0 (\boldsymbol{\gamma} _1\cdot \boldsymbol{\Gamma} _0)\boldsymbol{\Gamma} _0 - \lambda \boldsymbol{\Gamma} _1 \\ 
& \qquad -  \rho D_{\bGam} \bGam_0 \left(  u_0 \bGam_0 \cdot \left( \bgam_1 + u_1 \bGam_0 \right)  + 3  u_0^2 \bGam_0 \cdot \bGam_1\right) .
\end{aligned} 
\label{Rdef_fin} 
\end{equation} 

\paragraph{Angular momentum equation} The linearization of the last two terms of the angular momentum conservation law is given by 
\rem{ 
\[ 
\bgam \times \dede{\ell}{\bgam} + \bGam \times \dede{\ell}{\bGam}=
\bGam \times \left[ \frac{1}{2} \pp{A}{\bGam} \rho |\bgam + u \bGam|^2 - \lambda \left(\bGam - \bGam_0\right) \right] 
\] 
which gives upon linearization about the steady helical solution 
} 
\[
\begin{aligned} 
\left( \bgam \times \dede{\ell}{\bgam}  + \bGam \times \dede{\ell}{\bGam} \right)_1 &=
 -\bGam_0 \times \bGam_1
\left( \frac{1 }{2} \left(K_{\bGam}-D_{\bGam} \right) \rho u_0^2 + \lambda
 \right)  \, . 
\end{aligned} 
\]
We also need to compute the linearization of $Q=A(\bOm,\bGam) | \bGam|$ and its derivatives  as outlined in \ref{app:details}.

The angular momentum equation vanishes identically at the order $\epsilon^0$.  The linearization of angular momentum law, \textit{i.e.}, the term proportional to $\epsilon^1$, gives, using 
\eqref{linderiv},
\begin{equation} 
 \pp{}{t} \mathbb{I} \bom_1+ \left( \pp{}{s} + \bOm_0 \times \right) \left[  -\left( \left( \frac{1}{2} \rho u_0^2 - \mu_0 \right) K_{\bOm}  \, {\rm Id}_{3 \times 3} +\mathbb{J} \right) \bOm_1  \right]  
 -S \, \bGam_0 \times \bGam_1 
 =\mathbf{0} \, ,
\label{linangmom}
\end{equation}
where we have defined the constant $S$ according to 
\begin{equation} 
\label{Sdef} 
S :=    \lambda +  (K_{\bGam}  - D_{\bGam}) \left( \frac{1 }{2}  \rho u_0^2  - \mu_0 \right)  \, .
\end{equation}
Note that this equation is valid for a helical equilibrium, in which case $ \mu _0$ is given by \eqref{mu0}, and also for the straight equilibrium, in which case $ \boldsymbol{\Omega} _0=0$ and $ \mu _0$ can take an arbitrary constant value.

\paragraph{Linear momentum equation} This equation vanishes identically at the order $\epsilon^0$ with the choice of $\mu_0$ given by \eqref{mu0},  or for arbitrary value of $ \mu _0$  if $ \boldsymbol{\Omega} _0=0$. At the first order in $\epsilon$ we get 
\begin{equation} 
\label{linmomeq}
\begin{split} 
\pp{\mathbf{N}_1 }{t} & +  \boldsymbol{\omega} _1 \times \rho A_0 \boldsymbol{\Gamma} _0 u_0 
+  \bOm_1 \times  \left(  \frac{3}{2} \rho A_0 u_0^2 -\frac{1}{2} D_{\bGam} u_0^2 -\mu_0 (A_0 - D_{\bGam}) \right)   \bGam_0 
 + \left( \pp{}{s} + \bOm_0 \times \right) \\
&  \Big( \mathbf{R}_1 +  \mu_0  \big( \left( K_{\bGam}-A_0 \right)  \bGam_1 + \left(A_0 +2  D_{\bGam} \right) \bGam_0 \left(\bGam_1 \cdot \bGam_0\right)   \big)  - \mu_1 (A_0-D_{\bGam}) \bGam_0  \Big)= \mathbf{0} \, , 
\end{split}
\end{equation}
where $\mathbf{R}_1$ is defined in \eqref{Rdef_fin}.  Note that if $ \mu _0$ is given by \eqref{mu0}, the  term multiplying $\bOm_1 \times$ vanishes. 

\paragraph{Fluid momentum equation} In order to linearize the fluid momentum equation, it is useful to compute the linearization for $m$ as 
\[ 
m_1 = \left( \frac{1}{Q} \dede{\ell}{u} \right)_1 = \rho \left( \bGam \cdot \left( \bgam+\bGam u \right) \right)_1 = \rho  \left(   \bgam_1 \cdot \bGam_0 + 
2 \bGam_1 \cdot \bGam_0 u_0 + u_1 \right)  = \rho   V_1 \, , 
\] 
where $V_1$ is defined in \eqref{linderiv}. 
Therefore, the linearization of the fluid momentum equation is obtained as 
\begin{equation} 
\label{linfluidmom} 
\pp{   \rho V_1}{t} + \pp{}{s} \left(  \rho  \left( V_1 +u_1\right) u_0  - \mu_1  \right) = 0 \, . 
\end{equation} 

\paragraph{Conservation law} The linearization of the equation $Q_t+(Qu)_s=0$ gives 
\begin{equation} 
\label{lincons} 
\pp{}{t}  \left(  (A_0-D_{\bGam}) \bGam_1 \cdot \bGam_0 \right) + \partial _s (u _1 A_0 +  u_0   (A_0-D_{\bGam})  \bGam_1 \cdot \bGam_0)=0 \, . 
\end{equation}
While one might be tempted to investigate the case $D_{\bGam}=A_0$, we remind the reader that such values of parameters are explicitly excluded by the solvability condition for 
$\mu_0$ given by \eqref{mu0}. Thus, we shall set $D_{\bGam} \neq A_0$ in the remainder of the paper.

\paragraph{Compatibility conditions} Finally, the conditions \eqref{compatibility} linearize as 
\begin{align} 
\label{lincompatibility_om} 
 &\partial _t  \bOm_1 - \bOm_0 \times \bom_1   - \partial _s   \bom_1 =0  \, , 
\\
& \partial_t \bGam_1 + \bom_1 \times \bGam_0 -\partial_s \bgam_1 - \bOm_0 \times \bgam_1 =0 \,.
\label{lincompatibility_gam} 
\end{align} 

\rem{ 
\begin{remark}[On comparison of derived equations with previous studies] 
{\rm The equations derived in this section generalize the linear equations for the two-dimensional linear stability of a straight collapsible tube derived in \cite{FGBPu2015}. In that paper, it was shown that for the initial configuration of the tube being a straight line and the motion restricted to a plane, the linearized equations of motion were equivalent to the Timoshenko beam equations for non-moving fluid. For the moving fluid and the case of collapsible tube, the linearized equations form a novel system which augment the established equations in the literature \cite{Be1961a,GrPa1966a} by having better  stability properties and ability to describe collapsible tubes. The equations derived here coincide with the equations of \cite{FGBPu2015} when $\bOm_0 \rightarrow 0$ and form a corresponding Timoshenko-beam like extension of the equations of motion for the initially curved tube conveying fluid compared to the previous literature. 
}
\end{remark}
\todo{VP: Instead of above Remark, I am inserting the new section to compare our results with the previous studies including general $\mu_0$ and $D_{\bGam}\neq 0$. I used the blue color for general corrections and magenta for new terms compared to \cite{FGBPu2015}. }

} 

\section{Stability analysis of a straight tube and comparison to previous studies}
\label{sec:stab_straight}

\subsection{Derivation of linear stability for arbitrary $\mu_0$}

For the straight equilibrium $\bOm_0=\mathbf{0}$, the stability analysis simplifies as different modes of vibrations become independent, and the dispersion relation can be written with a lower-dimensional matrix. The stability analysis of equations \eqref{full_3D} around a straight tube equilibrium with $\mu_0=0$ and the deformation of cross-section \eqref{Aeq} only depending on $\bOm$, \textit{i.e.}, $D_{\bGam}=K_{\bGam}=0$, was undertaken in \cite{FGBPu2015}. We believe that a study of the more general case including the parameters $D_{\bGam}$ and $K_{\bGam}$ for the straight tube equilibrium is also of interest, as it further elucidates the relationship of our work to previous studies. 

In order to proceed, we notice that if $\bOm_0=\mathbf{0}$, and $\mathbf{E}_3$, the direction normal to the axis centerline, is parallel to the eigenvectors of both $\mathbb{I}$ and $\mathbb{J}$, then equations \eqref{full_3D} allow for an exact  reduction to the motion in the plane $( \mathbf{E} _1, \mathbf{E} _2)$. More precisely, we write $ \mathbf{r} (s,t)=(v(s,t), w(s,t))$ and $ \Lambda (s,t)= \operatorname{exp}( \phi (s,t) \mathbf{E} _3 )$, \textit{i.e.}, $\Lambda (s,t)$ is a rotation about the axis $\mathbf{E} _3 $ perpendicular to the plane of motion.
In terms of $u, v, \phi $, the reduced variables $(\bom,\bOm,\bgam,\bGam)$ read
\begin{equation}\label{2d_variables}
\begin{aligned} 
\boldsymbol{\omega}& = \omega \mathbf{E}  _3= \dot{\phi} \mathbf{E} _3\\
\boldsymbol{\Omega}&= \Omega \mathbf{E}  _3 = \phi ' \mathbf{E}_3\\
\boldsymbol{\gamma}&= \gamma _1 \mathbf{E}_1 + \gamma _2 \mathbf{E}_2= (\dot v\cos \phi+\dot w \sin \phi) \mathbf{E}_1 +(-\dot v\sin \phi+\dot w \cos \phi) \mathbf{E} _2 \\
\boldsymbol{\Gamma}& = \Gamma _1\mathbf{E}_1 + \Gamma _2 \mathbf{E}_2=(v'\cos \phi+ w '\sin \phi) \mathbf{E}_1 +(- v'\sin \phi+ w' \cos \phi) \mathbf{E}_2.
\end{aligned}
\end{equation} 
From their definition, the reduced variables verify the compatibility conditions \eqref{compatibility} which reduce here to
\begin{equation}\label{compatibility} 
\partial _t \Omega = \partial _s \omega , \quad  \partial _t \Gamma _1 - \omega \Gamma _2 = \partial _s \gamma _1 - \Omega \gamma _2 , \quad \partial _t \Gamma _2 +\omega \Gamma _1 = \partial _s \gamma _2 + \Omega \gamma _1.
\end{equation} 
While the equations of motion were derived for an arbitrary Lagrangian, we shall now focus on exact solutions for the concrete Lagrangian given in \eqref{Lagrangian_fluid_tube}. In the two-dimensional case, this Lagrangian reduces to
\begin{equation}
\label{2D_lagr}
\ell ( \omega  , \boldsymbol{\gamma} ,\Omega , \boldsymbol{\Gamma} ,u)
= \frac{1}{2} \int_0 ^L\Big( \alpha |\bgam|^2 + I \omega ^2 + \rho A(\Omega  , \boldsymbol{\Gamma} ) |\bgam + \bGam u |^2  - J \Omega ^2 
- \lambda |\bGam- \boldsymbol{\chi} |^2 \Big) |\bGam| \mbox{d} s, 
\end{equation} 
where $I$ and $J$ are now scalars, $ \boldsymbol{\chi} = \mathbf{E} _1$, and $A(\Omega, \boldsymbol{\Gamma} )$ is given in \eqref{Aeq}. Note that the bending rigidity $J=E I_A$ is the product of Young's modulus $E$ with the corresponding moment of area $I_A$. As before, we will introduce the function $f$ such that $\ell ( \omega  , \boldsymbol{\gamma} ,\Omega , \boldsymbol{\Gamma} ,u)= \int_0^L f( \omega  , \boldsymbol{\gamma} ,\Omega , \boldsymbol{\Gamma} ,u) |\bGam| \mbox{d} s$. 
The full nonlinear equations of motion for the two-dimensional motion are then 
\begin{equation}\label{full_2D} 
\left\lbrace
\begin{aligned} 
&I \partial_t  (\dot \phi | \boldsymbol{\Gamma}  |) - \partial _s \left( (J+ \big( \frac{1}{2} \rho | \boldsymbol{\gamma}+ \boldsymbol{\Gamma} u| ^2 - \mu ) K_{ \boldsymbol{\Omega} }\big) \phi' | \boldsymbol{\Gamma}  | \right) \\
& \qquad \qquad + ( \boldsymbol{\Gamma}\cdot \mathbf{E} _2 ) \left( (D_{ \boldsymbol{\Gamma}} - K_{ \boldsymbol{\Gamma}} ) \big( \frac{1}{2} \rho | \boldsymbol{\gamma}+ \boldsymbol{\Gamma} u| ^2 - \mu \big)- \lambda \right) | \boldsymbol{\Gamma} |=0
\\
\vspace{2mm} 
&\displaystyle\lp \prt_t + \dot \phi \mathbf{E}_3 \times\rp\dede{\ell}{\bgam} + \lp\prt_s + \phi' \mathbf{E}_3 \times\rp\left( \dede{\ell}{\bGam}-\frac{\partial Q}{\partial \boldsymbol{\Gamma}  } \mu   \right)=0   \\
&\displaystyle\vspace{0.2cm}\pp{m}{t}  + \partial _s\left(mu-\mu \right) =0 \, , \quad m:=\frac{1}{Q} \dede{\ell}{u}   \\
&
\partial _t \boldsymbol{\Gamma} +\dot \phi \mathbf{E}_3  \times \boldsymbol{\Gamma} = \partial _s \boldsymbol{\gamma} + \phi' \mathbf{E}_3  \times \boldsymbol{\gamma}
, \quad 
\bgam \cdot \mathbf{E}_3=0\, , \quad \bGam \cdot \mathbf{E}_3=0
\\
&\displaystyle  Q(\Omega , \boldsymbol{\Gamma} )=(Q_0  \circ \varphi ^{-1} )(\partial _s \varphi  ^{-1} )\; \Rightarrow \; \partial _t Q+ \partial _s (Qu)=0, \quad \Omega= \phi' \, . 
\end{aligned}\right.
\end{equation}
These equations generalize the exact two-dimensional dynamics obtained in \cite{FGBPu2015} by allowing the cross-section to depend on the extension/contraction of the tube through \eqref{Aeq}. The goal of this section is to focus on the linear stability of \eqref{full_2D} while the nonlinear behavior of this system will be considered in our future work.

To illustrate the comparison of the results produced by our methods with previous works, consider the equilibrium corresponding to a straight tube 
\[
\mathbf{r} _0 (s,t)=(s,0,0), \quad  \Lambda_0 (s,t) = \mathbf{I}, \quad u(s,t)= u _0, \quad \mu _0 (s,t)=\mu_0 
\]
so $ \boldsymbol{\omega} _0 (s,t)=0$, $ \boldsymbol{\gamma} _0 (s,t)=0$, $ \boldsymbol{\Omega} _0(s,t) =0$, $ \boldsymbol{\Gamma} _0(s,t)= \mathbf{E} _1 $ with $\mu_0$ being an arbitrary parameter. We assume small deformations of the form $ \mathbf{r} _\varepsilon (s,t)=(s+ \varepsilon v(s,t), \varepsilon w(s,t),0)$ and $ \Lambda _\varepsilon (t,s)= \operatorname{exp}( \varepsilon \phi  (s,t) \widehat{\mathbf{E}} _3 )$. Here and below, just like in previous section, we have defined the motion in the $(\mathbf{E}_1, \mathbf{E}_2)$ plane, and $\operatorname{exp}( \varepsilon \phi  (s,t) \widehat{\mathbf{E}} _3 )$ is the rotation about the $\mathbf{E}_3$ axis by the angle $\phi(s,t)$.  The infinitesimal deformations are then 
\begin{align*} 
\boldsymbol{\omega} _\varepsilon (s,t)&= \Lambda _\varepsilon ^{-1} \dot \Lambda _\varepsilon = \varepsilon \dot{\phi }(s,t)\mathbf{E}  _3 \\
\boldsymbol{\gamma} _\varepsilon (s,t)&= \Lambda _\varepsilon ^{-1} \dot  {\mathbf{r}} _\varepsilon = \operatorname{exp}( -\varepsilon \phi  (s,t) \widehat{\mathbf{E} } _3 )( \varepsilon \dot v (s,t), \varepsilon \dot w (s,t),0)\\
\boldsymbol{\Omega} _ \varepsilon (s,t)&= \Lambda _\varepsilon ^{-1} \Lambda _\varepsilon'=  \varepsilon \phi'(s,t)\mathbf{E}  _3 \\
\boldsymbol{\Gamma} _ \varepsilon (s,t)&=  \Lambda _\varepsilon ^{-1} \mathbf{r} _\varepsilon '=\operatorname{exp}( -\varepsilon \phi  (s,t) \widehat{\mathbf{E} } _3 )(1+\varepsilon v'(s,t), \varepsilon w '(s,t),0).
\end{align*} 
The perturbations in the first order of $\epsilon$ are given by 
\begin{align*} 
\boldsymbol{\omega} _1 (s,t)&= \omega _1 (s,t) \mathbf{E}  _3 = \dot \phi (s,t) \mathbf{E}  _3 \\
\boldsymbol{\gamma} _1 (s,t)&= \dot v(s,t) \mathbf{E}  _1+\dot w(s,t) \mathbf{E}  _2 \\
\boldsymbol{\Omega} _1 (s,t)&= \Omega _1 (s,t)\mathbf{E}  _3 = \phi '(s,t) \mathbf{E}  _3 \\
\boldsymbol{\Gamma} _1(s,t) &=  -\phi  (s,t) \mathbf{E}  _3 \times \mathbf{E}  _1 + v'(s,t) \mathbf{E}  _1+ w'(s,t) \mathbf{E}  _2 =  v'(s,t) \mathbf{E}  _1+( -\phi(s,t)+w'(s,t)) \mathbf{E}  _2 .
\end{align*} 
Under these approximations, the linearized angular momentum equation \eqref{linangmom} becomes 
\begin{equation} 
\begin{aligned} 
 I \ddot \phi   &   -\left( \left( \frac{1}{2} \rho u_0^2 - \mu_0 \right) K_{\bOm} + J \right) \phi^{\prime \prime}
 =S ( w'-\phi) \, ,
\label{linangmom2D}
\end{aligned}
\end{equation}  
where $S$ is defined earlier in \eqref{Sdef}. 
As it turns out, for the two-dimensional motion we consider here, computation of the $\mathbf{E}_2$ component of \eqref{linmomeq} is sufficient to close the system. Multiplying that equation by  $\mathbf{E}_2$, we obtain:
\begin{equation} 
\begin{aligned} 
\partial_t & \big( (\alpha+ \rho A_0 ) \dot w + \rho A_0 u_0 (w'-\phi)  \big)   +  \rho A_0 u_0 \dot \phi\\
& \quad +\phi' \left(  \frac{1}{2} \rho u_0^2(3A_0-   D_{\bGam} ) -\mu_0 (A_0 - D_{\bGam}) \right)\\
&\quad + \partial_s \left( \rho A_0 u_0 \dot w +  (w'-\phi) \left( \frac{1}{2} \rho u_0^2(3 A_0 -   K_{\bGam} )
-\mu _0 (A_0-K_{ \boldsymbol{\Gamma} })
-\lambda
 \right)  \right)=0\,,
\end{aligned} 
\label{linmomE2_0} 
\end{equation}
which can be rewritten as
\begin{equation}\label{linmomE2}  
(\alpha+ \rho A_0 ) \ddot w +  2 \rho A_0 u_0 \dot w' =\widetilde{S}  w^{\prime \prime} -S \phi' \,,
\end{equation} 
where $S$ is defined in \eqref{Sdef} and $\widetilde{S}$ is defined by
\begin{equation}\label{tildeSdef} 
\widetilde{S} := 
\lambda-\mu _0 \left(K_{ \boldsymbol{\Gamma} }-A_0 \right)
-
\frac{1}{2} \rho u_0^2 \left( 3 A_0 -  K_{\bGam} \right) \,.
\end{equation} 
Remarkably, when expression \eqref{mu0} for the constant $\mu_0$ is used, constant $\widetilde{S}$ defined in \eqref{linmomE2}  is given by the same expression as defined in \eqref{Sdef}: $\widetilde{S}=S$. However, for a tube that is initially straight, there is no requirement on $\mu_0$ and thus it can be chosen as an arbitrary parameter. In this case we have $\widetilde{S} \neq S$, in general. In what follows, we take $\mu_0$ to be an additional parameter. We believe that the physical meaning of this parameter $\mu_0$ is pressurizing the pipe in the equilibrium position. However, caution must be taken in such physical interpretation, as $\mu$ has the meaning of the Lagrange multiplier for incompressibility condition. While the units of $\mu$ coincide with the pressure, and our derivation of contribution due to $\mu$-terms is quite similar to the derivation of the pressure contribution in the incompressible Euler equation,  the exact physical meaning of $\mu$ is yet to be determined. 
 \begin{remark}[Connection to earlier results on straight tube stability \cite{FGBPu2014,FGBPu2015}]
 {\rm 
 In the previous works by two of the authors \cite{FGBPu2014,FGBPu2015} we have set $\mu_0=0$ as a particular case of a possible choice for $\mu_0$, and only considered the tilt deformations, corresponding to the choice of $K_{\bGam}=0$ and $D_{\bGam}=0$ in \eqref{Aeq}. With that choice of parameters, equations \eqref{Sdef} and \eqref{tildeSdef} give 
 \begin{equation}
 \label{S_Stilde}
 S= \lambda\, , \quad 
 \widetilde{S}=\lambda - \frac{3}{2} \rho u_0^2 A_0 \, .
 \end{equation}
Equations \eqref{linangmom2D} and \eqref{linmomE2} with $S$ and $\widetilde{S}$ given by \eqref{S_Stilde} reduce exactly to the linearized equations (8) in \cite{FGBPu2014} and the first two equations in (6.6) in \cite{FGBPu2015}. Therefore, the considerations in this section extend the   stability analysis for a straight tube obtained before, for a more general expression for the area \eqref{Aeq} and arbitrary $\mu_0$. Even though the main focus of this paper is on the demonstration of the prowess of the method for helical tubes, we believe that extension of the linear  study for more general parameter regime, undertaken in this section, is also of interest. 
 }
 \end{remark} 
 \smallskip
 \rem{ 
Remarkably, using \eqref{mu0}, rewritten as $A_0 \left( \frac{3}{2} \rho u_0 ^2 - \mu _0 \right) = D_{ \boldsymbol{\Gamma} } \left( \frac{1}{2} \rho u_0 ^2 - \mu _0 \right) $,  $S$ can be rewritten as
\[
S= \lambda +( K_{\boldsymbol{\Gamma} }- D_{ \boldsymbol{\Gamma} }) \left( \frac{1}{2} \rho u_0 ^2 - \mu _0 \right) .
\]
\todo{VP: That is really remarkable, Francois! Well seen. This shows that two coefficients in the equations come out to be the same \emph{for all $\mu_0$} satisfying appropriate conditions. } 
} 

We shall also note that the $\mathbf{E}_1$ component of the linear momentum equation couples with the fluid momentum and the conservation law to give the equations propagation of disturbances along the tube. While these instabilities are interesting in themselves, we believe that a thorough analysis of such disturbances  will digress too much from the core goal of the paper, and will not be performed here. 

Let us turn our attention to equations \eqref{linangmom2D} and \eqref{linmomE2}. If there is no flow then $u_0=0$ and $\mu_0=0$ so the equations of motion become
\begin{equation}
\left\{
\begin{array}{l}
\vspace{0.2cm}I \partial _t ^2 \phi -J \partial _s ^2 \phi  = \lambda (\partial _s w- \phi  ),\\
(\alpha+\rho A_0)  \partial _t ^2w= \lambda \partial _s \left(  \partial _s w - \phi \right) ,
\end{array}
\right.
\label{Timoshenko_no_fluid}
\end{equation} 
which are exactly the dynamical equation for the  Timoshenko beam with $J=E I$, $E$ the Young modulus, $I$ the second moment of area of the beam, $\lambda=k A G$, with $k$ being the Timoshenko coefficient, $A$ the cross-sectional area of elastic part, and $G$ the shear modulus. For a constant fluid velocity $u_0$ and non-changing cross-section, \textit{i.e.}, $K_{\bOm}=0$, $K_{\bGam}=0$, $D_{\bGam}=0$,  but arbitrary $ \mu _0$, we have $S=\lambda$ \ and $\widetilde{S}= \lambda + \left( \mu _0- \frac{3}{2} \rho u_0 ^2\right) A_0 $ , so equations \eqref{linangmom2D} and \eqref{linmomE2} give 
\begin{equation}
\left\{
\begin{array}{l}
\vspace{0.2cm}I \partial _t ^2 \phi -J \partial _s ^2 \phi  = \lambda (\partial _s w- \phi  ) ,\\
\left( \alpha+ \rho A_0\right)  \partial _t ^2w +2 \rho A_0 u_0 w_{st}    =\widetilde{\lambda} w_{ss}  - \lambda \phi_s \, , \quad \widetilde{\lambda}:=\lambda +\left(\mu _0 - \frac{3}{2} \rho  u_0^2 \right) A_0 . 
\end{array}
\right.
\label{Timoshenko_fluid}
\end{equation} 
These equations form the analogue of Timoshenko beam equations for the tube conveying fluid at a constant velocity. For $\mu_0=0$ as taken in \cite{FGBPu2014,FGBPu2015}, $\widetilde{\lambda} = \lambda - \frac{3}{2} \rho  u_0^2A_0$. The difference between $\widetilde{\lambda}$ and $\lambda$ forms a departure from the classical Timoshenko beam theory and may lead to an interesting novel results for the stability theory to  be investigated in future studies.

\rem{
For the particular choice $\mu_0 = \textcolor{green}{\frac{1}{2} }u_0^2$, we still have $S =\lambda$
\todo{FGB: I don't see this. For the choice $\mu_0 = \frac{3}{2} u_0^2$ (which corresponds to \eqref{mu0} with $D_{\boldsymbol{\Gamma}}=0$), we don't have $S= \lambda $ in \eqref{Sdef}. \\ 
VP: My bad, I meant to write $\frac{1}{2}$ instead of $3/2$. }
and, (perhaps somewhat surprisingly), the equations describing the stability of the tube are identical to those of the tube with a constant cross-section.

For more general values of $\mu_0$, $u_0$, $K_{\bGam}$, $K_{\bOm}$ and $D_{\bGam}$, the linearized equations of motions are 
\begin{equation}
\left\{
\begin{aligned}
&I \partial _t ^2 \phi -J \partial _s ^2 \phi  =S (\partial _s w- \phi  ),\\
&(\alpha+\rho A_0)  \partial _t ^2 w +2 \rho A_0 u_0 w_{st} = S  \partial _s \left(  w_s - \phi \right)  \\ 
\end{aligned}
\right.
\label{Timoshenko_fluid_gen}
\end{equation} 
where $S$ is defined by \eqref{Sdef}. 
\todo{FGB: This is just equations \eqref{linangmom2D} and \eqref{linmomE2}, or is there something more here? Why has the term $\left( \frac{1}{2} \rho u_0^2 - \mu_0 \right) K_{\bOm} $ disappeared?}
} 
\rem{ 
We can express the equation \eqref{Timoshenko_fluid} as a single equation for the variable $w$ by applying the operator 
$I \partial_t^2 - J \partial_s^2 + \lambda$ to the second equation to obtain: 
\begin{equation} 
\begin{aligned} 
\left( I \partial_t^2 - J \partial_s^2 + \lambda \right) &\Big[ 
\left( \alpha+ \rho A_0\right) w_{tt} +2 \rho A_0 u_0 w_{st} 
 -P_0 w_{ss}  \Big] 
+ \lambda P_0 w_{ss} =0 \, . 
\end{aligned} 
\label{Timoshenko_fluid_single} 
\end{equation} 
Equation \eqref{Timoshenko_fluid_single} represents the Timoshenko beam equations for an elastic tube with a constant cross-section. 
} 
For more general values of $\mu_0$, $u_0$, $K_{\bGam}$, $K_{\bOm}$ and $D_{\bGam}$, the linearized equations of motions are given by equations \eqref{linangmom2D} and \eqref{linmomE2}. Let us compare this equation with the classic model forming the basis of previous works on the subject, see \cite{Be1961a,GrPa1966a} and the related papers, which is written in our notation as follows: 
\begin{equation} 
(\alpha + \rho A_0 ) w_{tt} + \rho A_0 u_0^2 w_{ss}+2 \rho A_0 u_0 w_{st} + J w_{ssss}=0. 
\label{Pa_lin}
\end{equation}
This model can be derived by computing the Euler-Lagrange equations for the following Lagrangian \cite{Be1961a}: 
\begin{equation} 
L(w, w_t)=\frac{1}{2} \int_0^L  \left( \alpha w_t^2 + \rho A_0 \left( w_t + w_s u_0 \right)^2 - J w_{ss}^2 \right) \mbox{d}s \, .  
\label{Benjamin_eq}
\end{equation} 
We shall note that care must be taken in treating the boundary terms while taking the variations as the mechanical system is not closed \cite{Be1961a,FGBPu2015}. 
The potential energy of the rod in \eqref{Benjamin_eq} is that of an Euler beam, not a Timoshenko beam as in \eqref{Timoshenko_fluid}, so  it is natural that \eqref{Timoshenko_fluid} is an improvement over \eqref{Pa_lin}, since the Timoshenko beam equation possesses better dispersion properties compared to the Euler beam. Indeed, as one can easily conclude from the dispersion analysis of the equation \eqref{Pa_lin}, waves of the type $w(s,t)=e^{i \varpi t - i k s}$ lead to the dispersion relation $\varpi=\varpi(k)$ that is ill-defined in the limit $k \rightarrow \infty$, as both the  phase $\varpi(k)/k$ and group $\varpi'(k)$ velocities of the waves diverge in that limit of short wavelengths. Thus, while \eqref{Pa_lin} is useful in computing the long-wave instabilities of a tube with moving fluid, one cannot hope to simulate it directly on a computer since the results will depend on the numerical implementation of the derivatives, in a similar way with the case of Euler beam. In contrast, the system\eqref{Timoshenko_fluid} has no difficulties with its dispersion relation $\varpi=\varpi(k)$, similarly with the case of the Timoshenko beam \cite{FGBPu2014,FGBPu2015}. 


Let us now turn our attention to the study of the complete system \eqref{linangmom2D} and \eqref{linmomE2} and derive a single equation for $w(s,t)$ as follows. We write this system of equations in operator form as 
\begin{equation} 
\begin{aligned} 
{\mathcal D}_1 \phi &= S w_s \, , \quad {\mathcal D}_1:= I \partial^2_t - \left( \left( \frac{1}{2} \rho u_0^2 - \mu_0 \right) K_{\bOm} + J \right) \partial^2_s + S \\ 
{\mathcal D}_2 w &=  - S \phi_s \, , \quad {\mathcal D}_2:=( \alpha+ \rho A_0 ) \partial^2_t + 2 \rho A_0 u_0 \partial^2_{st}-  \widetilde{S} \partial^2_s .
\end{aligned} 
\label{D_def} 
\end{equation}
{Using the fact that ${\mathcal D}_1$ and ${\mathcal D}_2$ have constant coefficients and commute with each other and with the spatial derivatives, 
we can  write a single equation in $w(s,t)$ as 
\begin{equation} 
{\mathcal D}_1 \mathcal{D}_2 w= -S^2 w_{ss} \, , 
\label{w_straight_tube} 
\end{equation}
where ${\mathcal D}_1$ and ${\mathcal D}_2$ are defined in \eqref{D_def} and $S$ is given in \eqref{Sdef}. Equation \eqref{w_straight_tube} extends the equation for the instability analysis for a straight tube conveying fluid to the case when tube's cross-section depends on both the bend and the stretch of the tube through \eqref{Aeq}, and arbitrary pressure $\mu_0$ inside the tube.}
{
\paragraph{Dimensionless equations and parameters} 
Let us now derive a dimensionless version of equation \eqref{w_straight_tube}. Here and below, $\overline{a}$ denotes the non-dimensionalised variable $a$. Let us choose the length scale based on the length of the tube, and time scale based on the characteristic frequency of bending motion of beam with no fluid as in \cite{RiPe2015}, which in our model is given as $T= L \sqrt{I/J}$ from \eqref{Timoshenko_no_fluid}. Then, under the substitution $s \rightarrow L \overline{s} $ and $t \rightarrow T \tau$, the dimensionless parameters of the problem are defined as 
\begin{equation} 
\begin{aligned} 
& \overline{\lambda}=\frac{L^2 \lambda}{J}
\, \quad 
\overline{u_0} = \frac{u_0 T}{L},  
\quad \left( \overline{K_{\bGam}} ,  \overline{D_{\bGam}}\right) =\frac{1}{L^2} \left(  K_{\bGam}, D_{\bGam}\right)  \, , \quad 
\overline{K_{\bOm}} = \frac{K_{\bOm}}{L^4} \, , 
 \\ 
&  \left( \overline{\mu_0}, \overline{S}, \overline{\lambda} \right) =  
\frac{L^2}{J} \left( \mu_0, S, \lambda \right)\, , \quad 
\overline{\alpha}= \frac{\alpha}{I}, \quad \overline{\rho}=\frac{\rho L^2}{I}, \quad \overline{J}=1, \quad \overline{I}=1. 
\end{aligned} 
\label{param_non_dim}
\end{equation} 
The unknown variable $\phi$ is already dimensionless, and $w$ needs to be scaled as $w = L \overline{w}$. The dimensionless version of \eqref{D_def} and \eqref{w_straight_tube} is then  
\begin{equation} 
\begin{aligned} 
\overline{{\mathcal D}}_1 \phi &= \overline{S} \overline{w}_{\overline{s}} \, , \quad {\mathcal D}_1:=  \partial^2_\tau - \left( \left( \frac{1}{2}\overline{ \rho} \overline{u_0}^2 - \overline{\mu_0} \right) \overline{K_{\bOm}} + 1 \right) \partial^2_{\overline{s}} + \overline{S} \\
\overline{{\mathcal D}}_2 \overline{w} &= -\overline{S} \phi_{\overline{s}} \, , \quad \overline{{\mathcal D}}_2:=( \overline{\alpha}+ \overline{\rho} \overline{A_0} ) \partial^2_\tau + 2 \overline{\rho} \overline{A_0} \overline{u_0} \partial^2_{\overline{s} \tau} - \overline{\widetilde{S}} \partial^2_{{\overline{s}}}
\\ 
\overline{{\mathcal D}}_1 \overline{\mathcal{D}}_2 \overline{w}&=  - \overline{S}^2 \overline{w}_{\overline{s}\overline{s}} \,.
\end{aligned} 
\label{D_def_non_dim} 
\end{equation} 
In what follows, we shall drop the bars above the variables while analyzing \eqref{D_def_non_dim} in order not to make the notation excessively complex. 
}
{\subsection{Critical velocity corresponding to the loss of stability for  equations \eqref{D_def_non_dim} }}
While the focus of this paper is on the linear stability of helical tubes, we believe it is important to outline some results of the numerical solutions of the equations \eqref{D_def_non_dim} for initially straight tubes. In this paper, in order to conform to the next Section~\ref{sec:num_sol} on helical tubes, we shall only consider boundary conditions that are fixed on both ends, \textit{i.e.}, $\phi=0$ and $w=0$ at $s=0$ and $s=1$ (we remind the reader that we use dimensionless coordinates and drop the overline above the variables to make the notation more compact). The algorithm of computation of solutions for the straight tube essentially follows the next section, albeit being substantially less algebraically complex. The solution algorithms proceeds as follows. 
\begin{enumerate} 
\item Since \eqref{D_def_non_dim} is an equation with constant coefficients, the dependence of solutions on time can only be in the form $e^{i \varpi t}$, and the dependence of solutions on $s$ can be of the form $e^{i k s}$, for some complex numbers $\varpi$ and $k$. The $s$-dependence can only break down for the multiple eigenvalue case which is a set of measure zero in parameters (albeit important for bifurcations) and can be considered separately. Thus, we can substitute $(\phi,w) = (\Phi_0,W_0) e^{i (k s - \varpi t)}$, with $\Phi_0$ and $W_0$ being constants, into \eqref{D_def_non_dim}. 
\item  Given complex numbers $\varpi$, compute the algebraic equation connecting $\varpi$ and $k$ corresponding to the vanishing of the determinant in \eqref{D_def_non_dim}, 
\begin{equation} 
\begin{aligned} 
F(\varpi,k)= &\left(- \varpi^2 + 
k^2 \left( \left( \frac{1}{2}  \rho  u_0^2 -  \mu_0 \right)  K_{\bOm} + 1 \right) + S \right) 
\\ 
& \times 
\left(- \varpi^2 \left( \alpha+\rho A_0 \right) - 2 \rho A_0 u_0 k \varpi + S k^2 \right)- S^2 k^2 =0 . 
\end{aligned} 
\label{disp_rel_straight}
\end{equation} 
Recall that we have dropped the bars on the dimensionless variables in order not to make the notation excessively complex. 
\item For a given complex number $\varpi$, the characteristic equation $F(\varpi,k)=0$ in \eqref{disp_rel_straight} is a fourth order polynomial equation in $k$ and therefore has 4 roots $k=k_j(\varpi)$, $j=1,\ldots,4$ with corresponding eigenvectors $(\Phi_0,W_0) = (\Phi^j_0,W^j_0)$.  As it turns out, the eigenvectors are never normal to the $w$-coordinate, so without loss of generality, we can set $W^j_0=1$. Alternatively, we can normalize the eigenvectors in some other way. We shall keep  eigenvectors' coefficients to be general with the understanding that a normalization should be chosen; in simulations, we have chosen $W^j_0=1$.  The general solution of the equations \eqref{D_def_non_dim} is given by 
\begin{equation} 
\phi=\sum_{j=1}^4 C_j \Phi^j_0 e^{i (k_j (\varpi) s- \varpi t )}\, , 
\quad 
w=\sum_{j=1}^4 C_j W^j_0 e^{i (k_j (\varpi) s- \varpi t) }\, , 
\label{gen_sol_straight} 
\end{equation} 
for some constants $C_j$, $j=1,\ldots, 4$. More generally,  a solution generalizing \eqref{gen_sol_straight} for any eigenvalues can be obtained by rewriting the equation \eqref{D_def} as a first-order ODE in $s$ with constant coefficients and solving that equation using matrix exponentiation. 
\item In order to conform to the analysis for helical tubes undertaken in Section~\ref{sec:num_sol} below, we only use fixed boundary conditions, when both $w$ and $\phi$ vanish at the boundaries. From the boundary conditions $\phi(0,t)=\phi(1,t)=0$, $w(0,t)=w(1,t)=0$ we obtain the condition of vanishing determinant for a non-trivial solution to exist
\begin{equation} 
\Delta(\varpi)= \left| 
\begin{array}{llll} 
\Phi_1^0 & \Phi_2^0 &\Phi_3^0 & \Phi_4^0 
\\ 
\Phi_1^0 e^{i k_1(\varpi) } & \Phi_2^0 e^{i k_2(\varpi)} &\Phi_3^0 e^{i k_3(\varpi)} & \Phi_4^0 e^{i k_4(\varpi) }
\\ 
W_1^0 & W_2^0 &W_3^0 & W_4^0 
\\ 
W_1^0 e^{i k_1(\varpi) } &W_2^0 e^{i k_2(\varpi)} &W_3^0 e^{i k_3(\varpi)} &W_4^0 e^{i k_4(\varpi) }
\end{array} 
\right| =0 \, . 
\label{zero_det_straight}
\end{equation} 
\end{enumerate}
 As it turns out, for the values of parameters we have tried, the bifurcations leading to the loss of stability occur when two real roots split away into the imaginary axis from $\varpi=0$ for $u>u_c$. 
 For $u<u_c$, all roots of the equation \eqref{zero_det_straight} are real, and for $u>u_c$ there are one, or more,  
 roots with \revision{R3Q6c}{${\rm Im}(\varpi)>0$, corresponding to the instability.  There is also another corresponding set of roots with ${\rm Im}(\varpi)<0$ which are stable.} In order to numerically study this loss of stability, and elucidate the physical nature of the pressure-like term $\mu_0$, we perform a series of simulations using the algorithm outlined above, using the following expression for $\mu_0$ 
 \begin{equation} 
 \mu_0 = \beta \rho u_0^2 \, , 
 \label{mu_cond}
 \end{equation} 
 with $\beta$ being a dimensionless parameter held constant. We choose a set of $\beta$ ranging from $-1$ to $1$, and for each $\beta$, we compute the first bifurcation value of the system \eqref{D_def_non_dim}. The results of the simulations are shown on Figure~\ref{fig:CriticalVelocity}, computed to a soft rubber tube, see  Section~\ref{sec:num_sol} for the exact values of the material parameters and dimensions. 
 \revision{R3Q19}{For calculations presented below on Figure~\ref{fig:CriticalVelocity}, the approximate values of terms in the dimensionless variables are $\rho \sim 10^7$, 
$\mu_0 \sim \rho u_0^2 K_{\bOm} \sim 10^{-2}$, $S \sim 10^4$, $\alpha \sim 6 \cdot 10^3$. While the relative change brought about by the term $\rho u_0^2 K_{\bOm}$ may appear small compared to $1$, when multiplied by $k^2$, it gives a noticable change for the critical velocity at the bifurcation.  We shall also caution that no terms should be dropped from \eqref{disp_rel_straight}, since this may destroy the variational nature of the equations and thus introduce artificial effects in the stability analysis.}

 As we see, the critical velocity increases with the increase of $\beta$. We have chosen to present the dimensional results for critical velocity, which comes out to be about $2-4$ m/s, which is a very reasonable number for the instability threshold for a soft rubber tube.  Thus, we believe, it is reasonable to think of $\mu_0$ as some kind of equilibrium pressure, having a stabilizing effect. More studies are definitely needed to elucidate the physical nature of $\mu_0$, which we will undertake in our future work, especially in view of the novel contribution to the Timoshenko-like equation we have outlined above. In spite of this question being of interest and importance, we believe that a deeper study of the stability of a straight tube for arbitrary $\mu_0$ may  distract the reader from the main point of the paper, and therefore we proceed now to the question of stability for helical tubes. 
 
 \begin{figure}[h]
\centering
\includegraphics[width=1\textwidth]{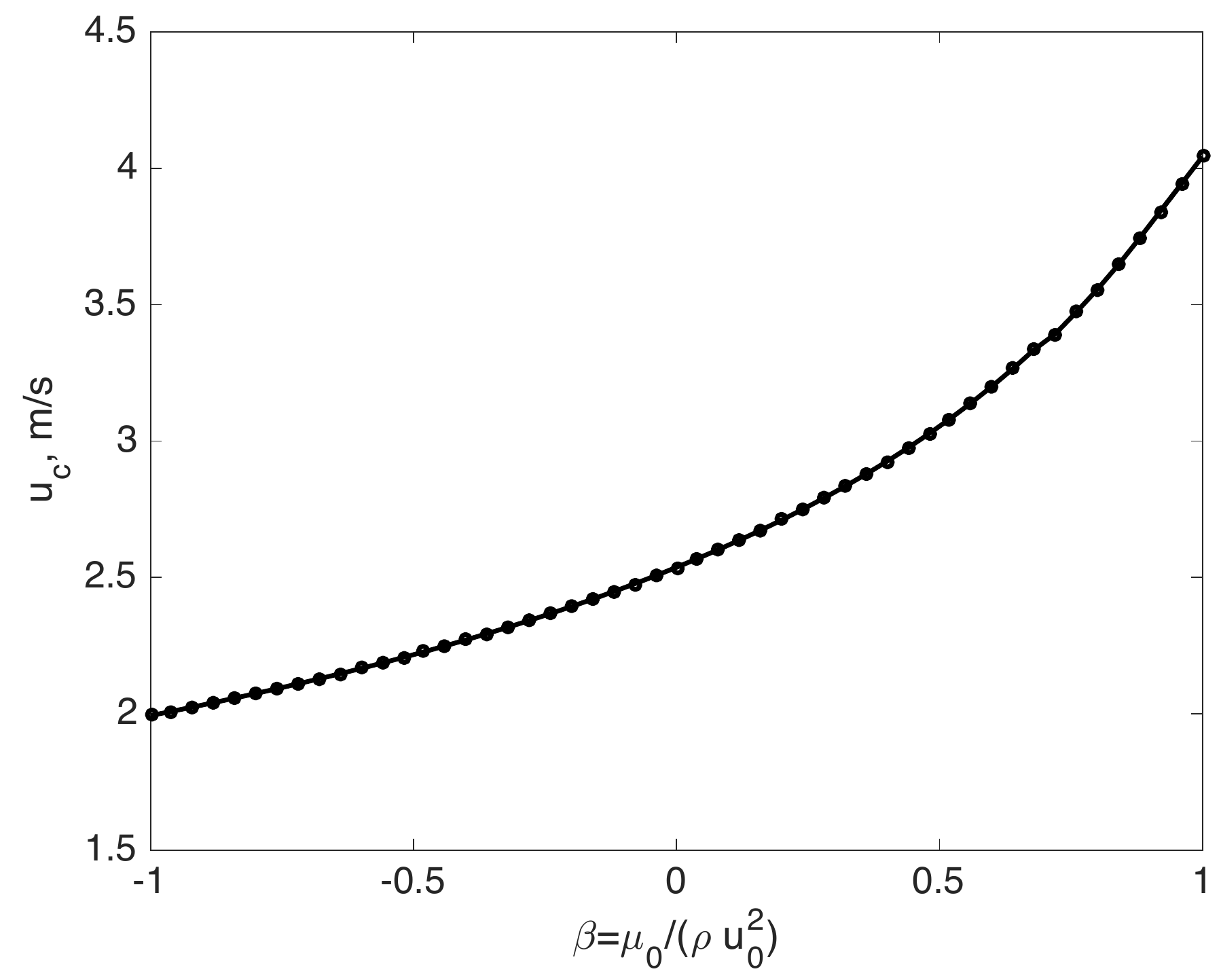}
\caption{Critical velocity (in m/s) corresponding to the loss of stability of a tube with the same characteristics as the tube analyzed below in Figure~\ref{fig:Stable_scan} and Figure~\ref{fig:Unstable_scan} for a fixed value of the parameter $\beta$ defined by \eqref{mu_cond}. The graph $u_c=u_c( \beta )$ is understood as follows: given $\beta $, the equilibrium characterized by $(u_0, \mu _0= \beta\rho u_0 ^2)$ is unstable when $u_0 >u_c(\beta )$.}
\label{fig:CriticalVelocity} 
\end{figure} 
\color{black} 
\section{Numerical solution of the stability problem  for helical tubes}
\label{sec:num_sol}
We shall now turn our attention to the numerical solution of the linear stability of the helical equilibrium. In order to make the method more clear and connect to the standard literature, we define the general solution vector $ \mathbf{Y} (s,t)$ of length 14 as
\begin{equation} 
\label{gensol} 
\mathbf{Y}=\left( \bom_1^T, \bOm_1^T,\bgam_1^T,\bGam_1^T,u_1,\mu_1 \right)^T
\end{equation} 
and formulate the linearized system in the general form as 
\begin{equation} 
\label{geneq} 
\boldsymbol{A}   \pp{\mathbf{Y}}{t} +\boldsymbol{B} \pp{ \mathbf{Y}}{s}+\boldsymbol{F}  \mathbf{Y} = \mathbf{0} ,
\end{equation} 
with $\boldsymbol{A}$, $\boldsymbol{B}$, and $\boldsymbol{C}$ being constant $14 \times 14$ matrices defined below. Equations \eqref{geneq} define a 14-dimensional system for 14 unknowns $\mathbf{Y}$. 
That system is found by assembling the equations \eqref{linangmom}, \eqref{linmomeq}, \eqref{linfluidmom}, \eqref{lincons},  \eqref{lincompatibility_om} and \eqref{lincompatibility_gam}. The ordering of the equations in \eqref{geneq} is arbitrary. For convenience, we have choosen the following ordering of equations defining \eqref{geneq}: 
\begin{center} 
\begin{tabular}{|c|c|}
\hline 
Equations in \eqref{geneq} & Originating equations 
\\
\hline
1-3 & \eqref{linangmom} 
\\
\hline 
4-6 &  \eqref{lincompatibility_om} 
\\
\hline 
7-9 &  \eqref{linmomeq}
\\
\hline 
10-12 & \eqref{lincompatibility_gam}
\\
\hline 
13 & \eqref{linfluidmom}
\\
\hline 
14 & \eqref{lincons}
\\
\hline 
\end{tabular} 
\end{center} 
With these definitions, the matrices $\boldsymbol{A} $, $\boldsymbol{B} $ and $\boldsymbol{F} $ are defined as follows. For the sake of brevity, we denote a $3\times 3$ matrix of  zeros as $\underline{0}$ and combine the equations according to the notation above. We deduce that the matrix $\boldsymbol{A} $ multiplying the time derivative in \eqref{gensol} is
\begin{equation} 
\boldsymbol{A} \! =\! 
\left( 
\begin{array}{cccccc} 
\mathbb{I} & \underline{0} & \underline{0} & \underline{0} & \mathbf{0} & \mathbf{0} 
\\
\underline{0} &{\rm Id}_{3 \times 3}  & \underline{0} & \underline{0} & \mathbf{0} & \mathbf{0} 
\\
\underline{0}  &  \underline{0}  &  \left( \alpha +  \rho A_0\right) {\rm Id}_{3 \times 3}  & A_{3,\bGam}& \rho A_0 \bGam_0 & \mathbf{0} 
\\
\underline{0} &\underline{0}  & \underline{0} & {\rm Id}_{3 \times 3}   & \mathbf{0} & \mathbf{0} 
\\
 \mathbf{0} ^T & \mathbf{0} ^T  & \rho \bGam_0^T &  2 \rho u_0 \bGam_0^T   & \rho &  0 
\\
 \mathbf{0} ^T&  \mathbf{0} ^T &  \mathbf{0} ^T  & (A_0-D_{\bGam})\bGam_0^T  &  0& 0
\end{array} 
\right) \, . 
\label{Amatr_def}
\end{equation}
\!\!The matrix $\boldsymbol{B} $ associated to the spatial derivatives is then 
{\fontsize{10pt}{12pt}\selectfont
\begin{equation} 
\boldsymbol{B}  = 
\left( 
\begin{array}{cccccc} 
\underline{0} &  -\tilde{\mathbb{J} }  & \underline{0} & \underline{0} & \mathbf{0} & \mathbf{0} 
\\
-{\rm Id}_{3 \times 3} & \underline{0}   & \underline{0} & \underline{0} & \mathbf{0} & \mathbf{0} 
\\
\underline{0} &\underline{0}  & \boldsymbol{B} _{3,\bgam} &  \boldsymbol{B} _{3,\bGam} &  \rho (3A_0- D_{\boldsymbol{\Gamma} })  u_0 \bGam_0 & (D_{\bGam}-A_0) \boldsymbol{\Gamma} _0
\\
\underline{0} &\underline{0}  & -{\rm Id}_{3 \times 3} & \underline{0} & \mathbf{0} & \mathbf{0} 
\\
\mathbf{0}^T & \mathbf{0}^T  & \rho u_0 \bGam_0^T & 2 \rho u_0^2 \bGam_0^T   &  2  \rho u_0  &  -1 
\\
\mathbf{0}^T  &\mathbf{0}^T   & \mathbf{0}^T  &  (A_0-D_{\bGam}) u_0 \bGam_0^T  & A_0 & 0 
\end{array} 
\right) \, ,
\label{Bmatr_def}
\end{equation}}
\!\!where we have defined the $3 \times 3$ matrices 
\begin{equation} 
\hspace{-5mm} 
\begin{split} 
&A_{3,\bGam}= \rho   u_0 \left( A_0 {\rm Id}_{3 \times 3} +(A_0 -D_{\bGam})  \bGam_0 \otimes \bGam_0 \right) 
\\
&\tilde{\mathbb{J}} =  \mathbb{J}+ \left( \frac{1}{2} \rho u_0^2 - \mu_0 \right) K_{\bOm}  \, {\rm Id}_{3 \times 3} \\ 
&\boldsymbol{B} _{3,\bgam} = \rho A_0 u_0 \left({\rm Id}_{3 \times 3} + \bGam_0 \otimes \bGam_0  \right)-\rho D_{\bGam} u_0 \bGam_0 \otimes \bGam_0  =A_{3,\bGam} 
 \\ 
&\boldsymbol{B} _{3,\bGam} =  
-S \cdot  {\rm Id}_{3 \times 3}   + 
\left[
\frac{3}{2} \rho u_0^2 \left( A_0- 2 D_{\bGam} \right) 
+ 
\mu_0 \left( A_0+2 D_{\bGam} \right) 
\right] 
\bGam_0 \otimes \bGam_0
\end{split}
\label{Bcoeff_def} 
\end{equation} 
with the scalar $S$ given by \eqref{Sdef} and $ \mu _0$ given by \eqref{mu0}.
Finally, the matrix $\mathbf{F}$  associated to the non-differentiated terms in $\eqref{geneq}$ is
{\fontsize{10.5pt}{11pt}\selectfont
\begin{equation} 
\boldsymbol{F}  = 
\left( 
\begin{array}{cccccc} 
\underline{0} &-\widehat{\bOm}_0 \tilde{\mathbb{J} } & \underline{0} &  -\widehat{ \boldsymbol{\Gamma} }_0 S
 & \mathbf{0} & \mathbf{0} 
\\
-\widehat{\bOm}_0 & \underline{0}   & \underline{0} & \underline{0} & \mathbf{0} & \mathbf{0} 
\\
\!\!\!\!- \rho A_0 u_0 \widehat{\bGam}_0 &\underline{0}  & \widehat{\bOm}_0\boldsymbol{B} _{3,\bgam} & \widehat{\bOm}_0 \boldsymbol{B} _{3,\bGam} & \widehat{\bOm}_0 \rho (3A_0- D_{\boldsymbol{\Gamma} }) u_0 \bGam_0 &  \widehat{\bOm}_0 (D_{ \boldsymbol{\Gamma} }- A_0)\boldsymbol{\Gamma} _0\!\!\!\!
\\
-\widehat{\bGam_0} &\underline{0} &  -\widehat{\bOm_0} & 
 \underline{0} & \mathbf{0} & \mathbf{0} 
\\
\mathbf{0}^T  &\mathbf{0}^T   &\ \mathbf{0}^T  & \mathbf{0}^T & 0  &  0 
\\
\mathbf{0}^T & \mathbf{0}^T & \mathbf{0}^T & \mathbf{0}^T   & 0 & 0
\end{array} 
\right) \, , 
\label{Fmatr_def}
\end{equation}}
where, again, we have used the scalar $S$ defined in \eqref{Sdef} 
and  used the hat map between $3$-vectors $\mathbf{a}$ and $3 \times 3$ antisymmetric matrices $\widehat{\mathbf{a}}$ introduced in Section~\ref{sec:hatmap}.
{Next, equations \eqref{geneq} can be non-dimensionalized using the rescaling of the parameters we have introduced above in \eqref{D_def_non_dim}. The length scale is still chosen to be the length of the tube, so $s=\overline{s} L$, but we need to be a bit more careful with the time scale. Since $\mathbb{J}$ and $\mathbb{I}$ are now tensors, we need to choose a characteristic value of these tensors to select the time scale $T$. For now, we assume that $\mathbf{E}_3$ is an eigenvalue direction for both of these tensors, and the characteristic time scale for the bending motion is then $T=L \sqrt{I_3/J_3}$. With that in mind, the rescaling of the variables is 
\begin{equation} 
\label{rescaling_gen} 
\overline{\bom_1} = T \bom_1 \, , \quad 
\overline{\bgam_1} = \frac{T}{L} \bgam_1\, , \quad 
\overline{\bOm_1} = L \bOm_1 \, , \quad 
\overline{\bGam_1} = \bGam_1 \, , \quad 
\overline{u_1} = \frac{T}{L} u_1 \, , \quad 
\overline{\mu_1} = \frac{L^2}{J_3} \mu_1 \, . 
\end{equation} 
Equation \eqref{geneq} becomes 
\begin{equation} 
\label{geneq_non_dim} 
\overline{\boldsymbol{A} }  \pp{\overline{\mathbf{Y}}}{\tau} +
\overline{\boldsymbol{B} } \pp{ \overline{\mathbf{Y}}}{\overline{s}}+\overline{\boldsymbol{F}  } \, \overline{\mathbf{Y}} = \mathbf{0} ,
\end{equation} 
with the matrices $\overline{\boldsymbol{A} }$, $\overline{\boldsymbol{B} }$, and $\overline{\boldsymbol{F} }$ obtained from the matrices $\mathbf{A}$, $\mathbf{B}$ and $\mathbf{F}$ by multiplying each column of the matrix by the coefficient derived from \eqref{rescaling_gen} and additionally dividing matrix $\mathbf{A}$ by $T$ and $\mathbf{B}$ by $L$. We do not present these rescaled matrices here for brevity. 
\\
The dimensionless parameters of the problem \eqref{param_non_dim} remain the same, with one correction that $\overline{\mathbb{J}}= 
\mathbb{J}/J_3$ and $\overline{\mathbb{I}}=\mathbb{I}/I_3={\rm diag}(I_1/I_3,I_2/I_3,1)$. As with \eqref{D_def_non_dim}, we shall drop the bars from the variables in the following computations as to not make the notation excessively complex, and assume that all variables are dimensionless. 
 } 

To find the dispersion relation from \eqref{geneq}, we look for solutions of the form
\begin{equation} 
\label{Y_gensol}
 \mathbf{Y}(k, \varpi ;s,t)=e^{i (k s - \varpi \tau)} \mathbf{V}_{k, \varpi }.
\end{equation} 
{Note that  equation \eqref{geneq} has constant coefficients. If we assume $\bY=\mathbf{V}(s) e^{- i \varpi t}$, then  equations \eqref{geneq} reduce to a homogeneous ordinary differential equations for $\mathbf{V}(s)$ with constant coefficients: 
\begin{equation} 
\boldsymbol{B}   \mathbf{V}'(s)+\left( \boldsymbol{F}  - i\varpi \boldsymbol{A} \right) \mathbf{V} = \mathbf{0} ,
\label{geneq_subst} 
\end{equation} 
If all roots of characteristic  equations $k_j$ obtained by substitution $\mathbf{V}(s) = \mathbf{V}_0 e^{i k s}$ into \eqref{geneq_subst} are distinct, then $\mathbf{V}(s)=\sum_j \mathbf{V}_j e^{i k_j s}$ represents the most general form of the solution. We must note that this simple form fails at the points of bifurcations when two roots of characteristic equations become equal. Because \eqref{geneq_subst}  has constant coefficients, the dispersion relation valid for arbitrary $k_j(\varpi)$ can be obtained using the fundamental solution of \eqref{geneq_subst} written as $\mathbf{V}(s) = \exp \big( -\boldsymbol{B} ^{-1} \left( \boldsymbol{F}  - i\varpi \boldsymbol{A} \right)s \big) \mathbf{V}_0$, provided the matrix $\boldsymbol{B}$ is non-degenerate. 
}
\\
{It is also interesting to remark that the helical steady state \emph{guarantees} that equation \eqref{geneq} has constant coefficients because of the symmetry with respect to rotations and translations. For any other base state, the linearization \eqref{geneq} will not be a constant coefficient equation and therefore a more general form of the solution for $\mathbf{V}(s)$ must be sought, leading to the solution of a boundary-value eigenvalue problem. This path was undertaken in \cite{RiPe2015} where the stability of several base configurations were studied. }
\\
Substitution in \eqref{geneq} gives a linear system $\left( - i \varpi \boldsymbol{A}  + i k \boldsymbol{B}  + \boldsymbol{F}  \right)  \mathbf{V} _{k, \varpi }= \mathbf{0}$. The system allows nontrivial solutions $ \mathbf{V} _{k, \varpi }$ if
\begin{equation} 
{\rm det} \left( - i \varpi \boldsymbol{A}  + i k \boldsymbol{B}  + \boldsymbol{F}  \right) =0.
\label{disp0} 
\end{equation}  
Clearly, $\boldsymbol{A} $ is degenerate, so trying to solve for $\varpi=\varpi(k)$ is difficult. On the other hand, if $\det \boldsymbol{B}  \neq 0$, the solution $k=k(\varpi)$ can be found for all values $\varpi \in \mathbb{C} $ from \eqref{disp0}.
For each $\varpi \in \mathbb{C}$, we obtain 14 (typically distinct) eigenvalues $k_j (\varpi)$ with corresponding eigenvectors $ \mathbf{V}_{j, \varpi }:= \mathbf{V} _{ k_j(\varpi ), \varpi }$, for $j=1, \ldots, 14$. 

Before we proceed, let us consider  whether it is possible for $\det \boldsymbol{B}$ to vanish. As we can see from \eqref{Bmatr_def}, 
\begin{equation} 
\det \boldsymbol{B}  = 0 \; \Leftrightarrow \; {\rm det} \tilde{\mathbb{J}} =0 \quad \mbox{or} \quad 
\left| 
\begin{array}{ccc} 
\boldsymbol{B} _{3,\bGam}  & \rho  (3  A_0 -D_{\bGam}) u_0 \bGam_0  & (D_{\bGam}-A_0)  \bGam_0 \\ 
2 \rho u_0^2 \bGam_0^T  & 2 \rho u_0 & -1 \\ 
 (A_0-D_{\bGam}) u_0 \bGam_0^T  &  A_0   & 0 
\end{array} 
\right| =0,
\label{det_calc}
\end{equation} 
where $ \boldsymbol{\Gamma} _0= (1,0,0)^T$, which is the default value of $\bGam_0$ in all our calculations.
There are two possibilities for $\det \boldsymbol{B}$ to vanish. Either 
\begin{equation} 
\det \widetilde{\mathbb{J}}=0 \quad \Leftrightarrow  \quad u_0= u_{*,i}=\sqrt{\frac{A_0-D_{\bGam}}{A_0} \frac{J_i}{\rho K_{\bOm}}} \,, \;\; \text{for some $i$}\,, 
\label{detJTildeVanish}
\end{equation} 
where $J_i$ is the $i$-th eigenvalue of $\mathbb{J}$,  or the second determinant in \eqref{det_calc} vanishes, which is computed as
\begin{equation}\label{u_star_Gam}
\begin{aligned} 
& S=0 \;\;\;\mbox{(double root)} \qquad  \text{or} \qquad 
C A_0+\rho u_0 ^2 ( D_ { \boldsymbol{\Gamma} } -A_0 )(3 A_0 + D_{\bGam}) =0, 
\\
&C:= -S+ \frac{3}{2} \rho u_0 ^2 (A_0-2D_{\boldsymbol{\Gamma} })+ \mu _0(A_0+2 D_{ \boldsymbol{\Gamma} }),
\end{aligned}
\end{equation}  
where we recall that $\mu_0$ is given by \eqref{mu0} and $S$ given by \eqref{Sdef}.
Thus, for a tube with constant cross-section, 
the effect of critical velocities described by \eqref{detJTildeVanish} and \eqref{u_star_Gam} is not present. 
As it turns out, the points $\det \boldsymbol{B}=0$ are important for the loss of stability in the sense that for all our numerical simulations, the loss of stability happened after the minimal value of $u_{*,i}$ defined in \eqref{detJTildeVanish}. 
}

We proceed by writing the general solution for a given $\varpi \in \mathbb{C}$ as 
\begin{equation}\label{Y_formula} 
\mathbf{Y}( \varpi ;s,t)= \sum_{j=1}^{14}C _j \mathbf{Y} (k_j(\varpi ),\varpi ;s,t)=e^{-i \varpi t} \sum_{j=1}^{14} C_{j}  \mathbf{V}_{j, \varpi } e^{i k_j (\varpi) s} \, . 
\end{equation} 
The value of $\varpi$ is obtained from the dispersion relation associated to the boundary conditions at $s=0$ and $s=1$. Remember that the coordinate $s$ is dimensionless so $0\leq s \leq 1$.

Let us demonstrate how to write this dispersion relation for the fixed boundary conditions at the extremities, given by prescribing the values of $ \mathbf{r} (s_0,t)$, $ \Lambda (s_0,t)$, and $u(s_0,t)$ at $s_0=0,1$ compatible with the steady helical solution. One deduces the following boundary conditions for the linearised system
 \begin{equation}
 \bom_1(0,t) =\bom_1 (1,t) =\mathbf{0}\, , \;\; \bgam_1(0,t) = \bgam_1(1,t) =\mathbf{0}, \;\; 
 u_1(0,t) = u_1(1,t) =0 \, ,  
\label{fixedBC}
\end{equation} 
which in our notation is written as
\begin{equation} 
\label{BC}
\mathbf{Y}^m( \varpi , 0,t) = \mathbf{Y}^m(  \varpi, 1,t) =0, \;\; \forall \;t, \quad \text{for }\quad  m\in J=\left\{ 1,2,3,7,8,9,13 \right\},
\end{equation} 
where $ \mathbf{Y} ^m$ denote the $m$-component of $ \mathbf{Y} $. Thus, the boundary conditions are written as the linear system 
\begin{equation} 
\begin{split}
& \sum_{j=1}^{14}   V_{j,\varpi }^m C_j  =0 \, , \quad m\in J \in \left\{ 1,2,3,7,8,9,13 \right\} \\ 
&  \sum_{j=1}^{14}   V_{j, \varpi }^m e^{i k_j(\varpi) } C_j  =0, \quad m\in J= \left\{ 1,2,3,7,8,9,13 \right\} .
\end{split} 
\label{omegaEq0} 
\end{equation}
The condition of existence of non-trivial solutions $C_j= C_j( \varpi )$, $j=1,...,14$ to \eqref{omegaEq0} can be written in terms of the determinant of a $14 \times 14$ matrix.

Alternatively, we can simplify this expression in terms of basis vectors $\mathbf{e}_j$ of $\mathbb{R}^{14}$ spanning the 14-dimensional space of boundary conditions. Indeed, defining the matrix composed of the basis vectors $\boldsymbol{U}(  \varpi  )=\left( \mathbf{V}_{1,  \varpi }, \ldots, \mathbf{V}_{14,\varpi }\right)$,
 and the  matrix $\boldsymbol{D}( \varpi ; s,t)=\exp \big(i \boldsymbol{K} (\varpi ) s\big)$, where $\boldsymbol{K}(\varpi )$ is the diagonal matrix consisting of eigenvalues $k_j(\varpi)$, we can write the solution \eqref{Y_formula} in terms of the fundamental matrix $\boldsymbol{\Phi} (\varpi ; s,t)$ as 
\begin{equation} 
\label{FundamentalMatrix}
\mathbf{Y}(\varpi ; s,t) = \boldsymbol{\Phi} (\varpi ; s,t) \mathbf{Y}(   \varpi ; 0,t)  \, , \quad 
 \boldsymbol{\Phi} (\varpi ; s,t) = \boldsymbol{U}(  \varpi ) \boldsymbol{D}( \varpi ; s,t) \boldsymbol{U}(  \varpi )^{-1} \, . 
\end{equation}
Suppose the boundary conditions are formulated as vanishing of the vector $\mathbf{Y}^m( \varpi ; s=0,1,t)$ , $m \in J$ as in \eqref{omegaEq0}. 
Let us choose the boundary conditions at $s=0$ and integrate to the right  $s=1$. One can always choose boundary conditions on the left satisfying $\mathbf{Y} ^m( \varpi ;0,t)= 0$, when $m \in J$. In order to continue the solution to  $s=1$, we need to specify  $\mathbf{Y} ^m( \varpi ;0,t)$ when $m \notin J$.  Defining the complementary set $\tilde{J}$ which in our case is $\tilde{J}=\left\{ 4,5,6,10,11,12,14 \right\}$, we denote this undetermined set of boundary conditions at $s=0$ as $\mathbf{Y} ^{\tilde{J}}( \varpi ;0,t)$.  Then, the value of $\mathbf{Y} ^m( \varpi ;1,t)$ at the right boundary is given by 
$\mathbf{Y} ^J( \varpi ;1,t)= \Phi _{J, \tilde J}( \varpi ; 1,t) \mathbf{Y} ^{\tilde J}( \varpi ;0,t)$, with the fundamental matrix $\Phi(\varpi;1,t)$ defined in  \eqref{FundamentalMatrix} and the indices $\Phi _{J, \tilde J}$ denote sub-matrix of $\Phi$ with the elements $\Phi_{k,m}$ with all $k \in J$  and  $m \in \tilde{J}$. Since the values  $\mathbf{Y} ^m( \varpi ;1,t)$ are set to vanish at the right boundary due to the boundary conditions \eqref{BC}, we have
\begin{equation}
\label{BC2} 
\mathbf{Y} ^J( \varpi ;0,t)= \mathbf{0} ,\quad  \Phi _{J, \tilde J}( \varpi ; 1,t)\mathbf{Y} ^{\tilde J}( \varpi ;0,t)= \mathbf{0} .
\end{equation} 
Since we have chosen the corresponding part of boundary conditions at $s=0$ to vanish, \textit{i.e.}, 
$\mathbf{Y} ^J( \varpi ;0,t)= \mathbf{0}$, then we need to find the complementary vector $\mathbf{Y} ^{\tilde{J}}( \varpi ;0,t)$ such that the boundary conditions at  $s=1$ are verified. A non-trivial solution for $\mathbf{Y} ^{\tilde{J}}( \varpi ;0,t)$ enforcing vanishing of the boundary conditions at  $s=1$ in \eqref{BC2} exists if and only if the corresponding determinant vanishes, \textit{i.e.}, 
\begin{equation} 
F(\varpi) :={\rm det} \big( \boldsymbol{\Phi} _{J,\tilde{J}} ( \varpi ; 1,t) \big) ={\rm det} \left( \boldsymbol{U}( \varpi )  \boldsymbol{D}( \varpi ; 1,t) \boldsymbol{U}^{-1}( \varpi ) \right)_{J,\tilde{J}}   =0 .
\label{FundMatr_det} 
\end{equation}

Equation \eqref{FundMatr_det} represents a condition for finding the complex frequency $F(\varpi)=0$ for a given set of parameters of the tube. Given a reasonable approximation to the roots at $u=0$, the roots at $u>0$ can be found, for example, by tracking the roots while $u$ is increasing from a given value. We shall note that while this method has been widely used in the literature from the earliest works  on the subject, \emph{e.g.}, \cite{GrPa1966a}, alternative methods have been used to compute the eigenvalues of the problem in its classical setting, such as the Generalized Differential Quadrature (GDQ) method  \cite{ToMaViEl2010}, allowing for direct computation of eigenvalues of $\varpi$ by defining a certain approximation matrix. However, we shall note that numerical challenges exist for GDQ method even for the stability analysis for a one-dimensional deflection for a single straight tube with no change of the cross-section, see \cite{ToMaViEl2010} for details. Our complex matrix problem \eqref{geneq} is substantially more challenging, with the most difficult conceptual issues coming from the existence of the pressure-like variable $\mu$ appearing without any time derivatives, and leading to the degeneracy of the matrix $\boldsymbol{A}$.  We shall thus use the direct method of computation of eigenvalues and postpone the study of a  possible use of GDQ method for further work. 

To be more precise, our method operates as follows. For a given value of $u_0=u_i$, $i=1, \ldots N$, we identify all the roots in a rectangle of the complex plane satisfying $| {\rm Re}(\varpi)|\leq A_R, | {\rm Im}(\varpi)|\leq A_I$ and write these roots in a data file. We then identify the evolution of each root as a function of $u$ by scanning through the data file and finding the nearby root location. This method allows, first,  to exclude the possibility of a root that is not in the initial tracked set to become unstable before the tracked set of roots do, second, an easy handling of the bifurcation points, which would need special care when tracked roots collide or get close to each other, and third, to avoid computation of the roots which are too large in absolute value so the loss of accuracy may occur.  We shall also note that there may be other roots outside the search area in the complex plane. The question of appropriate search domain needs to be considered separately using analytical estimates for roots position, something that we will undertake in our future studies.

More generally, we can formulate the following result for arbitrary set of Dirichlet-type boundary conditions for the system \eqref{geneq}. 
\begin{lemma}[On the general form of the dispersion relation]{\rm Suppose that each boundary $s=0$ and $s=1$ has exactly 7 boundary conditions $\mathbf{Y}^m=0$, with $m \in I$ at $s=0$  (inlet) and $m \in O$ at  $s=1$ (outlet). Here $I$ and $O$ are two sets of 7 integers chosen from the set $\{ 1, 2, \ldots, 14\}$.  Then, the allowed  complex frequencies $\varpi$ are the roots of the zero determinant conditions
\begin{equation} 
{\rm det} \big(\boldsymbol{\Phi} _{O,\tilde{I}} ( \varpi ; 1,t)\big)={\rm det} \left( \boldsymbol{U}( \varpi ) \boldsymbol{D}(1,t) \boldsymbol{U}^{-1} (\varpi ) \right)_{O,\tilde{I}}   =0 \, . 
\label{FundMatr_det_gen} 
\end{equation} 
In other words, the frequencies are obtained by computing the determinant of the submatrix of the fundamental matrix at $s=1$, selecting the rows corresponding to the boundary conditions at  $s=1$, and the columns corresponding to the complement of the boundary conditions at $s=0$. 
}
\end{lemma}

\begin{remark}[On unevenly posed boundary conditions]
{\rm 
We shall note that the dispersion relation \eqref{FundMatr_det_gen} is only valid when there are exactly the same number of boundary conditions (seven) specified on the left and the right. For more general boundary conditions, when there are $k \neq 7$ conditions posed at  $s=0$ and $14-k$ conditions are posed at  $s=1$ the dispersion relation is more complicated compared to \eqref{FundMatr_det_gen}, albeit it is still possible to derive it from \eqref{omegaEq0} by using appropriate tools from linear algebra. However, it is not clear to us how to assign a physical meaning to such boundary conditions, as it seems that any realistic  boundary conditions for inlet and outlet of the tube should contain exactly the same number of equations. Therefore, we do not present the consideration of such general boundary conditions in this paper. 
}
\end{remark}
Let us now illustrate how this method applies to the computation of instability of a helical tube. The material of the tube  is taken to be a soft rubber with Young's modulus of $10^7$Pa and shear modulus of $6 \cdot 10^6$ Pa. The tube's cross-section is circular with the inner radius of $R_0=10$mm and wall thickness of $1$mm, roughly corresponding to a standard medical tube. The fluid is assumed to be water with density $\rho =10^3$kg/m$^3$. The tensors $\mathbb{I}$ (inertia) and $\mathbb{J}$ are computed using standard expressions for the inertia and torsion/twist stiffness. The coefficients $K_{\bOm}$ and $K_{\bGam}$ in \eqref{Aeq} are taken to be $K_{\bOm}=0.1 R_0^2$ defining the critical bend of the tube, and $K_{\bGam}=0.1 A_0$ corresponding to the typical diminishing of the cross-sectional area by 5\% if  the tube variable $\bGam$ is increased by a factor of $2$. 
These values are also typical of commonly used medical tubes. The initial configuration of the tube is helical with $\bGam_0 = \mathbf{E}_1= (1,0,0)^T$, \textit{i.e.}, in the initial configuration the parameter $s$ represents the arclength along the helix. 

\begin{figure}[h]
\centering
\includegraphics[width=1\textwidth]{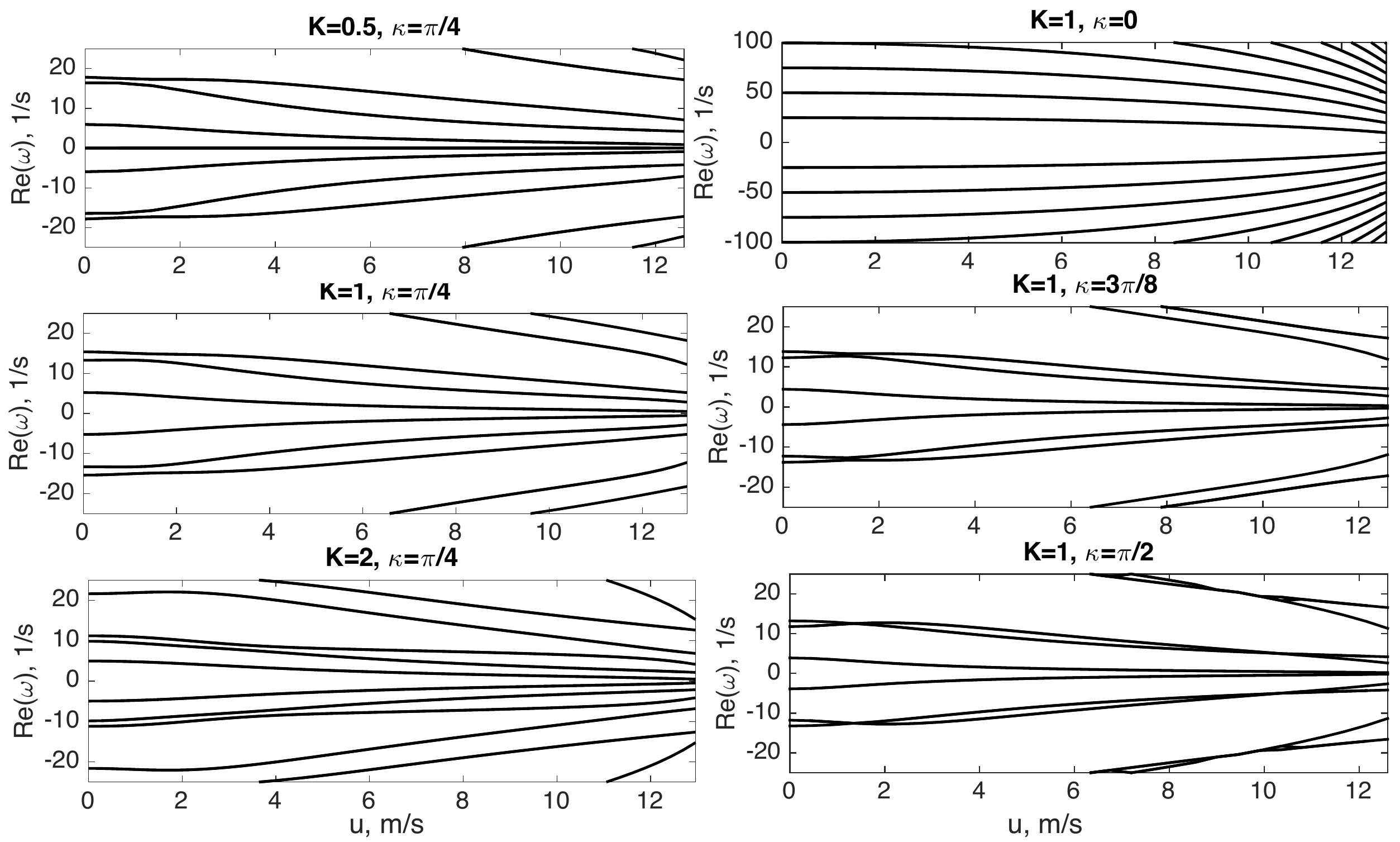}
\caption{Results of scan of eigenvalues $\varpi_i$ as a function of velocity $u_0$ for $\bOm_0=K\pi (\cos \kappa, \sin \kappa, 0)^T /L$. Left panels: fixed $\kappa=\pi/4$, varying $K=0.5$,$1$, and $2$ (top to bottom). Right panels: fixed $K=1$ and varying $\kappa=0$, $3 \pi/8$, and $\pi/2$ (top to bottom). The top right-hand picture corresponds to a straight line with rotating elastic frame. The bottom right-hand side corresponds to a half of a circle.  Only the real part of the eigenvalues is shown as the imaginary part remains identically zero, to the accuracy of the calculations, for the interval of velocity studied. 
}
\label{fig:Stable_scan}
\end{figure}

\begin{figure}[h]
\centering
\includegraphics[width=0.48\textwidth]{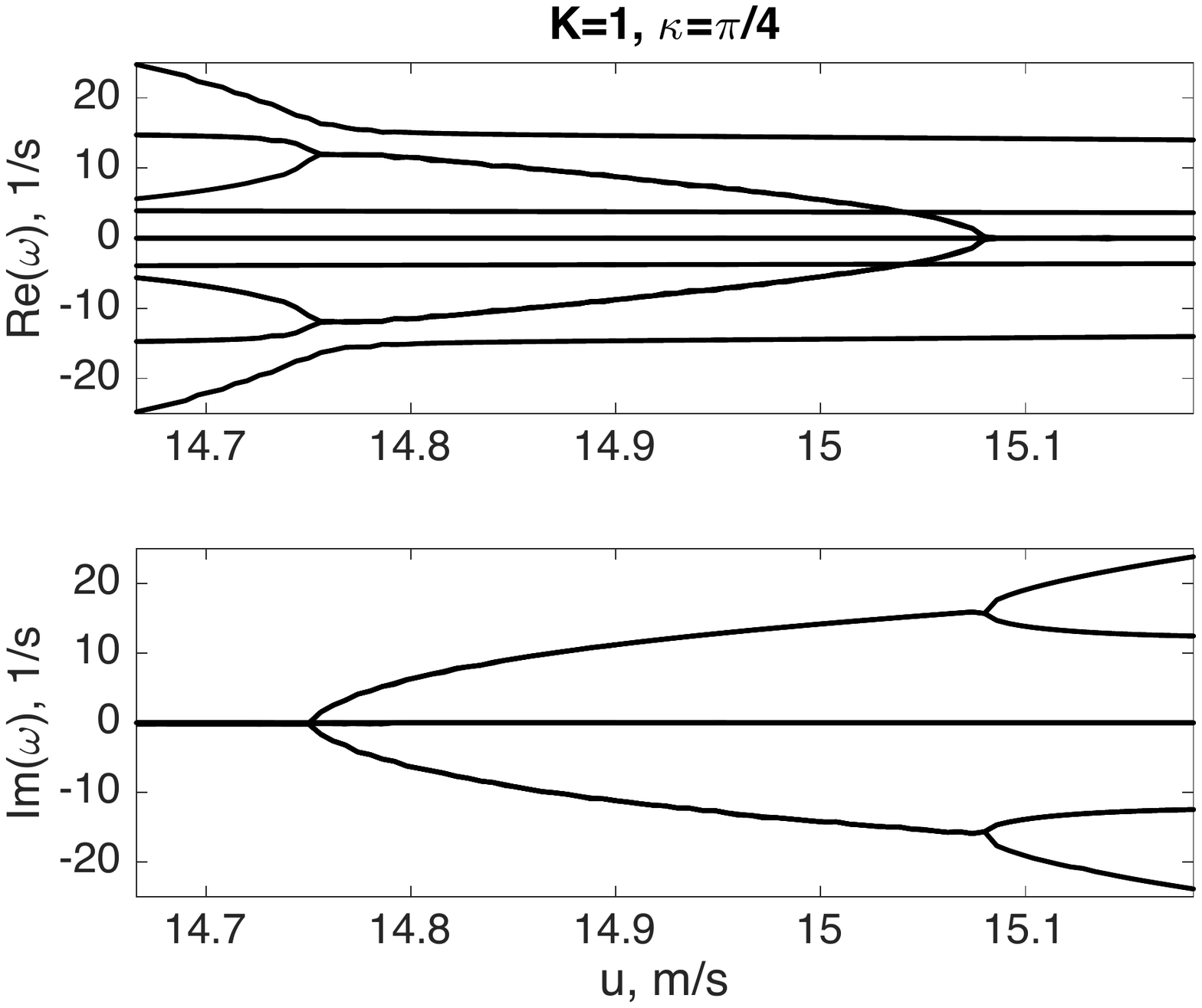}
\includegraphics[width=0.48\textwidth]{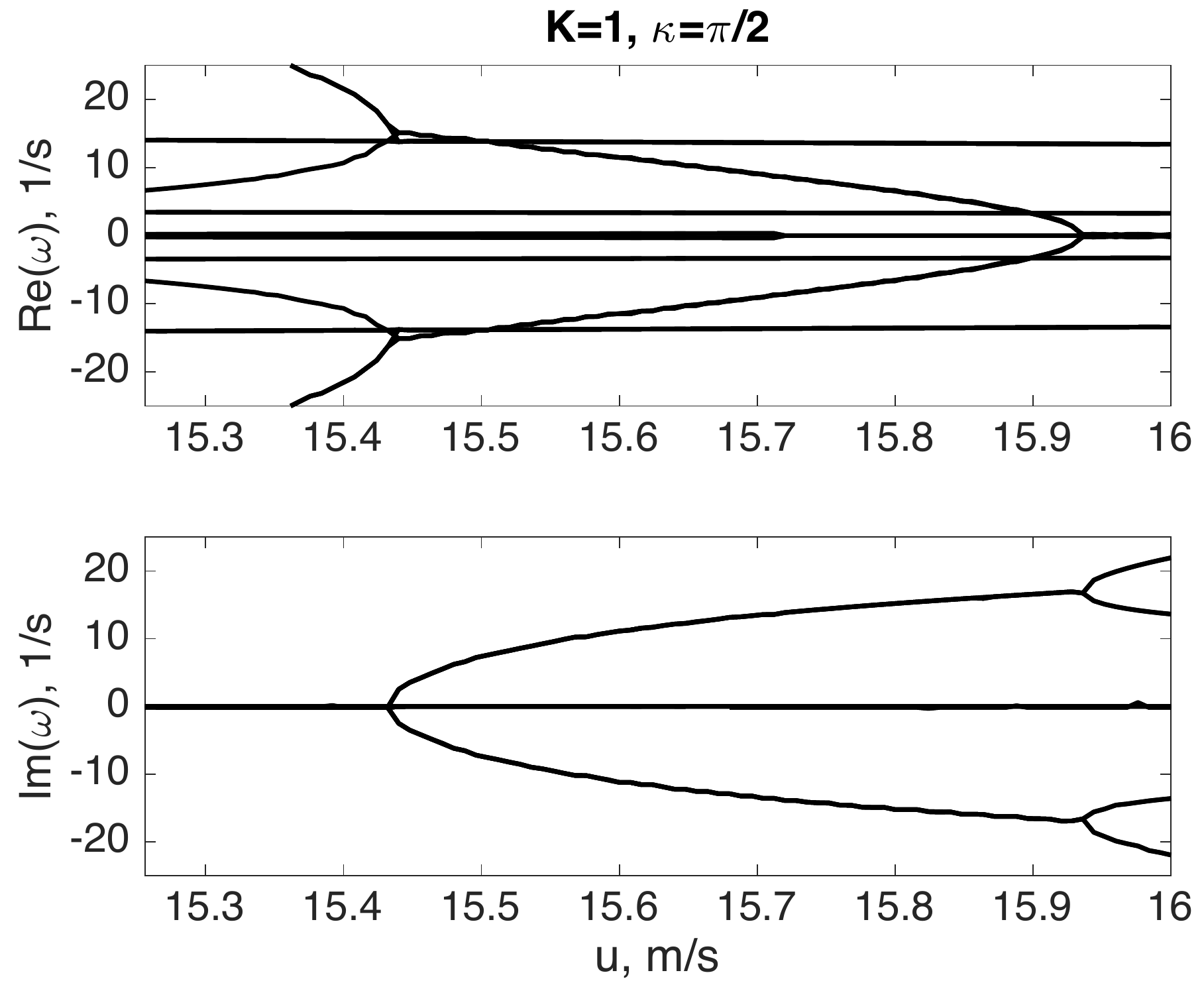}
\caption{Results of scan of eigenvalues $\varpi_i$ as a function of velocity $u_0$ for $\bOm_0=K\pi (\cos \kappa, \sin \kappa,0)^T/L$, for $K=1$, with the results for $\kappa=\pi/4$ shown on the left, and $\kappa=\pi/2$ on the right. Top panels: real part of the eigenvalues. Bottom panels: imaginary part of the eigenvalues.
\label{fig:Unstable_scan} 
}
\end{figure}

{
For each boundary $s=0$ and $s=L$, we consider  the  fixed boundary conditions 
$\bom_1=0$, $\bgam_1=0$ and $u_1=0$ on the boundary 
we discussed above, corresponding to the vanishing of $Y_j$ with $j$ belonging to the set $J=\left\{ 1,2,3,7,8,9,13\right\}$. 
} 
\rem{ 
Alternatively, we define the boundary conditions of vanishing $Y_j$ 
with $j$ belonging to the complementary set $\tilde{J}=\left\{ 4,5,6,10,11,12,14\right\}$.
Physically, this set of boundary conditions corresponds to vanishing of $\bOm_1$, $\bGam_1$ and $\mu_1$ on the boundary, so that we have $ \boldsymbol{\Omega} (t,S_f)= \boldsymbol{\Omega} _0$, $ \boldsymbol{\Gamma} (t,S_f)=\boldsymbol{\Gamma} _0$, $ \mu (t, S_f)= \mu _0$, where $S_f$ is the coordinate of the free boundary, $S_f=0$ and/or $S_f=L$. We shall call such boundary conditions the \textit{free boundary conditions}, with the reason being that they satisfy a lot of requirement for truly free boundary conditions that were considered above in \eqref{forcedef} and in much more details in \cite{FGBPu2015}. The detailed and rigorous treatment of boundary conditions is beyond the scope of the present paper and we refer the reader to \cite{FGBPu2015} for all necessary discussions.
} 
\revision{R3Q7}{We compute the eigenvalues $\varpi$ for a scan of $u$ increasing from $0$ to $ \lesssim 20$m/s. 
The instability corresponds to ${\rm Im}(\varpi)>0$.  There are many parameters to investigate, so we have chosen two essential ones related to the geometry of the helix. Since we are keeping the vector $\bGam_0=\mathbf{E}_1$ fixed, the important geometric quantities are the angle between $\bOm_0$ and $\bGam_0$, and the norm $|\bOm_0|$. Thus, we select the following parameterization: $\bOm_0=K \pi (\cos \kappa, \sin \kappa, 0)^T/L$. We have chosen a pre-factor $\pi$ in this formula so that $K=1$ corresponds to half of a rotation. Note also that for the angle $\kappa=\pi/2$ the steady configuration becomes a part of a circle since $\bOm_0 \cdot \bGam_0=0$.
On Figure~\ref{fig:Stable_scan}, we present the results of the stability diagram for a scan in helix amplitude, \emph{i.e.},  $K=0.5$, $K=1$, $K=2$ for the value $\kappa=\pi/4$ (left panel) 
and $\kappa=0$, $\kappa=3 \pi/8$, $\kappa=\pi/2$ for the value $K=0.5$ (right panel), with the last case corresponding to a half-circle. The first case $\kappa=0$ corresponds to a straight line with elastic frame that is rotating with a constant rate around the $\mathbf{E}_1$ axis. Physically, the case $\kappa=0$ can be realized for a tube that is produced by a composite material that is wound around the axis of the tube. Note also that while we use $\mu_0$ given by \eqref{mu0} for all cases shown on Figure~\ref{fig:Stable_scan}, technically speaking, for $\kappa=0$, the value of $\mu_0$ can be arbitrary since  $\bOm_0 \parallel \bGam_0$.
For the values of material parameters, equation \eqref{detJTildeVanish} gives $u_* \sim 14.1$m/s, and as long as $u \lesssim u_*$, the system is stable with ${\rm Im}(\varpi)=0$. Thus, we only show ${\rm Re}(\varpi)$ on Figure~\ref{fig:Stable_scan}.  Close to $u=u_*$, the calculation becomes challenging due to the singular nature of the matrix $\boldsymbol{B}$. 
\\
Next, on the Figure~\ref{fig:Unstable_scan}, we present two cases obtained for $u>u_*$. This case has to be computed  carefully, to avoid the numerical instabilities caused by the singularity of the matrix $\boldsymbol{B}$. As it turns out, the bifurcation structure is quite complex because of the sensitivity of the eigenvalues to parameter change close to $u=u_*$. The singularity of the matrix  $\boldsymbol{B}$ causes rapid motion of the eigenvalues, and we caution the reader that only the eigenvalues in the search area of the complex plane are presented.  The exact values for bifurcations depend on the parameters, but the nature of bifurcation for the region of parameters studies remains similar, namely, arising from the collision of eigenvalues on the real line as $u$ increases. 
In all our simulations, the system was stable up until, approximately, the minimum value of $u=u_*$ defined by \eqref{detJTildeVanish}, and lost the stability soon thereafter. The exact nature of the stability loss depends on the parameters of the helix, but in the first approximation, it seems that the minimum value defined by \eqref{detJTildeVanish} could serve as a good estimate for the critical velocity.  This  interesting fact will be studied later in more details, as it requires a thorough analysis of the singular perturbation of the equation \eqref{FundMatr_det_gen} around the value $u=u_*$ when the matrix $\boldsymbol{B}(u)$ becomes singular. 
\\
The Figure~\ref{fig:Unstable_scan} shows two representative results of eigenvalue scan for $\bOm_0 =K \pi (\cos \kappa, \sin \kappa, 0)^T/L$, for $\kappa=\pi/4$ (left) and $\kappa=\pi/2$ (right), for $K=1$. Notice that $\kappa=\pi/2$ corresponds to a piecewise circular tube, since $\bOm_0 \cdot \bGam_0=0$. }

\section{Conclusions and further studies} 
We have developed a fully three dimensional stability theory for a collapsible tube conveying fluid with a helical equilibrium configuration. A particular case of such tube is a part of the circle, arising when the vectors $\bOm_0$ and $\bGam_0$ are normal to each other. 
While the studies of instabilities of linear tubes have been quite extensive, we are not aware of any work addressing the instability of the helical, or even partially circular, tubes. We believe that this is due to the fact that it is almost impossible to derive a consistent 3D theory of helical instability from the standard approaches. On the other hand, the geometric approach of \cite{FGBPu2014,FGBPu2015} allows for a natural consideration of the stability of helical equilibria without substantial difficulty. 
In our opinion, the geometric theory presented here provided a much more natural and straightforward path to the description of the linear stability. 

Concerning the results presented in this paper, we consider the bifurcation structure of eigenvalues shown in  Figures~\ref{fig:Stable_scan} \& \ref{fig:Unstable_scan} to be highly interesting. 
Furthermore, of particular interest to subsequent studies is the dependence of the results on the parameters $\bGam_0$ and $\bOm_0$. For normalization purposes, $s$ can be chosen to be initial arclength so $|\bGam_0|=1$. For tubes made out of uniform materials, one can generally take $\bGam=\mathbf{E}_1$, then because of the invariance of the problem with respect to reflections, translations and rotations, the relevant parameters are $|\bOm_0|$ and the angle between the vectors $\bGam_0$ and $\bOm_0$. 

{More generally, our theory allows to compute linear stability of tubes smoothly connected at the joints, for example, a U-shaped tubes consisting of a semi-circle jointed, at the ends, by two pieces of a straight line. The centerline for such shape remains smooth, and the cross-section does not change at the joint, while the curvature changes abruptly at the joint. More generally, arbitrary smooth connection of helical tubes may be treated by this method as well. In this case, the linear stability will be a generalization of the boundary conditions \eqref{FundMatr_det_gen} with the perturbations in nearby sections coupled due to appropriate continuity relations. This problem may have additional complications, such as the nature of the elastic juncture itself, and the way the fluid transitions from the straight to circle line in the above example, which may affect the internal flow of fluid and the boundary conditions for before/after juncture transitions.  These considerations are beyond the scope of present paper. In addition, very little analytic progress can be done in the case of flows with a juncture even in the framework of physical approximations employed here, and we will postpone the studies of this type of problems for further investigations.
}

{In conclusion, we see the methods developed in this paper as an essential step forward towards treating more difficult problems in fluid-structure interaction. For the treatment of increasingly complex problems, such as changing cross-sections considered here, varicose instability of walls with the tube's radius or cross-sectional shape having its own dynamics, flow of compressible gas and split in tubes, variational methods are unparalleled in the their ability in incorporating the most complex interactions.  As we have seen,  variational methods provide an exact balance of torques and forces by definition and so yield a shortcut that can circumvent the difficulty in accounting for all terms in the force/torque balance by direct calculation.  More complex problems involving non-potential forces such as friction will require a combination of variational and force balance laws, and present an interesting challenge for future research.    
}

\section*{Acknowledgements} 
We gratefully acknowledge useful discussions with Mitchell Canham,  Darryl Holm,  Tudor Ratiu, Stephan Llevellyn Smith, and Cesare Tronci. FGB is partially supported by the ANR project GEOMFLUID 14-CE23-0002-01. DG acknowledges the support of RFFI (the Russian Foundation for Fundamental Research) grant 15-01-00848a. VP acknowledges  support from NSERC Discovery Grant and the University of Alberta Centennial Fund.

{\footnotesize 
\bibliographystyle{unsrt}
\bibliography{Garden-hoses}
}

\appendix 


\section{Equivalence of exact geometric and Cosserat rod equations}
\label{app:rod}

It is interesting to compare our results to the classical case of the purely elastic rod, particularly in terms of the available conservation laws. We shall use the notation of  \cite{DiLiMa1996} for forces and torques for easy comparison. For simplicity, we now assume that $s$ is the arclength in order to avoid extra multipliers of $\big| \bGam(s) \big|$ in the expressions. Our approach closely follows that of \cite{ElGBHoPuRa2010} to which we refer the reader for details. It was also demonstrated in \cite{ElGBHoPuRa2010} that the presence of non-local forces, \emph{e.g.}, electrostatic charges, can be incorporated into the force balance, which is hard to achieve \emph{a priori} with the force and torque balance approach. 

The equations of motion for exact geometric rod with no fluid motion \eqref{exact_rod_deriv}, written explicitly, read 
\begin{equation}\label{full_3D_rod}  
\left\lbrace\begin{array}{l}
\displaystyle\vspace{0.2cm}\lp \prt_t + \bom\times\rp\dede{\ell}{\bom}+\bgam\times\dede{\ell}{\bgam} +\lp\prt_s + \bOm\times\rp\! \dede{\ell}{\bOm} +\bGam\times \dede{\ell}{\bGam} =0\\
\displaystyle\vspace{0.2cm}\lp \prt_t + \bom\times\rp\dede{\ell}{\bgam} + \lp\prt_s + \bOm\times\rp \dede{\ell}{\bGam}=0  
\\
\displaystyle\vspace{0.2cm} \partial _t \boldsymbol{\Omega} = \boldsymbol{\Omega} \times \boldsymbol{\omega} +\partial _s  \boldsymbol{\omega}, \qquad  \partial _t \boldsymbol{\Gamma} + \boldsymbol{\omega} \times \boldsymbol{\Gamma} = \partial _s \boldsymbol{\gamma} + \boldsymbol{\Omega} \times \boldsymbol{\gamma}\,.
\end{array}\right.
\end{equation}

In the Cosserat rod approach, the linear momentum and angular momentum equations  are computed with respect to an orthonormal frame $\Lambda(s,t) =\left\{ \mathbf{d}_1,  \mathbf{d}_2,  \mathbf{d}_3 \right\} \in SO(3)$ that evolves with the rod. The transformation of momenta $(\boldsymbol{\pi}, \mathbf{p})$ and torques and forces $\lp  \mathbf{m} \, , \, \mathbf{n} \rp$ in Cosserat rod equations to our coordinates is computed as 
\begin{equation}
\begin{aligned}
\lp  \boldsymbol {\pi} \, , \, \mathbf{p} \rp
=
\lp
\Lambda \frac{\d \ell}{\d \bom} + \br \times \pp{\ell}{\bgam} \, , \,
\Lambda \frac{\d \ell}{\d \bgam}
\rp
\, ,
\\ 
\lp  \mathbf{m} \, , \, \mathbf{n} \rp=
\lp
\Lambda \frac{\d \ell}{\d \bom} + \br \times \pp{\ell}{\bgam} \, , \,
\Lambda \frac{\d \ell}{\d \bgam}
\rp
\, .
\end{aligned} 
\label{Mom_Torque_transform}
\end{equation}
One can simply guess the transformation formulas \eqref{Mom_Torque_transform} from general geometric ideas about transformation of vectors. There is also a more consistent way to compute these formulas based on coadjoint actions which we outline in \ref{app:Ad_actions} below. 
The balance of linear and angular momenta in the Cosserat approach gives (cf. equations (2.5.7) and (2.5.8) of \cite{DiLiMa1996})
\begin{align}
&\dot{\mathbf{p}} + \mathbf{n}'=\mathbf{f}
\,,
\label{linmom}
\\
&\dot{\boldsymbol{\pi}} + \mathbf{m}'+\br' \times \mathbf{n}
=\mathbf{T}
\,,
\label{angmom}
\end{align}
where $\mathbf{f}$ and $\mathbf{T}$ are external momenta and torques, respectively. {Similar transformation maps the equations of motion \eqref{full_3D} to the Cosserat rod equations with a moving fluid \emph{with a constant cross-section} described in \cite{BeGoTa2010,RiPe2015}. }

Our equations \eqref{full_3D_rod} are obtained by substituting \eqref{MomKirch} and \eqref{TorqKirch} into \eqref{linmom} and \eqref{angmom}, respectively, and computing the derivatives of the ${\rm Ad}^*$ terms. The potential part of external forces $\mathbf{f}$ and torques $\mathbf{T}$, if it exists, enters as the appropriate derivative of the Lagrangian $\pp{L}{\br}$; the non-potential, \emph{e.g.}, friction, forces can be added using the Lagrange-d'Alembert approach for non-conservative forces. We refer the reader to \cite{ElGBHoPuRa2010} for details.

\section{General formulas for momentum transformation in arbitrary coordinates} 
\label{app:Ad_actions}
In order to transfer the momenta from the stationary frame to the moving frame, one might be tempted to use the traditional approach of dyadic vector transformations. However, in our opinion, such approach is highly cumbersome and can easily lead to an error. In the simplest case, as we mentioned above, one can simply guess the transformation formulas. In general, for a more complex cases, such explicit guess may not be possible. Fortunately, there is a way to compute the transformation formulas in a consistent and well-defined way using the modern language of adjoint and coadjoint operators as follows. 

The momenta $\pp{L}{\dot \Lambda}$ and $\pp{L}{\dot \br}$ are covectors defined in the cotangent space at the point $( \Lambda , \br)$ of the configuration space which is the Lie group or rotations and translations $SE(3)$. The reduced momenta $\pp{\ell}{\bom}$ and $\pp{\ell}{\bgam}$ are in the cotangent space to the unity of $SE(3)$. The tangent space to the identity element of $SE(3)$ is its Lie algebra 
$\mse(3)$ and so the reduced momenta $\pp{\ell}{\bom}$ and $\pp{\ell}{\bgam}$ are in the dual of Lie algebra $\mse(3)$, denoted $\mse(3)^*$.   

Each element of the Lie group $SE(3)$ consists of rotations and translations $(\Lambda,\br)$, with the multiplication law 
\[ 
(\Lambda_1, \br_1) \cdot (\Lambda_2, \br_2) = (\Lambda_1 \Lambda_2 , \Lambda_1 \br_2+\br_1) 
\]  
and identity element $({\rm Id}_{3 \times 3}, \mathbf{0})$. An element in the Lie algebra of this group is thus a pair of two $\mathbb{R}^3$ vectors defined as $(\bom,\bgam)=\big( (\Lambda^T \dot \Lambda)^\vee, \Lambda^T \dot \br  \big)$. 

The adjoint action for any Lie group is defined as follows \cite{Ho2008}. Take a Lie group $G$ and two elements $g,h \in G$ and consider the conjugation in the Lie group defined by
\begin{equation}
{\rm AD}_g h := g h g^{-1} \, . 
\label{AD_def}
\end{equation} 
The Lie algebra $ \mathfrak{g}  $ of $G$ is defined as the tangent space to $G$ at the identity element $e$ of $G$.
If we now take a smooth curve $h(t)$ in $G$, with $h(0)=e$ (the identity element) and $\dot h(0)=a \in \mathfrak{g}  $,  and differentiate \eqref{AD_def}, we get the definition of the adjoint action 
\begin{equation} 
{\rm Ad}_g a =  \left. \frac{ \mbox{d} }{\mbox{d} t} \left( g\, h(t)\, g ^{-1} \right) \right|_{t=0}\,,
\label{Ad_def} 
\end{equation} 
valid for all elements $a$ of the Lie algebra $ \mathfrak{g}$. The coadjoint action is defined using \eqref{Ad_def} and a pairing (scalar product) between elements of the Lie algebra $a \in \mathfrak{g}$ and elements of its dual $\alpha \in \mathfrak{g}^*$ as 
\begin{equation} 
\big< a, {\rm Ad}_g^* \alpha \big> = \big< {\rm Ad}_g a, \alpha \big>\,,  \quad \mbox{for any} \quad a \in \mathfrak{g},\; \alpha \in \mathfrak{g}^*.
\label{Ad_star_def} 
\end{equation} 
Using \eqref{AD_def}, \eqref{Ad_def} and \eqref{Ad_star_def}, we can compute the conjugation, adjoint, and coadjoint actions for the group $G=SE(3)$ of rotations and translations \cite{Ho2008}: 
\begin{equation}
\begin{aligned} 
{\rm AD}_{(\Lambda,\br} (A,\mathbf{v})  &= \big( \Lambda A \Lambda^{-1}, \br + \Lambda \mathbf{v} -  \Lambda A \Lambda^{-1} \mathbf{r} \big) 
\\ 
{\rm Ad}_{(\Lambda,\br)^{-1}} (\omega, \bgam) &= \big( \Lambda^{-1} \omega \Lambda, \Lambda^{-1} \bgam + \Lambda^{-1} \omega \bgam 
\big) 
\\
{\rm Ad}^*_{(\Lambda,\br)^{-1}} (\boldsymbol{\mu}, \boldsymbol{\beta}) &= \big( \Lambda \boldsymbol \mu + \br \times \Lambda \boldsymbol \beta , \Lambda \beta \big).
\end{aligned} 
\label{Ad_SE3}
\end{equation} 
The transformation of vectors (velocities) is governed by the adjoint action ${\rm Ad}$, and covectors, such as forces and torques, by the coadjoint action ${\rm Ad}^*$ in \eqref{Ad_SE3}. 
The momenta $(\delta \ell/\d \bom,\,\delta \ell/\d \bgam) \in \mathfrak{se}(3) ^\ast $ arising from the variational principle and the Cosserat 
angular and linear momenta 
$(\boldsymbol {\pi},\,\mathbf{p})$ are connected by
\begin{equation}
\lp  \boldsymbol {\pi} \, , \, \mathbf{p} \rp
=
{\rm Ad^*}_{(\Lambda, \br)^{-1}}
\lp
\frac{\d \ell}{\d \bom} \, , \,
\frac{\d \ell}{\d \bgam}
\rp
\, ,
\label{MomKirch}
\end{equation}
where ${\rm Ad}^*$ is given by the last equation of \eqref{Ad_SE3}. 
Similarly, the internal torques and forces $\mathbf{n}$ and $\mathbf{m}$ are connected to ours as
\begin{equation}
\lp  \mathbf{m} \, , \, \mathbf{n} \rp
=
{\rm Ad^*}_{( \Lambda , \br) ^{-1} }
\lp
\frac{\d \ell}{\d \bOm} \, , \,
\frac{\d \ell}{\d \bGam}
\rp
\,.
\label{TorqKirch}
\end{equation}
Equations \eqref{MomKirch} and \eqref{TorqKirch} together with formulas \eqref{Ad_SE3} give exactly \eqref{Mom_Torque_transform}.

\section{Details of derivations of the linearized equations for helical tubes} 
\label{app:details} 

\paragraph{1) Derivation of the vector $\mathbf{R}_1$}
In order to obtain $ \mathbf{R}  _1 $, we consider the Lagrangian density $f$ in \eqref{Lagrangian_fluid_tube}, \textit{i.e.}, 
\[
f\!:=\!\frac{1}{2} \! \left(  \alpha | \bgam|^2 +  \mathbb{I} \bom\! \cdot\! \bom +\rho A( \boldsymbol{\Omega} , \boldsymbol{\Gamma} ) \left| \boldsymbol{\gamma} + \boldsymbol{\Gamma} u\right | ^2-  \mathbb{J} (\bOm-\bOm_0) \!\cdot \! (\bOm-\bOm_0 ) 
-\lambda |\bGam-\bGam_0|^2 \right)\,,
\]
with linearisation $f=f_0+ \epsilon f_1 + \ldots $ computed as 
\[
\begin{aligned} 
f_0&=\frac{1}{2}  \rho A_0 u_0^2, \\ 
f_1 &= \frac{1}{2} \rho A_0 \left( |\bgam+ \bGam u|^2\right)_1 -\frac{1}{2} \rho u_0^2 D_{\bGam}    \boldsymbol{\Gamma} _0\cdot\boldsymbol{\Gamma} _1\\
&= \rho A_0 \left( \bgam_1 \cdot \bGam_0 u_0 + \bGam_0 \cdot \bGam_1 u_0^2 + u_0 u_1 \right) -\frac{1}{2} \rho u_0^2 D_{\bGam}   \boldsymbol{\Gamma} _0\cdot\boldsymbol{\Gamma} _1 \, ,
 \end{aligned} 
\]
where we have used the simplified notation $(a)_1$ to denote the linearization of the variable $a$ around the equilibrium solution.
The vector $ \mathbf{R}  _1 $ reads
\begin{align} 
\label{Rdef} 
\mathbf{R}_1 &:= -\left( \frac{1}{2}  K_{\bGam} \rho u_0^2 + \lambda \right) \bGam_1   +
\rho A_0 \left( \bgam_1 u_0 + \bGam_1 u_0^2 + 2 \bGam_0 u_0 u_1 \right) \nonumber \\ 
& \qquad   -\frac{1}{2} \rho D_{\bGam} \mathbf{E}_1  \left( |\bgam+ \bGam u |^2 |\bGam| \right)_1   +\rho A_0 u_0 ^2 (\boldsymbol{\Gamma} _0\cdot \boldsymbol{\Gamma} _1)\boldsymbol{\Gamma} _0 +f_0 \left( \frac{\bGam}{|\bGam|} \right)_1 + f_1 \bGam_0 \, . 
\end{align}
After some rather tedious calculations using $| \bGam_0|=1$, the linearisation
\begin{equation} 
\label{Gam_lin}
\left( \frac{\bGam}{|\bGam|} \right)_1= \bGam_1 - \bGam_0 \left(\bGam_1 \cdot \bGam_0\right)
\end{equation}
and
\begin{equation} 
\left( |\bgam+ \bGam u |^2 |\bGam| \right)_1 = 2 u_0 \bGam_0\cdot   \left( \bgam_1 + u_1 \bGam_0 + u_0 \bGam_1 \right) + u_0^2 \bGam_0 \cdot \bGam_1 \, , 
\label{term_2} 
\end{equation} 
the equality $\bGam_0 = \mathbf{E}_1$, and the expressions for $f_0$ and $f_1$, we get the expression \eqref{Rdef_fin}. 

\paragraph{2) Derivation of the linearization of angular momentum equation} 
In the derivation of \eqref{linangmom}, we need the following linearization: 
\begin{equation} 
\label{linQderiv}
\left\{
\begin{array}{rl} 
\ds \pp{Q}{\bOm} &= \ds - \epsilon K_{\bOm} \bOm_1 + ... \vspace{2mm} \\ 
\ds \pp{Q}{\bGam} &= \ds \left( A_0 - D_{\bGam}  \right) \bGam_0 +  \epsilon \left( -K_{\bGam} \bGam_1 + A_0 \left( \frac{\bGam}{|\bGam|} \right)_1 
-D_{\bGam} \mathbf{E}_1 |\bGam|_1  + A_1 \boldsymbol{\Gamma} _0 \right) + ...  \, ,
\end{array} 
\right. 
\end{equation} 
where $A_1= - D_{ \boldsymbol{\Gamma} } \mathbf{E} _1\cdot \boldsymbol{\Gamma} _1$  and use $\bGam_0= \mathbf{E}_1$ to get  
\[ 
\left( \pp{Q}{\bGam}\right)_1 =   \left( A_0  -K_{\bGam} \right) \bGam_1 - \left( A_0+ 2 D_{\bGam} \right)  \bGam_0 \left(\bGam_1 \cdot \bGam_0\right) \, . 
\]
From these results, it is also useful to calculate the following quantities: 
\[ 
\left( \dede{\ell}{\bOm} - \mu \pp{Q}{\bOm}\right)_1 = \mathbf{P}_1 +   \mu_0  K_{\bOm} \bOm_1 
\] 
where $\mathbf{P}_1$ and $\mathbf{R}_1$ are defined in \eqref{linderiv} and \eqref{Rdef},
together with 
\begin{align*} 
\left( \bGam \times   \pp{Q}{\bGam}\right)_1 &= \bGam_0 \times \left( \pp{Q}{\bGam}\right)_1 
+ \bGam_1 \times \left( \pp{Q}{\bGam} \right)_0= 
\bGam_0 \times \bGam_1 \left[ \left( A_0 - K_{\bGam} \right) - \left( A_0 - D_{\bGam} \right) \right]\\
&= 
\bGam_0 \times \bGam_1 \left( D_{\bGam} - K_{\bGam} \right)\,,
\end{align*} 
\[ 
\left( \dede{\ell}{\bGam} - \mu \pp{Q}{\bGam}\right)_1 = \mathbf{R}_1 +  \mu_0  \left[  \left( K_{\bGam}-A_0 \right)  \bGam_1 + \left(A_0 +2  D_{\bGam} \right) \bGam_0 \left(\bGam_1 \cdot \bGam_0\right)   \right]- \mu_1 (A_0-D_{\bGam})  \bGam_0 \, , 
\] 
and
\[ 
\begin{aligned}
- \left( \mu \bGam \times \pp{Q}{\bGam} \right)_1 &= - \mu_1 \underbrace{\bGam_0 \times \left(  \pp{Q}{\bGam} \right)_0}_{=0} 
- \mu_0  \left(\bGam \times \pp{Q}{\bGam} \right)_1 = \mu_0 \left( \bGam_0 \times \bGam_1 \right)  \left( K_{\bGam} -D_{\bGam} \right)\,.
\end{aligned} 
\] 
\end{document}